\documentclass[useAMS,usenatbib]{mn2e}
\usepackage{color}
\usepackage{array} 
\usepackage{amsmath}
\usepackage{graphicx}
\usepackage{epstopdf}
\usepackage[labelfont=bf,labelsep=period,justification=raggedright,singlelinecheck=false,font=small]{caption}
\usepackage{wrapfig}
\usepackage{float}
\usepackage{amssymb}
\usepackage[normalem]{ulem}
\usepackage{hyperref}
\usepackage{undertilde}
\usepackage[T1]{fontenc}
\usepackage[utf8]{inputenc}
\usepackage[export]{adjustbox}
\usepackage[colorinlistoftodos]{todonotes}

\usepackage{scalerel}
\usepackage{tikz}
\usetikzlibrary{svg.path}

\usepackage{ctable}

\definecolor{orcidlogocol}{HTML}{A6CE39}
\tikzset{
  orcidlogo/.pic={
    \fill[orcidlogocol] svg{M256,128c0,70.7-57.3,128-128,128C57.3,256,0,198.7,0,128C0,57.3,57.3,0,128,0C198.7,0,256,57.3,256,128z};
    \fill[white] svg{M86.3,186.2H70.9V79.1h15.4v48.4V186.2z}
                 svg{M108.9,79.1h41.6c39.6,0,57,28.3,57,53.6c0,27.5-21.5,53.6-56.8,53.6h-41.8V79.1z M124.3,172.4h24.5c34.9,0,42.9-26.5,42.9-39.7c0-21.5-13.7-39.7-43.7-39.7h-23.7V172.4z}
                 svg{M88.7,56.8c0,5.5-4.5,10.1-10.1,10.1c-5.6,0-10.1-4.6-10.1-10.1c0-5.6,4.5-10.1,10.1-10.1C84.2,46.7,88.7,51.3,88.7,56.8z};
  }
}

\newcommand\orcidicon[1]{\href{https://orcid.org/#1}{\mbox{\scalerel*{
\begin{tikzpicture}[yscale=-1,transform shape]
\pic{orcidlogo};
\end{tikzpicture}
}{|}}}}

\def\apj{ApJ}
\def\aap{A\& A}
\def\apjl{ApJL}
\def\apjs{ApJS}
\def\mnras{MNRAS}
\def\jcap{JCAP}

\def\aj{AJ}
\def\araa{ARA\& A}
\def\nat{Nature}
\def\prd{Phys. Rev. D}

\def\na{New Astron}

\def\physrep{Phys. Rep.}
\def\bain{Bull. astr. insts Netherlds}
\newcommand\altaffilmark[1]{$^{#1}$}
\newcommand\altaffiltext[1]{$^{#1}$}

\newcommand{\msun}{{\rm M}_{\odot}}

\newcommand{\cmb}{{\sc cmb}}
\newcommand{\eor}{\hbox{\sc eor}}
\newcommand{\mini}{mini-halos}
\newcommand{\rt}{{\sc rt}}
\renewcommand{\rhd}{{\sc rhd}}
\newcommand{\lls}{{\sc lls}}
\newcommand{\dla}{{\sc dla}}
\newcommand{\pdf}{{\sc pdf}}
\newcommand{\ihm}{{\hbox{\sc ihm}}}
\newcommand{\relhics}{{\sc RElHIcs}}
\newcommand{\Ifront}{\hbox{I-front}}
\newcommand{\swift}{{\sc swift}}
\newcommand{\sph}{{\sc sph-m1rt}}
\newcommand{\rsl}{{\sc rsl}}
\newcommand{\vsl}{{\sc vsl}}

\restylefloat{table}
\begin{document}

\title[Mini-halo Radiation Hydrodynamics]{The impact and response  of 
\mini\ and the inter-halo medium on cosmic reionization}

\author[T. K. Chan et al.]
  {Tsang Keung ~Chan\altaffilmark{1,2,3}\thanks{Email: (TKC)tkchan@phy.cuhk.edu.hk}$^{\orcidicon{0000-0003-2544-054X}}$, Alejandro Ben{\'i}tez-Llambay\altaffilmark{3,4}$^{\orcidicon{0000-0001-8261-2796}}$, Tom Theuns\altaffilmark{3}$^{\orcidicon{0000-0002-3790-9520}}$,\newauthor Carlos Frenk\altaffilmark{3}$^{\orcidicon{0000-0002-2338-716X}}$, Richard Bower\altaffilmark{3}$^{\orcidicon{0000-0002-5215-6010}}$ \\
  \altaffiltext{1}{ Department of Physics, The Chinese University of Hong Kong,
Shatin, Hong Kong, China}\\
  \altaffiltext{2}{Department of Astronomy and Astrophysics, the University of Chicago, Chicago, IL60637, USA}\\
  \altaffiltext{3}{Institute for Computational Cosmology, Department of Physics, Durham University, South Road, Durham DH1 3LE, UK}\\ 
  \altaffiltext{4}{Dipartimento di Fisica G. Occhialini, Universit\`a degli Studi di Milano Bicocca, Piazza della Scienza, 3 I-20126 Milano MI, Italy}\\
}

\maketitle

\begin{abstract}
An ionization front (\Ifront) that propagates through an inhomogeneous medium is slowed down by self-shielding and recombinations. We perform cosmological radiation hydrodynamics simulations of the \Ifront\ propagation during the epoch of cosmic reionization. The simulations resolve gas in \mini\ (halo mass $10^4\lesssim M_h[{\rm M}_\odot]\lesssim 10^8)$ that could dominate recombinations, in a computational volume that is large enough to sample the abundance of such halos. The numerical resolution is sufficient (gas-particle mass $\sim 20{\rm M}_\odot$, spatial resolution $< 0.1\;{\rm ckpc}$) to allow accurate modelling of the hydrodynamic response of gas to photo-heating. We quantify the photo-evaporation time of \mini\ as a function of $M_h$ and its dependence on the photo-ionization rate, $\Gamma_{-12}$, and the redshift of reionization, $z_i$. The recombination rate can be enhanced over that of a uniform medium by a factor $\sim 10-20$ early on. The peak value increases with $\Gamma_{-12}$ and decreases with $z_i$, due to the enhanced contribution from mini-halos. The clumping factor, $c_r$, decreases to a factor of a few at $\sim 100\;{\rm Myr}$ after the passage of the \Ifront\ when the \mini\  have been photo-evaporated; this asymptotic value depends only weakly on $\Gamma_{-12}$. Recombinations increase the required number of photons per baryon to reionize the Universe by 20-100~per cent, with the higher value occurring when $\Gamma_{-12}$ is high and $z_i$ is low. We complement the numerical simulations with simple analytical models for the evaporation rate and the inverse Str\"omgren layer. The study also demonstrates the proficiency and potential of \sph\ to address astrophysical problems in high-resolution cosmological simulations.
\end{abstract}

\begin{keywords}
cosmology: theory --- dark ages, reionization, first stars --- large-scale structure of Universe --- intergalactic medium --- radiative transfer
\end{keywords}

\label{firstpage}

\section{Introduction}
\label{sec:introduction}

The epoch of reionization (\eor) refers to the time during which luminous sources (e.g. hot stars) transformed the inter-halo medium (hereafter \ihm)\footnote{ We use the term ``inter-halo medium'' for the medium outside of halos, rather than the more conventional \lq inter-{\em galactic} medium\rq. This is because 
our small high-redshift simulation volume contains few galaxies but many halos.}  from mostly neutral to mostly ionized (see reviews by, e.g. \citealt{Loeb01, Madau17} or \citealt{Wise19}). Following cosmic recombination at redshift $z\sim 1100$, the Universe remained neutral up to redshifts $z\sim 15-30$ when the first stars formed and emitted ionizing photons \citep{Tegm97firstobj,Abel02firststar,Brom02firststar}. These stars and the first galaxies ionized their immediate surroundings. With more and brighter galaxies forming, cumulatively more than one ionizing photon per hydrogen was emitted, causing these cosmological H{\sc II} regions \citep{Shap87cosHII} to percolate \citep{Gnedin00,Bark04ionbubble,Furl04ionbubble,Ilie06ionbubble}, signalling the end of the \eor. After reionization, the ionizing photons emitted by galaxies and quasars keep the \ihm\ highly ionized, with ionizations approximately balanced by recombinations in higher density regions
\citep{Haar12UVbackground,Bolt07extreion}.

Observations of the polarization of the cosmic microwave background (\cmb) due to Thomson scattering place the midpoint of the \eor\ (where about half the \ihm\ is highly ionized) at around $z\sim 8$ \citep{Planck2018CMB}. This is consistent with the evolution of the fraction of galaxies that are detected in 
Lyman-$\alpha$ emission \citep{Maso18reion}, and constraints on the neutral fraction in the \ihm\ inferred from the spectra of high-redshift quasars
\citep{Fan06EoR,Mortlock11,Bouw15emiss}. Measurements of the patchy kinetic Sunyaev-Zeldovich effect in the \cmb\ place limits on the duration of the \eor\ \citep{Zahn12kSZreion,Geor15kSZreion}, with \cite{Planck16} setting an upper limit to the width of the reionization period of $\Delta z\lesssim 2.8$. 

Improving constraints on the \eor\ is a major science driver for various observational projects, including {\sc lofar} \citep[e.g.][]{Greig21}, {\sc jwst} \cite[e.g.][]{Robertson21}, and in the near future, the {\sc ska} \cite[e.g.][]{Koopman15}. On the theory side, several groups have modelled the \eor\ with simulations in cosmological volumes (with linear extents\footnote{The c in cMpc is used to indicate that this is a co-moving size; we will use pMpc to indicate proper distances.} $\sim 10-100 \;{\rm cMpc}$, e.g. \citealt{Finl11VETgadget2,Ilie14reionvol,Gned14CROC,Ocvi16CoDa,Pawl17Aurora,Rosd18SPHINX,Dous19SCORCH,Kann21THESAN}). These simulations use \lq sub-grid\rq\ physics to model unresolved processes, which may introduce degeneracies in the interpretation.

There are three main challenges to our understanding of the \eor: ({\em i}) determining the rate at which sources emit ionizing photons, ({\em ii}) computing the escape fraction, $f_{\rm esc}$, of ionizing photons that contribute to ionizing the \ihm\; (rather than being absorbed in the immediate surroundings of where they were produced), and ({\em iii}) understanding the nature and evolution of photon sinks, where ionized gas recombines again. 

The precise nature of the dominant ionizing sources remains controversial. These could be stars in very faint galaxies below the detection limit of the Hubble Space Telescope deep fields (with 1500\AA\ magnitude $M_{\rm UV}\gtrsim-13$; \citealt{Robe13faintreion,Fink19lowescreion}). Another possibility is slightly brighter galaxies that have higher values of $f_{\rm esc}$ because they drive strong winds \citep{Sharma16, Shar17starburstreion, Naidu20}. {\sc jwst} promises to measure the slope of the faint-end luminosity function, which could constrain the contribution of faint galaxies to the emissivity. However, it will remain challenging for observations to constrain $f_{\rm esc}$ directly, although the UV-slope and the dominance of nebular lines in the spectra of galaxies may be good proxies for $f_{\rm esc}$ \cite[e.g.][]{Chrisholm18, Chrisholm22}.

In this paper, we focus on Challenge ({\rm iii}): understanding the nature and evolution of photon sinks. Clumpy gaseous structures impede reionization by boosting the opacity (often described in terms of the mean free path of ionizing photons) and the recombination rate (which consumes ionizing photons) of the \ihm.

The recombination rate in an inhomogeneous medium relative to that of a uniform medium is characterised by the \lq clumping factor\rq, $c_{l,{\rm all}}\sim \langle n_{\rm HII}^2\rangle/\langle n_{\rm H}\rangle^2$, where $\langle\rangle$ denotes a volume average and $n_{\rm HII}$ and $n_{\rm H}$ are the density of ionized and total gas. Calculating the clumping factor accurately is challenging. Its numerical value is uncertain, ranging from 2 to 30 (e.g. \citealt{Gned97reionclumping,Trac07rtreion,Ilie07srreion,Pawl09clumping,McQu11LLS,Raic11clumping,Finl12clumping,Shul12clump}), depending on its definition and the simulation outcome.

To further exacerbate the issue, a significant number of recombinations occurs in gas inside the smallest gravitationally bound structures: \mini. These are low-mass halos, $10^4\lesssim M_h/\msun\lesssim 10^8$, not massive enough to form a galaxy (without metals or molecules) but able to retain their cosmic share of baryons before reionization. 

\cite{Haim01minihalo} claimed that \mini\ boost the number of photons that are required to reionize the universe by a factor of $\sim 10$ (although they assumed \mini\ are optically thin). The semi-analytical model of \cite{Ilie05SSSIfront} and the sub-grid model of \cite{Ciar06mhreion} also suggest that \mini\ can boost reionization photon budget, based on high-resolution 2D radiation hydrodynamical (\rhd) simulations \citep{Shap04minihalo}. \cite{Embe13clumping} performed a suite of high-resolution cosmological hydrodynamics simulations that resolve all \mini\ with at least 100 particles. Post-processing these simulations with radiative transfer (hereafter \rt), they obtained
values of $c_{l,{\rm all}}\sim 10$, confirming the major impact of small-scale structures on reionization. But \cite{Embe13clumping} might have overestimated the clumping factor by not including photo-heating and photo-evaporation.

\cite{Park16clumping} performed cosmological \rhd\ simulations with photo-evaporation. These calculations resolve \mini, and model the ionizing background as a roughly uniform radiation field with Gadget-RT. But the simulated volumes are relatively small and may not capture the full mass range of \mini. \cite{DAlo20clumping} performed \rhd\ simulations in a simulation volume with linear extent $\sim 1 \;{\rm cMpc}$, using a uniform mesh with cell size $\sim 1\;{\rm ckpc}$ (which might be larger than the virial radius of low-mass \mini). Both \cite{Park16clumping} and \cite{DAlo20clumping} concluded that $c_{l,{\rm all}}\sim 10$ during the early stages ($\ll 100 {\rm Myr}$) of reionization, but then drops rapidly to lower values $\sim 2$, due to photo-evaporation. But these studies do not explicitly study {\it mini-halo photo-evaporation} and {\it its impact on reionization}, which we investigate here.

Improving upon previous studies, we perform high-resolution (dark matter particle mass $\sim100\;\msun$ and adaptive spatial resolution $<0.1\;{\rm ckpc}$) cosmological \rhd\ simulations in volumes of linear extent $\sim 1 \;{\rm cMpc}$ to investigate the response of gas in small-scale structures on the passage of an ionization front and its impact on reionization.

The simulations are performed with the cosmological simulation code {\sc swift} (\citealt{Scha18SWIFTascl}; \citealt{Scha23SWIFT})
\footnote{\url{https://www.swiftsim.com}} with smoothed particle hydrodynamics ({\sc sph}; \citealt{Lucy77SPH,Ging77SPH}). We model radiation hydrodynamics with \sph, a novel {\sc sph} two-moment method with local Eddington tensor closure \citep{Chan21SPHM1RT}. The simulated volume is large enough to sample the full mass range of \mini\ (see \S\ref{sec:suite}) and the resolution is high enough to resolve the small \mini\footnote{For example, a $10^6\;\msun$ minihalo has a virial radius around 0.3 pkpc (or 2 ckpc at $z\sim6$), compared to our gravitational softening length $\sim 0.1 \;{\rm ckpc}$.}. 

With this simulation suite, we investigate the minihalo photo-evaporation process and its dependence on redshift and photo-ionization rate. We study how high-density gas can stall the ionization front due to self-shielding. Secondly, we perform a detailed analysis of how the inhomogeneous medium impedes reionization, including the clumping factor, the number of recombination photons, and the role of \mini . We extend upon \cite{Chan23IAUS}, which presented only the clumping factor evolution of one of our representative simulations.

This paper is organized as follows. In \S\ref{sec:analytic}, we describe the stages of reionization. We introduce our simulation suite in \S\ref{sec:method}. The analysis of the simulations is described in \S\ref{sec:results}, focusing on the impact of an ionization front on gas in cosmic filaments and mini-halos. We use these results to gauge the impact of these structures on the progress of reionization (in terms of recombinations and the clumping factor). We put our results in context and discuss them in \S\ref{sec:discussion}. Finally, we conclude with a summary and an outlook for future research. A set of appendices provide the details of our numerical implementation and convergence tests.

This is a long paper, and we recommend the reader to start by
skimming through the following sections. First, the definition and discussion around mini-halos in \S\ref{sec:halocat} (Fig. \ref{fig:difm_z}) and the stages of reionization in the beginning of \S\ref{sec:overview} (Fig. \ref{fig:illustration}). The main results are Figs. \ref{fig:Htot_evap} and \ref{fig:tevmulti} in \S\ref{sec:mhevaporation}, which quantify the rate at which mini-halos photo-evaporate. Finally, the impact of small-scale structures on reionization is presented in \S\ref{sec:clumping} (Figs. \ref{fig:clumping}, \ref{fig:Recom_mbins}, and \ref{fig:recom}).

\section{Overview and theoretical estimates}
\label{sec:analytic}
\subsection{Halos and their role in reionization}
\label{sec:halocat}

\begin{figure}
\includegraphics[width={0.95\columnwidth}]{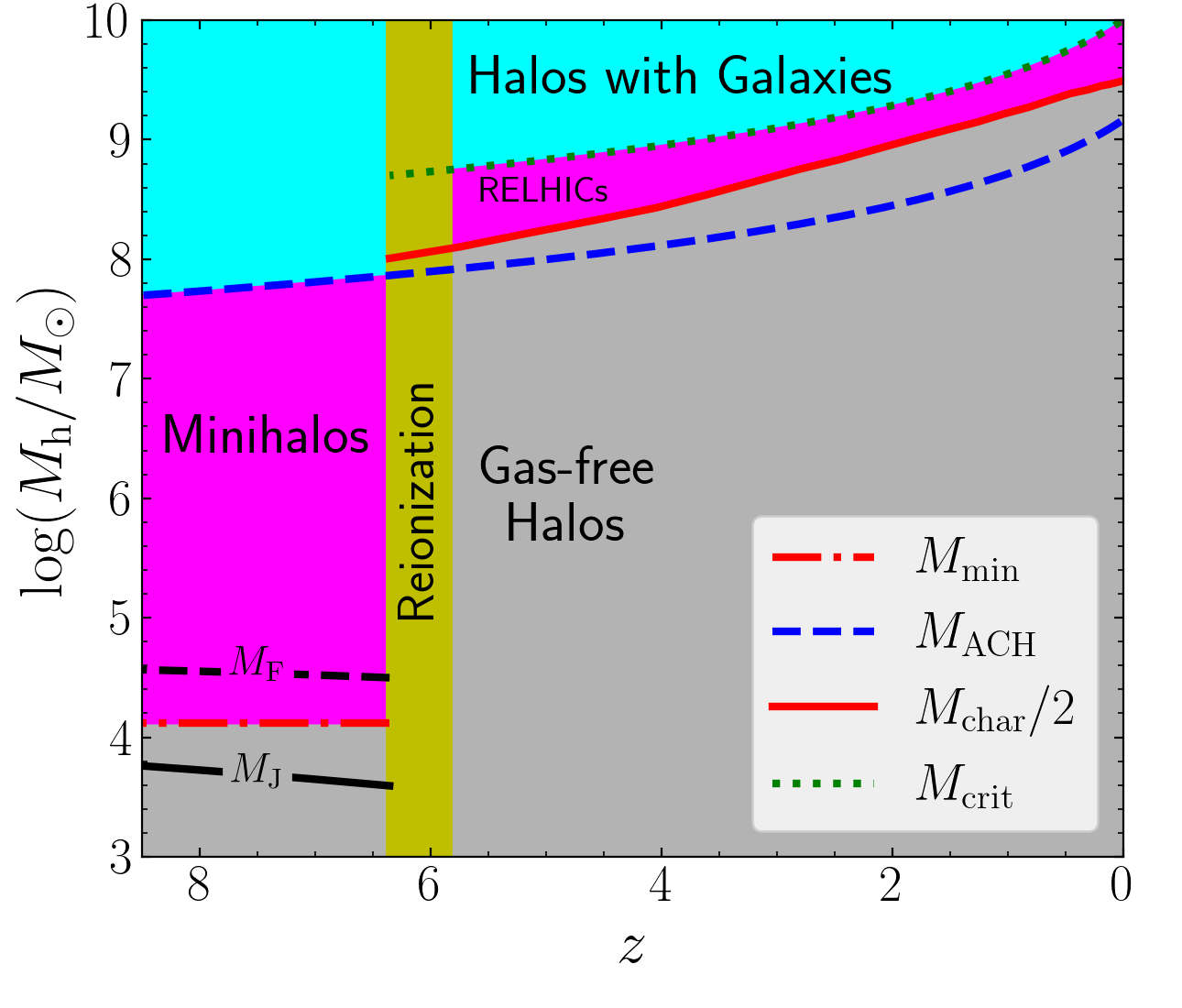}
\vspace{-0.5cm}
\caption{Properties of halos with different masses as a function of redshift, the {\em yellow vertical band} indicates the \eor, assumed to occur around $z\sim 6$. $M_{\rm J}$ is the Jeans mass ({\em black solid line}, Eq.~\ref{eq:MJ}), $M_F$ is the filtering mass ({\em black dashed line}, Eq.~\ref{eq:MF}), $M_{\rm min}$ is the mass above which halos contain 50\% of the cosmic baryon fraction on average as computed from the simulations ({\em red dot-dashed line}), $M_{\rm ACH}$ is the halo mass above which gas at the virial temperature can cool atomically ({\em blue dashed line}, Eq.~\ref{eq:MACH}). Halos with mass $M_{\rm min}<M_h<M_{\rm ACH}$ are \mini\ ({\em left magenta region}). After reionization, $M_{\rm char}$ is the mass above which halos contain 50\% of the cosmic baryon fraction on average as computed by \protect\cite{Okam08} ({\em red solid line}), and $M_{\rm crit}$ is the mass above which halos can host a galaxy according to \protect\cite{Beni20lmdwarf} ({\em green dotted line}). We moved the $M_{\rm char}$ line down since halos below $M_{\rm char}$ can still hold a small fraction of gas: such halos can host \relhics\ ({\em right magenta region}, \protect\citealt{Beni17RELHICs}). Halos in the grey region cannot retain a significant fraction of their baryons, whereas halos in {\it cyan region} could host a galaxy. Hence, the halos above the grey region are {\it photon sinks}, whereas those in cyan region host {\it photon sources}. The figure also shows that all \mini\ are photo-evaporated during and after the \eor.}
\label{fig:difm_z}
\end{figure}

\begin{figure}
\includegraphics[width={0.48\textwidth}]{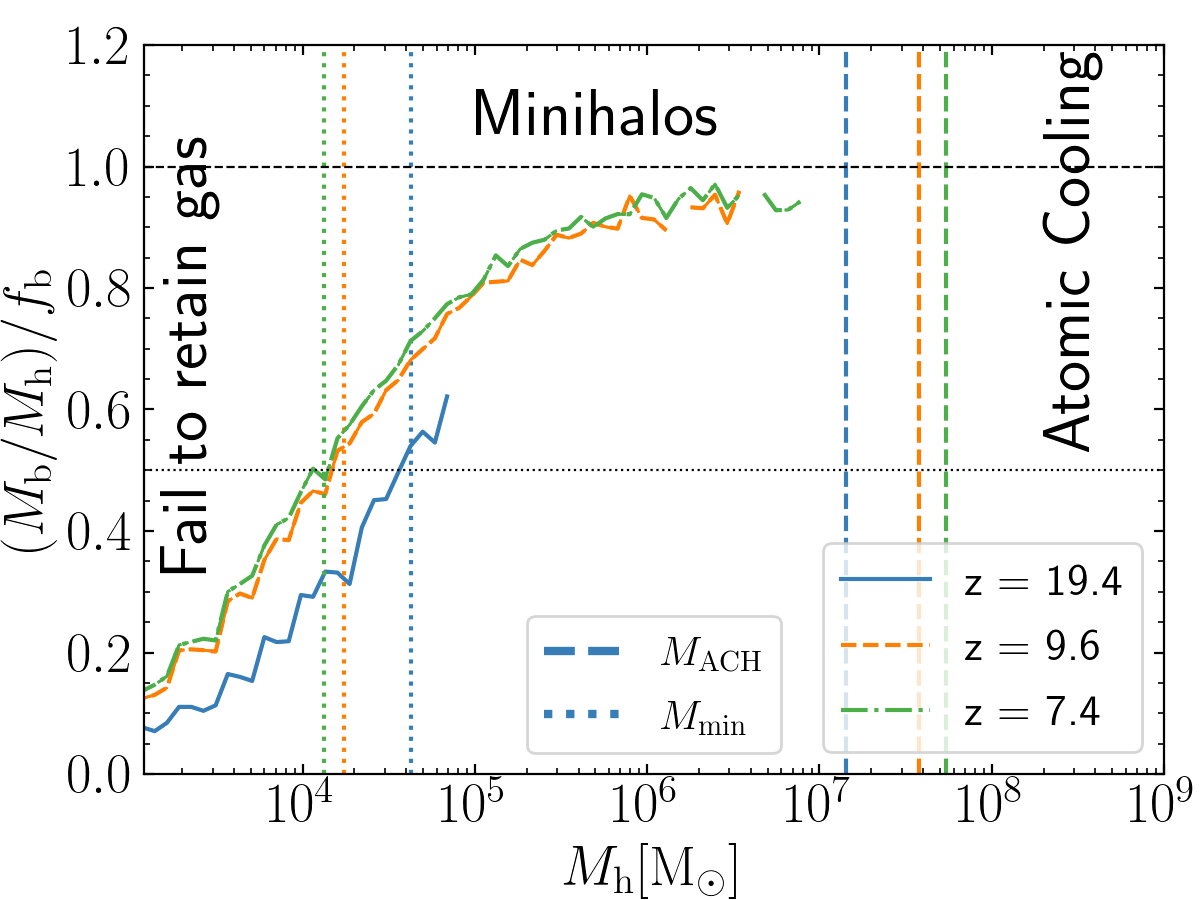}
\caption{The gas fraction inside halos at different redshifts before reionization in our highest resolution simulation (S512G00; see \S\ref{sec:method} for details), in units of the cosmic baryon fraction ($f_b\equiv \Omega_b/\Omega_m$). {\em Vertical dotted lines} indicate $M_{\rm min}$, the mass above which the baryon fraction of a halo is $f_b/2$ or more on average
(differently coloured lines from left to right, for $z=7.4$, 9.6 and 19.4).
{\em Vertical dashed lines} indicate the halo mass above which gas at the virial temperature can cool atomically, $M_{\rm ACH}$ (from right to left, for $z=7.4$, 9.6 and 19.4). Horizontal lines indicate a baryon fraction of $f_b/2$ (dotted) and $f_b$ (dashed). Mini-halos are halos with masses between $M_{\rm ACH}$ (vertical dashed) and $M_{\rm min}$ (vertical dotted).
}
\label{fig:gas_dmfrac}
\end{figure}

The impact of halos on the \ihm\ pre- and post-reionization depends on their mass. 
Below some minimum mass, halos have 
too shallow potential wells to accrete or retain cosmic gas. The value of this minimum mass depends on the mean temperature of the gas, $T_\ihm$. Prior to reionization, $T_\ihm$depends on redshift $z$ as \citep[e.g.][]{Furlanetto06}
\begin{eqnarray}
    T_\ihm&=&2.73\,(1+z_{\rm dec})\,\left(\frac{1+z}{1+z_{\rm dec}}\right)^2\,~{\rm K},\nonumber\\
    1+z_{\rm dec} &=& 134\,\left(\frac{\Omega_b\,h^2}{0.021}\right)^{2/5}\,.
\label{eq:Tihm}    
\end{eqnarray}
Here, $z_{\rm dec}$ is the redshift where
the cosmic gas decouples from the {\sc cmb} temperature.

We consider the following three estimates of the (pre-ionization)  minimum mass:
\begin{enumerate}
    \item the Jeans mass, $M_{\rm J}$, 
\begin{eqnarray}
M_{\rm J}
&=& 
\left(\frac{5 k_{\rm B}T_\ihm}{{\rm G}\mu m_{\rm H}}\right)^{3/2}\,
\left(\frac{3}{4\pi\Omega_M\,\rho_{c,0}(1+z)^3}\right)^{1/2}\nonumber\\
&\sim&
4.2\times10^3\msun \left(\frac{1+z}{9}\right)^{3/2}\,,
\label{eq:MJ}
\end{eqnarray}
\citep[e.g.][]{Mo10}, where $\mu\approx 1.22$ is the mean molecular weight per hydrogen atom of fully neutral primordial gas. We note that $M_{\rm J}$
{\em decreases} with cosmic time, therefore low-mass halos need to accrete baryons to reach the cosmic baryon fraction.

\item the \lq filtering mass\rq, $M_{\rm F}$, introduced
by \citealt{Gned00filtering} to account for the redshift dependence of $T_\ihm$,
\begin{align}
M_{\rm F}\sim M_{\rm J}\left[3\ln \left(\frac{1+z_{\rm dec}}{1+z}\right)-6+6\left(\frac{1+z}{1+z_{\rm dec}}\right)^{1/2}\right]^{3/2}\,;
\label{eq:MF}
\end{align}
see Appendix~A for more details

\item $M_{\rm min}$, the halo mass above which halos contain on average more than 50\% of the cosmic baryon fraction, as determined from the simulations described in \S \ref{sec:suite}\footnote{We use our highest resolution cosmological simulation \lq S512G00\rq, which has a dark matter particle mass of $\sim 10\;\msun$, see \S\ref{sec:method} for details.}.

\end{enumerate}

The \ihm\ temperature increases by almost three orders of magnitude during reionization, and hence so does the minimum mass.
We use the simulations of \cite{Okam08} to compute the (post-reionization) minimum mass and refer to it
below as $M_{\rm char}$ (the characteristic value of the post-reionization minimum mass). We further define $M_{\rm ACH}$ as the minimum halo mass in which cosmic gas can cool atomically ({\bf A}tomically {\bf C}ooling {\bf H}alos, ACHs)
\begin{align}
M_{\rm ACH} \sim 5\times 10^7 \left(\frac{1+z}{9}\right)^{-3/2}\, \msun\, ,
\label{eq:MACH}
\end{align}
and $M_{\rm crit}$ as the halo mass above which stars can form post-reionization according to \cite{Beni20lmdwarf}.

Figure~\ref{fig:difm_z} illustrates the evolution of these masses, with the vertical yellow band indicating the assumed \eor.
Pre-reionization, $M_J$ and $M_F$ decrease with cosmic time, whereas $M_{\rm min}$ remains approximately constant. 
Numerically, $M_{\rm min}\approx 2\times 10^4{\rm M}_\odot$, which is a factor of a few larger than $M_J$ and a factor of a few smaller than $M_F$. 

Average baryon fractions of halos (pre-reionization) in units of the cosmic mean are shown in Fig.~\ref{fig:gas_dmfrac} at various redshifts; the horizontal dotted line indicates 50 per cent. The figure also illustrates that neither $M_J$ nor $M_F$ provides accurate estimates of $M_{\rm min}$.

In Figure~\ref{fig:difm_z}, the left magenta region represents {\it mini-halos}, the main topic of this paper. They are halos with $M_{\rm min}\lesssim M_h\lesssim M_{\rm ACH}$, which contain their cosmic share of baryons but this gas cannot cool atomically \footnote{We have not indicated the mini-halos in which Pop III stars can form due to cooling by molecular hydrogen. However, this star formation pathway will likely be suppressed by the photo-dissociation photons from first stars/galaxies (see \S\ref{sec:caveats} and \citealt{Tren09H2PopIII}).}. After reionization, $M_{\rm char}>M_{\rm ACH}$ implies that there are no more halos that contain gas which cannot cool atomically: all mini-halos are evaporated during reionization. Such gas-free halos occupy the grey region. 

Finally, the right magenta region represents the \lq Reionization-Limited HI clouds\rq\ (\relhics) in the mass range $M_{\rm char}/2\lesssim M_h\lesssim M_{\rm crit}$ \cite{Beni17RELHICs}. They contain a significant amount of gas yet do not host a galaxy.

{\em The main point to take away from Fig.~\ref{fig:difm_z} is this}: before reionization, to capture all photon sinks requires resolving halos down to masses $\sim 10^4~\msun$ \footnote{However, X-ray preheating can relax this resolution requirement (see \S\ref{sec:discussion} for more discussions).}. {\it After} reionization, halos with masses $\gtrsim 10^8~\msun$ are the dominant photon sinks. This highlights the extended range in masses (from $\sim 10^4~\msun$ to $\gtrsim 10^8~\msun$) of photon sinks during and after \eor .

\subsection{The progression of the reionization process}
\label{sec:overview}

\begin{figure}
\includegraphics[width={0.48\textwidth}]{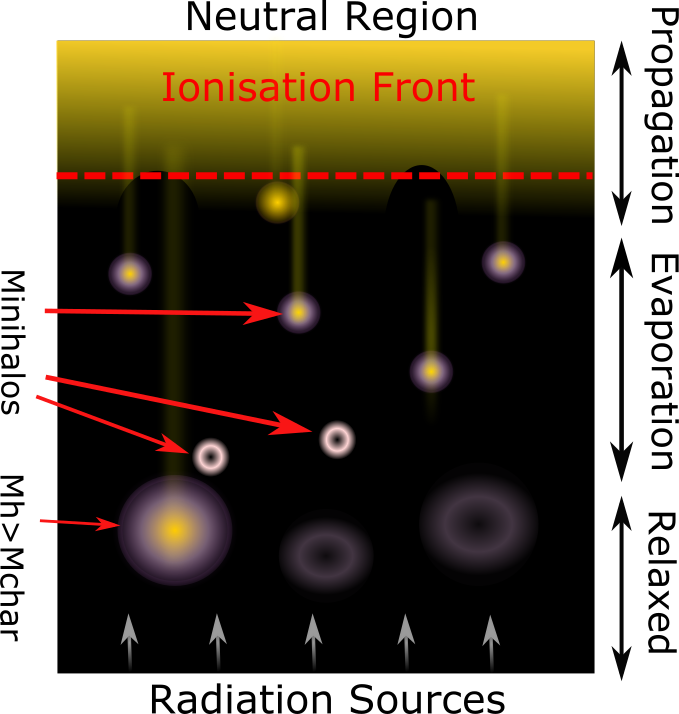}
\caption{A schematic picture of the three stages of reionization, in which an initially planar ionization front ({\em red dashed line}) moves upwards in the diagram. In the lower part of the diagram, the \ihm\ is highly ionized ({\em black region}) except in halos with mass above a characteristic mass, $M_{\rm char}$. The outskirts of the gas in these halos are highly ionized but the halos' potential wells are deep enough that the gas cannot photo-evaporate: these are Lyman-limit systems. The inner parts of the halos can self-shield, and the gas is neutral; these are damped Lyman-$\alpha$ systems ({\em yellow, solid discs}). In the middle part of the diagram, the ionization front has passed only recently, and the gas in \mini\ is still photo-evaporating ({\em hollow circles}). The ionization front has not yet reached the top of the diagram, where the \ihm\ is still neutral ({\em yellow region}). {\em Yellow vertical stripes} are regions where gas remains neutral due to the shadowing of the ionizing radiation by 
an intervening self-shielded region. The \Ifront\ is slowing down and becomes increasingly corrugated rather than planar as a result of the density inhomogeneities in the \ihm. 
}
\label{fig:illustration}
\end{figure}

We divide the \eor\ into three characteristic evolutionary phases, ({\em i}) the ionization front (\Ifront) {\em propagation} phase, ({\em ii}) the mini-halo {\em evaporation} phase, and ({\em iii}) the {\em relaxed} phase. These phases are illustrated in Fig. \ref{fig:illustration} and described below:\\ 

\noindent\underline{$\bullet$ \bf (I-front) Propagation Phase} : the
R-type\footnote{The terms \lq R-type\rq\ (rarefied) and \lq D-type\rq\ (dense) I-fronts were coined by \protect\cite{Kahn54ifront} (see, e.g., \protect\citealt{Oste06nebulae} for details).} \Ifront\ propagates at a small fraction of the speed of light, ionizing and heating the low-density \ihm. Halos more massive than a minimum mass, $M_{\rm min}$, contain a large fraction of their cosmic complement of baryons, and this gas is dense enough to self-shield and hence initially remains neutral. Dense gas in such halos significantly increases the recombination rate compared to that of a uniform \ihm. At the end of this phase, the neutral, high-density, self-shielded gas in halos is embedded in a highly ionized \ihm.\\

\noindent\underline{$\bullet$ \bf (Mini-halo)  Evaporation Phase}: the I-front propagates into halos, ionizing the gas in their outskirts.
In halos more massive than a minimum mass, $M_{\rm char}$, most of this ionized gas does not photo-evaporate, and their inner parts remain self-shielded. In contrast, the gas in lower mass halos photo-evaporates at a speed slower than the sound speed, taking 10-100 of Myrs to complete\footnote{The gas evaporates slower than the sound crossing time because I-fronts are trapped inside halos and propagate at subsonic speed \protect\cite{Shap04minihalo}.}. Such \mini\ are very numerous, and their photo-evaporation consumes a large number of ionizing photons, comparable to the number of photons needed to reionize the \ihm\ \cite[e.g.][]{Shap04minihalo,Ilie05minihalo}. At the end of this phase, all \mini\ are photo-evaporated. Only halos more massive than $M_{\rm char}$ contain significant amounts of gas. 

\noindent\underline{$\bullet$ \bf (Post-reionization) Relaxed Phase}: The \Ifront\ also propagates into halos more massive than \mini, but their deeper potential wells prevent them from photo-evaporating. These photon sinks correspond to Lyman-limit (the ionized outskirts of the halos) and damped-Lyman-$\alpha$ systems (the self-shielded inner parts, hereafter referred to as \lls's and \dla's, see, e.g. \citealt{Theuns21}). After reionization, these structures are mainly responsible for determining the opacity, the recombination rate and the clumping factor.

\subsubsection{(I-front) Propagation phase}
Consider the case of an initially planar \Ifront\ propagating through a (nearly) uniform \ihm\ which is fully neutral at time $t=0$. The speed of the \Ifront\ along its propagation direction\footnote{This also neglects collisional ionization and the presence of helium.}, which we assume to be the $Z$-axis, is 
\begin{equation}
    \frac{dZ}{dt} = \frac{F}{\langle n_{\rm H}\rangle} - \alpha_{\rm B}\,(1-x)^2\,\langle n_{\rm H}\rangle\,Z\,,
\label{eq:IFtrack}
\end{equation}
where $\langle n_{\rm H}\rangle$ is the (mean) hydrogen density by number, $x$ is the neutral fraction at coordinate $Z$ of the gas once it is ionized, and $\alpha_{\rm B}$ is the temperature-dependent recombination coefficient. The solution to this differential equation is
\begin{equation}
    Z = Z_S\,\left[1-\exp(-t/t_r)\right]\,,
\end{equation}
provided we assume that the gas downstream of the \Ifront\ is highly ionized, $x\ll 1$. Here, 
\begin{align}
t_r =& 
\frac{1}{\alpha_B(T_\ihm) \langle n_{\rm H}(z)\rangle}\approx2.1\,\left(\frac{1+z}{9}\right)^{-3}\,{\rm Gyr},\nonumber\\
\frac{dZ}{dt} =&  \frac{Z_S}{t_r}\,\exp(-\frac{t}{t_r})\nonumber\\
&\approx4.7\times 10^3\,\exp\left(-\frac{t}{t_r}\right)\,\left(\frac{\Gamma_{-12}}{0.1}\right)\left(\frac{1+z}{9}\right)^{-3}\,\,{\rm km}\,{{\rm s}}^{-1}\nonumber\\
Z_S =& \frac{F\,t_r}{\langle n_{\rm H}\rangle} \approx  11\,\left(\frac{\Gamma_{-12}}{0.1}\right)\left(\frac{1+z}{9}\right)^{-6}\,{\rm pMpc}\,,
\label{eq:IFspeed}
\end{align}
where $t_r$ is the recombination time and $Z_S$ is the one-dimensional analogue of the Str\"omgren radius\footnote{Using the value of $dZ/dt$ suggests that it would take of order $\gtrsim$10~Myr for an \Ifront\ to cross 1 Mpc. In fact, it takes considerably longer than this, because
of density inhomogeneities and recombination.}. The photo-ionization rate, $\Gamma$, is related to $F$ by
\begin{align}
    \Gamma = F\sigma_{\rm HI}&=1.62\times 10^{-13}\,\left(\frac{F}{10^5\,{\rm cm}^{-2}\,{\rm s}^{-1}}\right)\,{\rm s}^{-1}\nonumber\\
    &\equiv \Gamma_{-12}\,10^{-12}{\rm s}^{-1}\,,
    \label{eq:Gamma}
\end{align}
and the case-B recombination rate is
\begin{equation}
    \alpha_B(T)\approx 1.1\times 10^{-13}\,\left(\frac{T}{2.2\times 10^4~{\rm K}}\right)^{-0.7}\,{\rm cm}^3{\rm s}^{-1}\,.
\end{equation}
The numerical values in Eq.~(\ref{eq:Gamma}) assume that the ionizing spectrum is that of a black body with effective temperature $T_{\rm BB}=10^5~{\rm K}$, e.g. massive stars,
which yields a frequency-averaged photo-ionization cross section\footnote{Our value differs from that of \cite{Embe13clumping} because we use a different spectral shape.} of $\sigma_{\rm HI}=1.63\times 10^{-18}{\rm cm}^2$.
The temperature of the \ihm\ is $T_\ihm\approx 2.2\times 10^4~{\rm K}$ after being flash ionized by such an ionizing spectrum \cite[e.g.][]{Chan21SPHM1RT}.

The \Ifront's propagation speed is considerably slower than the speed of light. Note that the location of the Str\"omgren layer, $Z_S$, itself evolves as the Universe expands (see, e.g. \citealt{Shap87cosHII} for a more accurate calculation of the evolution of such H{\sc ii} regions). Eq.~(\ref{eq:IFspeed}) also shows that the recombination time at the mean density, $t_r$, is longer than the age of the Universe, $t_{\rm U}=2/(3H(z))=0.65\,(9/(1+z))^{3/2}~{\rm Gyr}$, at a redshift, $z \lesssim 8$. Gas in \mini\ is at a higher density and hence has a shorter recombination time, thus slowing down the \Ifront\ until the gas photo-evaporates.

\subsubsection{(Mini-halo) Evaporation phase}
\label{sec:ana_eva}
The dense gas in \mini\ traps the \Ifront. The response of the gas, when overrun by the \Ifront, depends on ratios of three time-scales: ({\em i}) the sound-crossing time $t_{\rm sc}$, ({\em ii}) the \Ifront\ crossing time $t_{\rm ic}$, and ({\em iii}) the recombination time $t_{\rm r, h}$.

The {\bf sound-crossing time} is the time for a sound wave to travel a distance equal to the virial radius, $R_h$, of the halo,
\begin{align}
t_{\rm sc} &= \frac{R_h}{c_s}\nonumber\\ &\sim 7.0\,\left ( \frac{1+z}{9} \right )^{-1}\left ( \frac{M_{h}}{10^5\msun} \right )^{\frac{1}{3}}\left ( \frac{T}{2.2\times 10^4\;{\rm K}}\frac{0.6}{\mu} \right )^{-\frac{1}{2}}\,{\rm Myr}\,,
\label{eq:tsc}
\end{align}
where $M_h$ is the virial mass of the halo and $c_s$ is the sound speed, which we evaluated at the temperature of the \ihm\ after flash ionization \citep{Chan21SPHM1RT}.

The {\bf  I-front propagating time}, $t_{\rm ic}$, is the time it takes for an I-front to cross a mini-halo, neglecting recombinations,
\begin{align}
t_{\rm ic} &= \frac{\Delta\langle n_{\rm H}\rangle R_h}{F}\nonumber\\
&\sim 6.6\, 
\left(\frac{\Delta}{200}\right)
\left(\frac{1+z}{9}\right )^2
\left(\frac{0.1}{\Gamma_{-12}}\right)\,
\left(\frac{M_{h}}{10^5\msun}\right )^{\frac{1}{3}}\,{\rm Myr}\,.
\label{eq:tic}
\end{align}

Finally, the mean {\bf recombination time} of the gas in a halo, $t_{\rm r, h}$, is:
\begin{align}
t_{\rm r, h}&=\frac{1}{c_{\rm mh}\Delta \langle n_{\rm H}\rangle\alpha_{\rm B}}\nonumber\\
&\sim 5.4\,\left ( \frac{1+z}{9} \right )^{-3}\left ( \frac{c_{\rm mh}}{2} \right )^{-1}\left ( \frac{\Delta}{200} \right )^{-1}\,{\rm Myr}\,.
\label{eq:trecom}
\end{align}
We evaluated the case-B recombination coefficient at a temperature, $T=2.2\times 10^4\,{\rm K}$, $\Delta$ is the gas over-density which we set to 200 (e.g. \citealt{Mo10}). Furthermore, $c_{\rm mh}$ is the clumping factor of the gas inside a mini-halo; in the simulations discussed below, we find typical values for $c_{\rm mh}$ in the range 2-4. The values of these three times are similar for our default choice of parameters, e.g. a halo mass of $M_h=10^5\msun$ and a photo-ionization rate with $\Gamma_{-12}\sim 0.1$.
The dependence on halo mass is the same for $t_{\rm ic}$ and $t_{\rm sc}$, but the redshift dependence differs; $t_{\rm r}$ does not depend on $M_h$.  Depending on the values of $z$ and $M_h$, we identify the three following regimes:\\

\noindent \underline{$\bullet$ \bf Sound-speed limited regime:} When $t_{\rm ic}\ll t_{\rm sc}$, the I-front races through a mini-halo so quickly that its gas cannot recombine nor react hydrodynamically. The gas is ionized and heated and will photo-evaporate\footnote{I-front trapping can still occur in the central dense regions of small halos where the recombination time is short, see \cite{Shap04minihalo}. If this core is small, it may not affect reionization significantly.} on a timescale, $t \sim t_{\rm sc}$. This is typically the case for low-mass haloes, at low redshift, and when $\Gamma_{-12}$ is large.\\

\noindent \underline{$\bullet$ \bf Ionization limited regime:} When $t_{\rm sc}\ll t_{\rm ic}$, the \Ifront\ moves slowly and the gas can photo-evaporate as soon as it ionizes: this also implies that the total gas density tracks the ionized gas density.\\

\noindent \underline{$\bullet$ \bf I-front trapping regime:} If $t_{\rm r}\ll t_{\rm ic}$, the gas is photo-ionized but recombines very quickly\footnote{These I-fronts are D-type (dense-type). See, e.g. \citealt{Draine11}.}. The \Ifront\ moves at sub-sonic speeds and eventually nearly stalls at the inverse Str\"omgren layer \citep{Shap04minihalo}. This regime occurs in halos of mass, $M_{h} \gtrsim 10^6\msun$, and in the central dense region of low-mass halos.

In principle, the duration of the photo-evaporation phase is not simply established by the time it takes to photo-evaporate the most massive \mini. It is because low-mass \mini\ may contribute more to the recombination rate since they are much more numerous. We will use the simulations described below to estimate the duration of this phase.

\subsubsection{(Post-reionization) Relaxed phase}
Long ($\gg 100\;{\rm Myr}$) after being overrun by the \Ifront, halos with $M_{\rm h}<M_{\rm char}$ are photo-evaporated and unable to accrete gas \citep{Okam08}: such halos no longer contribute to recombination.
Gas in the outskirts of more massive halos is highly ionized, with the inner parts self-shielded and neutral. These \lls's and \dla's determine the attenuation length of ionizing photons, and thereby the relationship between the emissivity of ionizing photons and the photo-ionization rate \cite[e.g.][]{Fauc09, McQu11LLS, Haar12UVbackground}. 

Cosmological \rhd\ simulations are required to capture the propagation and evaporation phases and the transition to the post-reionization phase. To resolve photon sinks during these stages, such simulations must resolve \mini\ above the Jeans mass in a computational volume that is large enough to sample the rarer \lls's and \dla's that determine the mean free path of ionizing photons in the post-reionization phase. This is a tall order, even if we neglect the even more challenging calculation of resolving the nature of the ionizing sources and the thorny issue of determining the fraction of those photons that can escape their natal cloud. In this paper, we focus on photon sinks during the \eor: we simply inject ionizing photons into our computational volume at a specified rate and follow how these photo-evaporate gas out of small halos.

\section{Simulations}
\label{sec:method}
\begin{table*}
\centering
\begin{tabular}{llllllllll}
\hline
\hline
Sim & $L_{\rm box}$ & $z_i$ & $\Gamma_{-12}$ & $N$ & $m_{\rm gas}$& $m_{\rm DM}$ & $l_{\rm soft}$&$\tilde{c}$&$\langle n_{\rm H}\rangle$\\
 & [ckpc] &  &     &  & [$\msun$]& [$\msun$] & [kpc]& [c]&[${\rm cm}^{-3}$]\\
\hline
Validation\\
S128z8G03       &400&7.9     & 0.3  &$128^3$&160&960&0.3&0.15&$1.2\times10^{-4}$\\
S256z8G03c001   &400   &7.9  & 0.3  &$256^3$&20&120&0.1&0.01&$1.2\times10^{-4}$\\
S256z8G03c005   &400   &7.9  & 0.3  &$256^3$&20&120&0.1&0.05&$1.2\times10^{-4}$\\
S256z8G03c01    &400   &7.9  & 0.3  &$256^3$&20&120&0.1&0.1&$1.2\times10^{-4}$\\
S256z8G03c02    &400   &7.9  & 0.3  &$256^3$&20&120&0.1&0.2&$1.2\times10^{-4}$\\
S256z8G03c05    &400   &7.9  & 0.3  &$256^3$&20&120&0.1&0.5&$1.2\times10^{-4}$\\
S256z8G003c001  &400   &7.9  & 0.03 &$256^3$&20&120&0.1&0.01&$1.2\times10^{-4}$\\
S256z8G003c005  &400   &7.9  & 0.03 &$256^3$&20&120&0.1&0.05&$1.2\times10^{-4}$\\
S256z8G003c01   &400   &7.9  & 0.03 &$256^3$&20&120&0.1&0.1&$1.2\times10^{-4}$\\
S256z8G003c02   &400   &7.9  & 0.03 &$256^3$&20&120&0.1&0.2&$1.2\times10^{-4}$\\
S512z8G00       &400   &7.9  & 0.0  &$512^3$&2.5&15&0.05&-&$1.2\times10^{-4}$\\
M128z8G03       &800   &7.9  & 0.3  &$128^3$&1300&7700&0.4&0.15& $1.2\times10^{-4}$\\
M256z8G03       &800   &7.9  & 0.3  &$256^3$&160&960&0.3&0.15& $1.2\times10^{-4}$\\
L512z8G03       &1600  &7.9  & 0.3  &$512^3$&160&960&0.3&0.15& $1.2\times10^{-4}$\\
L512z8G01       &1600  &7.9  & 0.1  &$512^3$&160&960&0.3&0.05& $1.2\times10^{-4}$\\
\hline
Production\\
M512z6G03      & 800  & 6.0  & 0.3    &$512^3$&20&120&0.1&0.15& $6.0\times10^{-5}$ \\
M512z6G003     & 800  & 6.0  & 0.03   &$512^3$&20&120&0.1&0.05& $6.0\times10^{-5}$ \\
M512z8G03      & 800  & 7.9  & 0.3    &$512^3$&20&120&0.1&0.15& $1.2\times10^{-4}$\\
M512z8G015     & 800  & 7.9  & 0.15   &$512^3$&20&120&0.1&0.075& $1.2\times10^{-4}$\\
M512z8G003     & 800  & 7.9  & 0.03   &$512^3$&20&120&0.1&0.05& $1.2\times10^{-4}$\\
M512z10G03     & 800  & 10.2 & 0.3    &$512^3$&20&120&0.1&0.15& $2.5\times10^{-4}$\\
 \hline
 \hline
\end{tabular}
\caption{Details of the simulation suite. {\em From left to right:} simulation identifier, linear extent of the simulated volume $L_{\rm box}$, redshift of reionization $z_i$, photo-ionization rate in units of $10^{-12}{\rm s}^{-1}$, $\Gamma_{-12}$ , number of gas and dark matter particles, $N$, gas particle masses $m_{\rm gas}$, dark matter particle masses $m_{\rm DM}$, gravitational softening lengths $l_{\rm soft}$, reduced speed of light at the mean density, $\tilde{c}$, and mean hydrogen density, $\langle n_{\rm H}\rangle$, at $z_i$.
The simulation identifier encodes the extent of the simulated volume, the number of particles, the value of $z_i$ and the value of $\Gamma_{-12}$.}
\label{table:sim}
\end{table*}

\subsection{Code and numerical set-up}
We simulate a periodic cubic cosmological volume using the publicly-available\footnote{\url{http://www.swiftsim.com}} {\sc sph} code {\small SWIFT} \citep{Scha16SWIFT,Scha18SWIFTascl}. Among the various implementations of the {\sc sph} algorithms (e.g. \citealt{Borr20SPHENIX}) included in the code,
we select the entropy-based version described by \cite{Spri02esph} and \cite{Spri05Gadget2}. 

Radiation hydrodynamics is solved with the \sph\ two-moment method with a modified {\sc m1} closure\footnote{ \cite{Wu21M1accuracy} suggested that the M1 method, the approach here, over-ionizes absorbers with idealized calculations, assuming uniform radiation coming from infinity. However, their argument is not applicable to the case here where radiation is plane-parallel. The accuracy of our method with a plane-parallel radiation field is demonstrated in Appendix \ref{sec:minihalotest}.}, as described by \cite{Chan21SPHM1RT}. A uniform, constant flux of ionizing radiation is injected into the computational volume from two opposing faces of the cubic volume. We consider a simulation suite in which we vary the redshift when we start injecting photons, $z_i$, and the intensity of the radiation, $\Gamma_{-12}$. The spectrum of the radiation is that of a black body of temperature, $T_{\rm BB}=10^5{\rm K}$, and is treated in a single frequency bin using the grey approximation; this also means that we neglect any spectral hardening of the radiation. The optically thin direction of the Eddington tensor is taken to be along the initial direction of propagation of the \Ifront; this improves the ability of the method to cast shadows and handle self-shielding. To reduce the computational cost, we propagate radiation at a reduced speed of light, $\tilde c$ \citep{Gned01OTVET}. In our implementation, $\tilde c$ scales with the value of the smoothing lengths of the {\sc sph} particles\footnote{See Appendix \ref{sec:vsl} for more details on this \lq variable\rq\ speed of light approximation.}. The interaction of radiation with matter is calculated with a non-equilibrium thermo-chemistry solver with hydrogen only (as in \citealt{Chan21SPHM1RT}). We include helium when calculating the heat capacity, but we do not consider its interaction with radiation. Note that we also neglect molecular hydrogen and other elements (see \S\ref{sec:caveats} for a discussion on the caveats of our approach). Our original \rt\ implementation did not account for the cosmological redshifting of radiation. We describe and test in Appendix \ref{sec:comoving} our choice of co-moving variables. Our implementation accounts for the decrease in the proper density of photons as the Universe expands, but it does not account for the increase in the wavelengths of these photons. The mean free path of ionizing photons is short in the case we simulate here. Therefore, this is a reasonable approximation.

Our simulation suite does not include feedback from evolving stars. However, as very dense gas particles severely limit the simulation time-step, and since we do not include the correct physics for these high-density regions anyway, we simply convert gas particles into stars once their density exceeds a physical density of 10 hydrogen atoms per ${\rm cm}^{3}$ and an over-density of  $\Delta=\rho/\langle\rho\rangle=10^3$. These criteria are similar to the \lq quick-Ly$\alpha$\rq\ approximation used by, e.g., \cite{Viel04Lyalpha}, who pointed out that the impact of this approximation on the Ly$\alpha$ flux power spectrum is small (less than 0.2\%). Moreover, the density of gas particles that turn into stars is higher than that of the regions that give rise to Lyman-limit systems. This indicates that our approximation is unlikely to affect the \Ifront\ speed or the photo-evaporation timescales of \mini\ (see further discussions on this in \S\ref{sec:caveats}).

We generate the initial conditions at redshift $z=127$, using the publicly-available {\small music} code \citep{Hahn11MUSIC}. The adopted cosmological parameters are: $h=0.678$, $\Omega_m=0.307$, $\Omega_\Lambda=0.693$, and $\Omega_b=0.0455$, $\sigma_{8} = 0.811$, and $n_{s}=0.961$, where symbols have their usual meaning. The hydrogen and helium mass fractions are $X=0.752$ and $Y=1-X$, respectively. We use Eq.~(\ref{eq:Tihm}) to compute the temperature at the mean density, $T_\ihm$, which gives the normalization of the adiabat describing the temperature-density relation of the particles in the initial conditions, $T_{\rm ini}(\rho)$:
\begin{align}
    T_{\rm ini}(\rho) = T_\ihm\,\left(\frac{\rho}{\langle\rho\rangle}\right)^{2/3}\equiv   T_\ihm\,\Delta^{2/3}\,.
\label{eq:entropyfloor}    
\end{align}

\subsection{Simulation Suite}
\label{sec:suite}
\begin{figure}
\includegraphics[width={0.48\textwidth}]{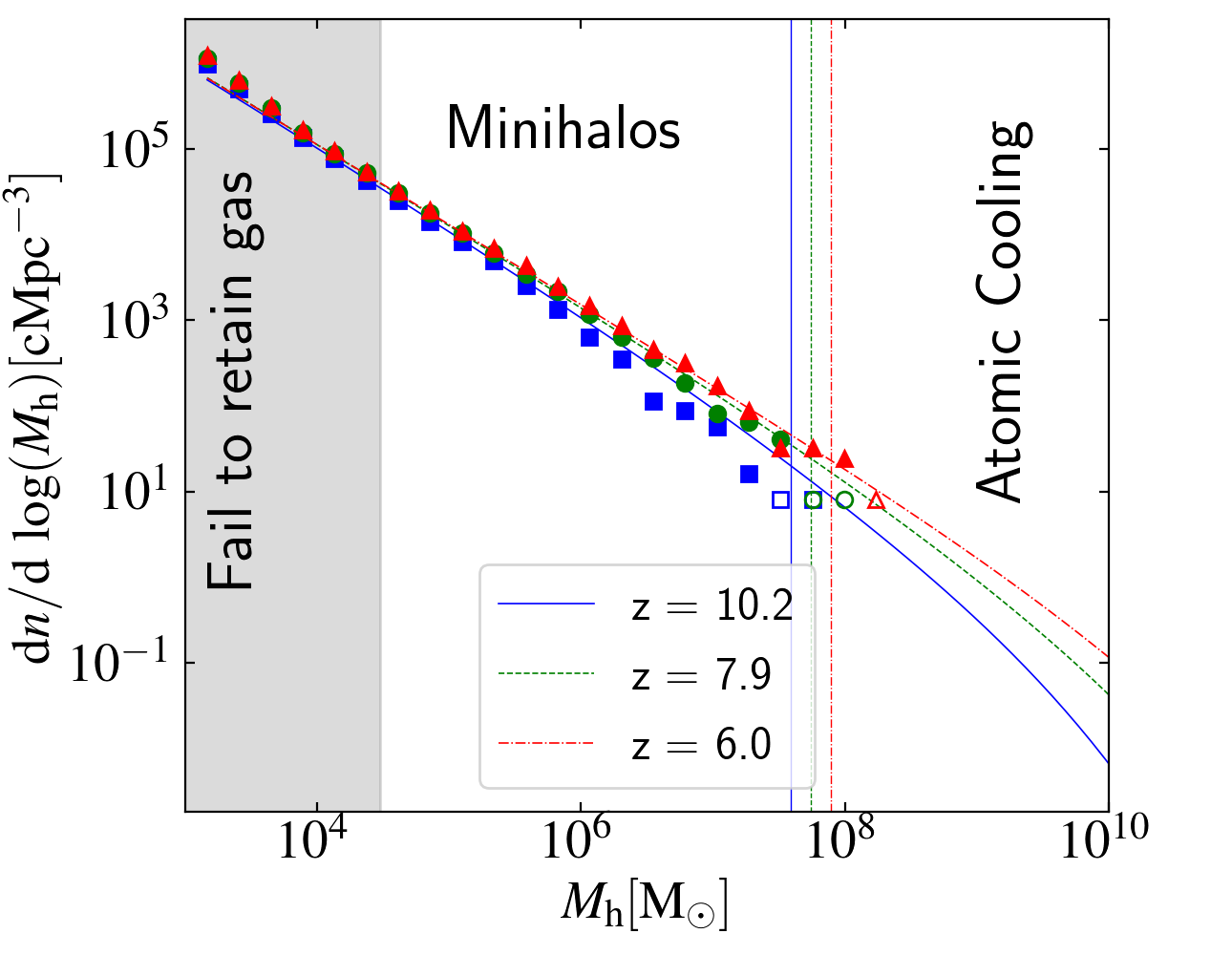}
\caption{{\em Symbols:} halo mass function of the fiducial simulation (M512) at three different redshifts, {\em lines} show the fitting functions from \protect\cite{Reed07}. Colours correspond to $z=10.2$ ({\em blue squares}, and {\em solid line}), $z=7.9$ ({\em green circle} and {\em dotted line }), and $z=6.0$ ({\em red triangles} and {\em dashed line}).
Empty symbols show mass bins with fewer than 5 halos per decade in halo mass. The thin vertical lines on the right show the minimum mass of atomic cooling halos (Eq. \ref{eq:MACH}); halos in the grey region are not well resolved, containing fewer than 300 dark matter particles. Our fiducial simulation resolves \mini\ with more than 300 particles in a volume sufficiently large to contain a few halos in which gas can cool atomically.}
\label{fig:halocatunheated}
\end{figure}

Our main objective is to study the photo-evaporation of \mini\ and how this impacts the progression of reionization. This requires sampling and resolving \mini\ from the Jeans mass, $M_{\rm min}$, to the atomic cooling limit, $M_{\rm ACH}$, (see Fig.~\ref{fig:difm_z}). We vary the redshift at which we start injecting radiation and the flux of the injected ionizing photons and run individual simulations until several 100~Myrs after the radiation injection. Simulation parameters are listed in Table \ref{table:sim}. We motivate the choices as follows.

We set the dark matter particle mass of the fiducial simulations to $m_{\rm DM} \sim 100 \;\msun$, 
so that a halo of mass $M_{\rm min}$ (see \S\ref{sec:analytic}) is resolved with $\sim 300$ particles. The baryonic content of such halos is resolved with roughly
30\% accuracy \citep{Naoz09mhgf} and the overall clumping factor of the simulated volume is accurate to $\sim 10-20\%$ \citep{Embe13clumping}. 

To include massive halos of mass $\sim M_{\rm ACH}$, we consider a fiducial linear extent of $800 \;{\rm ckpc}$. We demonstrate in Fig. \ref{fig:boxsizeNhalo} that such a volume contains more than one halo of mass $M_{\rm ACH}$ at $z=8$, and 10 at $z=6$.
This linear extent also yields approximately converged values for the mean free path of ionizing photons at the end of the \eor\ { (for unrelaxed gas; see \citealt{Embe13clumping})}\footnote{\protect\cite{Embe13clumping} did not consider the effect of photo-heating and thus photo-evaporation.
However, if photo-heating is included, halos with virial mass below the characteristic mass found by \cite{Okam08} will eventually photo-evaporate, with more massive halos determining the clumping factor. An accurate calculation of the clumping factor then requires an even larger volume. To study this situation, our simulation suite includes larger volumes simulated at lower resolution.}. 

We perform simulations with three choices of the \lq reionization redshift\rq,  $z_i\sim$ 10, 8, and 6, ($z_i$ is the redshift where we start injecting ionizing photons from two opposing faces of the cubic volume at constant flux). This range covers approximately current observational estimates for the start and tail-end of the \eor\ \citep[e.g.][]{Fan06EoR,Planck2018CMB}. The values of $F$ correspond to photo-ionization rates of $\Gamma_{-12}=0.03-0.3$
(where $F$ and $\Gamma_{-12}$ are related by the frequency-averaged photo-ionization rate, as in  Eq.~\ref{eq:Gamma}). This range in $\Gamma_{-12}$ is motivated by observational estimates \citep[e.g.][]{Calv11UVbg,Wyit11UVbg,DAlo18UVbg} as well as simulation results \citep[e.g.][]{Rosd18SPHINX}.

We take the value of the reduced speed of light to be proportional to that of the \Ifront: this allows us to use a lower value of $\tilde{c}$ at lower $\Gamma_{-12}$, decreasing the wall-clock time of the simulations (see Appendix \ref{sec:ressizecred} for tests of numerical convergence). Once the majority of the \ihm\ is ionized, {\em i.e.} at the start of the evaporation phase of the \eor, most I-fronts are expected to be D-type and hence propagate locally at a speed comparable to the sound speed, which is much slower than the speed of light. Therefore, we 
set ${\tilde c}=c/100$ once the mass-weighted neutral hydrogen fraction drops below $5\%$ (see the similar approach and convergence tests in \citealt{DAlo20clumping}). 

Finally, note that the simulated volumes are all relatively small and are not necessarily representative cosmological volumes during the \eor. Rather we think of them as selected patches of the Universe that are overrun by an \Ifront\; due to sources outside of the simulated volume. These patches are ionized at various redshifts, $z_i$, and with a range of values of the photo-ionization rate, $\Gamma_i$.

\subsection{Halo identification}
We use {\small nbodykit} \citep{Hand18nbodykit} to identify halos using the 
friend-of-friend algorithm \citep{Davi85fof}, as simply connected regions with
a mean density of $\sim 200$ times the average density of the Universe. The halo mass function of our simulations at redshifts before $z_i$ is compared
to that computed with the  {\sc colossus} python package\footnote{\url{https://bdiemer.bitbucket.io/colossus/}} \citep{Diemer18COLOSSUS} in Fig.~\ref{fig:halocatunheated} (we selected
the \citealt{Reed07} fit). The agreement between the simulation results and the fit indicates that our small volume contains the expected number of mini-halos up to the mass of atomic-cooling halos, as desired. We have performed additional runs in which we vary the box size and numerical resolution, see Appendix \ref{sec:ressizecred}.

\section{Results}
\label{sec:results}
\subsection{Overview}
\label{sec:simoverview}

We illustrate the first two reionization phases - \Ifront\ propagation and mini-halo evaporation -  using the M512z8G03 run. In this simulation, ionizing photons are injected after redshift $z_i=7.87$, with a constant flux equivalent to a photo-ionization rate of $\Gamma_{-12}=0.3$. The value of $z_i$ falls around the midpoint of the \eor\ as inferred from the Thompson optical depth \citep{Planck16}, and the value of $\Gamma_{-12}$ is typical of the expected mean value during this time  ($\Gamma_{-12}=0.3\cdots 0.6$, \citealt[e.g.][]{DAlo18UVbg}). Therefore, this setup simulates the history of a typical patch of the universe during the \eor. The evolution of the other simulations is qualitatively similar to that of M512z8G03. Simulations with higher values of $z_i$ have fewer self-shielded clouds since structure formation is less advanced, and those at lower $z_i$ or higher values of $\Gamma_{-12}$ have more self-shielded clouds.

\begin{figure*}
\includegraphics[width={1.0\textwidth}]{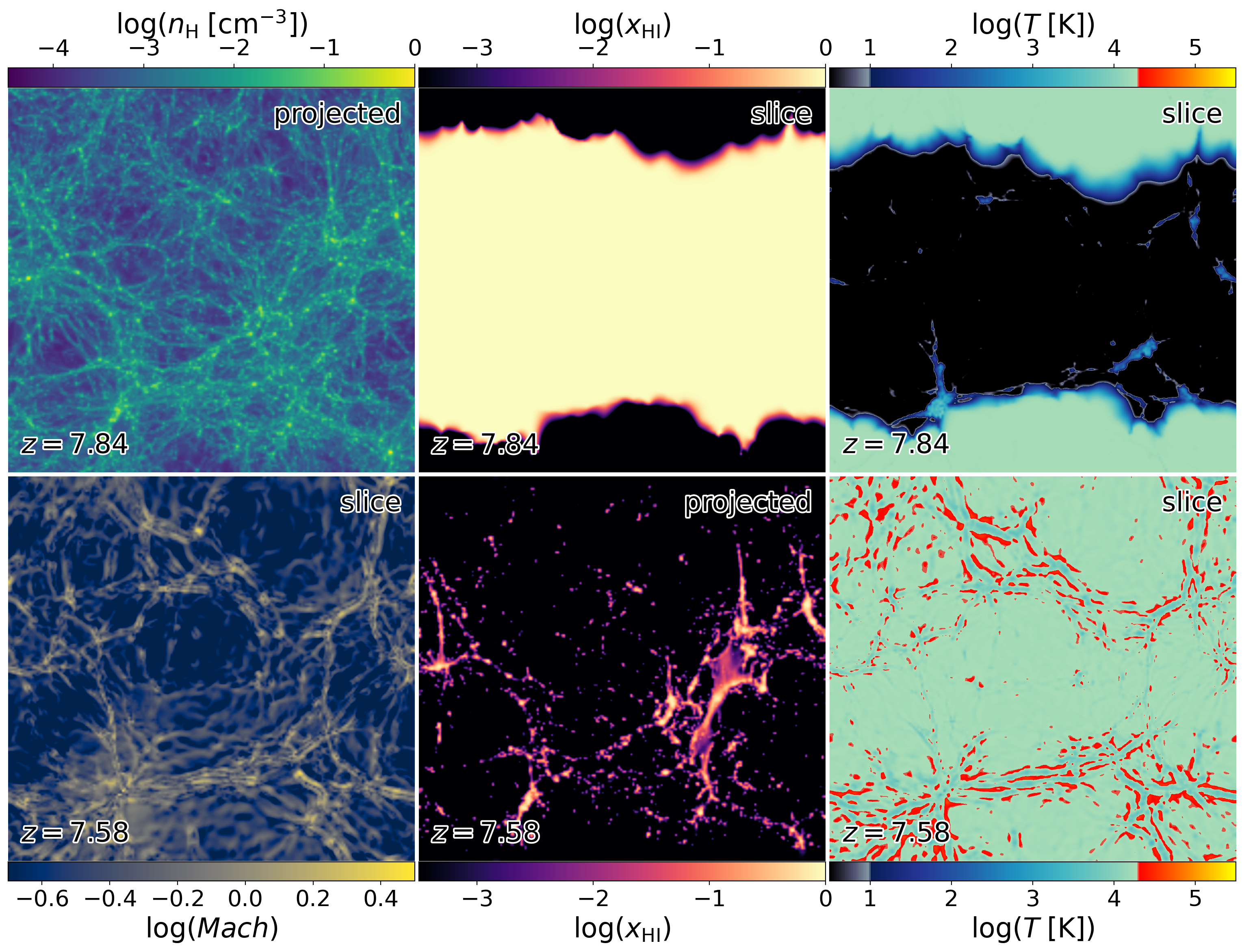}
\caption{Impact and response of gas structures to the passage of ionization fronts, entering the computational volume simultaneously from the top and the bottom.
Individual panels visualize slices/projections of the M512z8G03 simulation box at $z=7.84$ ({\em upper panels}) and $z=7.58$ ({\em lower panels}). Radiation is injected at redshift $z_i\approx 8$. Panels {\em from left to right} are hydrogen gas density ($n_{\rm H}$) (Mach number in the lower left), the neutral fraction ($x_{\rm HI}$), and the gas temperature ($T$). Panels labelled \lq projected\rq\ correspond to projections of the computational volume; the others correspond to infinitesimally thin slices through the mid-plane. As the \Ifront\ propagates through the volume, gas becomes ionized and photo-heated, and \mini\ photo-evaporate. Photo-heating causes the expansion of the filaments.}
\label{fig:simoverview}
\end{figure*}

Fig.~\ref{fig:simoverview} shows the simulated volume at $z=7.84$, which is $\sim 3\;{\rm Myr}$ after ionizing photons were first injected ({\em propagation phase}, upper row), when the \Ifront\ has propagated over approximately 1/6$^{\rm th}$ of the simulation volume. Using Eq.~(\ref{eq:IFtrack}) with $\alpha_B=0$ for the speed of the \Ifront\ in a homogeneous medium, we would expect an \Ifront\ to cross the computational volume in a time $\sim \langle n_{\rm H}\rangle\,L\,/F\sim 6~{\rm Myr}$ (where $L$ is the proper box size and $F$ is photon flux) if recombination could be neglected. Clearly, density inhomogeneities, and in particular \mini, slow down the \Ifront\ by a factor $\sim 2-3$ on average. The top row shows how the initial planar \Ifront\ becomes corrugated. This is because high-density regions locally slow down the front, as opposed to low-density regions which increase the front speed. Sufficiently dense gas in halos may even stop the \Ifront\, casting a shadow of gas that remains neutral behind them. Downstream of the \Ifront, the highly ionized gas is also photo-heated, reaching a temperature of $T\sim 2\times 10^4{\rm K}$ in a timescale of order $\Gamma^{-1}$ (photo-ionization rate; see Eq.\ref{eq:Gamma}). The middle panel shows that the \Ifront\ is sharp, with the distance from where the gas is mostly ionized to where it is mostly neutral of around 
\begin{align}d_{\rm IF}\sim1/(\langle n_{\rm H}\rangle \sigma_{\rm HI})\sim 1.4 \;{\rm pkpc} \left(\frac{1+z}{9}\right)^{-3}.
\end{align}
Therefore the \ihm\ during the \eor\ is either highly ionized or mostly neutral, to a good approximation.

The lower row of Fig.~\ref{fig:simoverview} corresponds to $z=7.58$ ($\sim 30~{\rm Myr}$ from the start of photon injection) when more than 99\% of the volume has been highly ionized.  When the \Ifront\ has already crossed the computational volume, it leaves behind filaments and halos with mostly neutral and cold gas. The lower-left panel shows that these filaments are photo-evaporating, with gas expanding out of the shallow potential well once it is ionized and heated \citep[see also][]{Bryan99}. The flow velocity is comparable to the local sound speed. These expansion waves compress and heat the gas in the filaments' outskirts, as seen in the lower right panel, where hotter gas surrounds the expanding cooler filaments. The denser gas in \mini\ takes longer to fully photo-evaporate, and sufficiently massive halos retain most of their baryons.

This discussion suggests assigning gas to three categories: (1) the low-density \ihm, e.g. voids; (2) the filaments of the cosmic web; (3) collapsed halos including \mini. These structures stand out in the top left panel of Fig. \ref{fig:simoverview}. To investigate how these structures are impacted by reionization and vice versa how they affect reionization, we proceed as follows.  In \S\ref{sec:gasreion}, we investigate the impact of \Ifront\ on the \ihm\ and filaments, and in \S\ref{sec:shielding}, study how self-shielding keeps the central parts of \mini\ neutral. In \S\ref{sec:mhevaporation}, we will turn to the photo-evaporation of \mini. In \S\ref{sec:clumping}, we will quantify how these small-scale structures impede reionization and the role of photo-evaporation/relaxation.

\subsection{Response of the \ihm\ to reionization}
\label{sec:gasreion}

\begin{figure*}
\includegraphics[width={1.0\textwidth}]{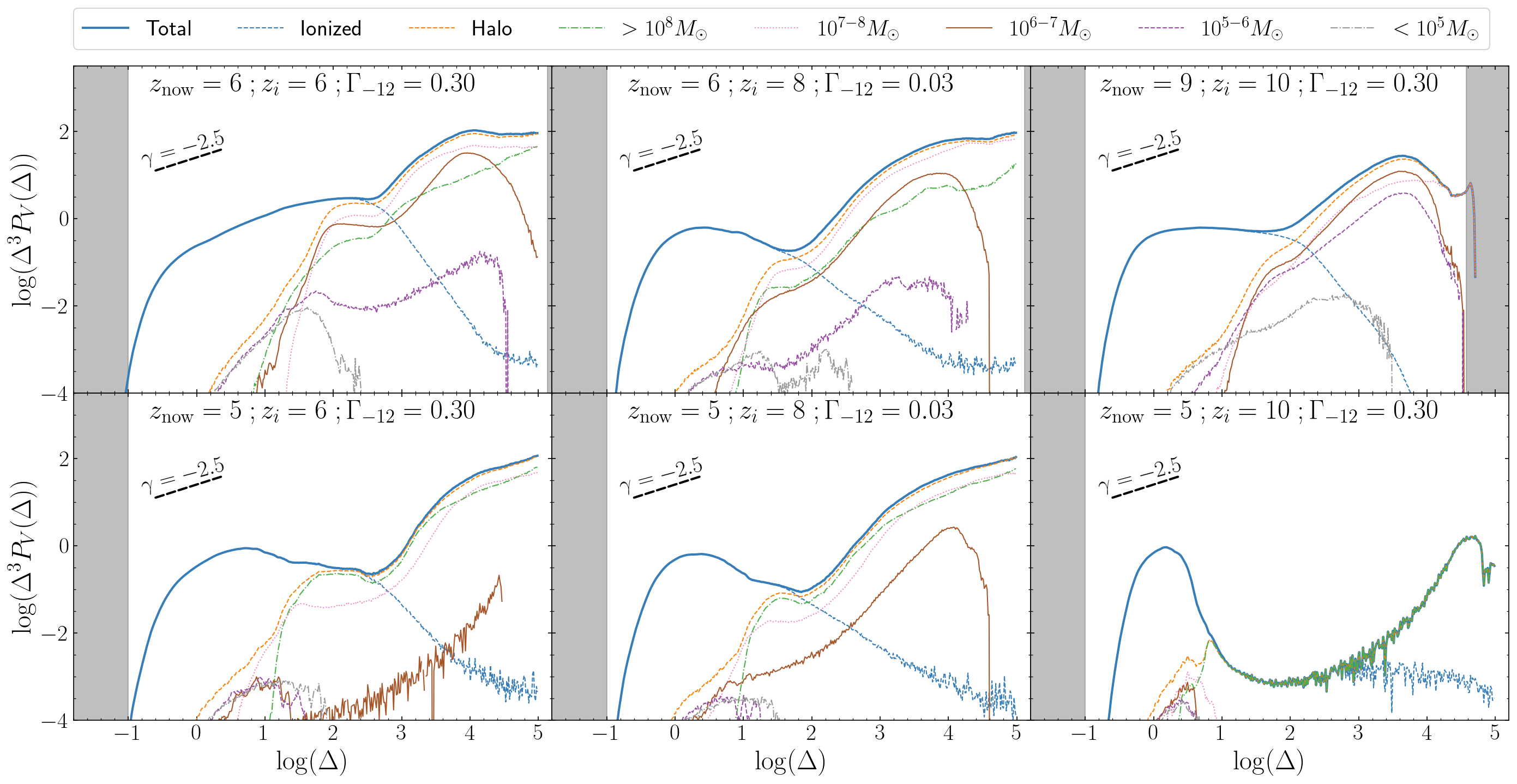}
\caption{Probability distribution by volume multiplied by $\Delta^3$, $\Delta^3{\cal P}_V$, as a function of the logarithm of the over-density, $\log\Delta$. In the top row, the \Ifront\ is still sweeping the computational volume. The lower row shows post-reionization (in practice, we show the last simulation snapshot at $z_{\rm now}\sim 5$). The \lq Total\rq\ lines refer to all gas in the volume, whereas the \lq Halo\rq\ lines refer only to gas in halos. We also distinguish the contribution of gas in halos according to halo mass (differently coloured lines). Those labelled \lq Ionized\rq\ correspond to $\Delta^3{\cal P}_V(\Delta)x_{\rm HII}^2$ such that the area under the curve is proportional to the recombination rate per unit volume per decade in $\Delta$.
The {\em left shaded regions} correspond to densities too low to resolve accurately, and the {\em right shaded regions} indicate gas above our artificial \lq star formation threshold\rq\ where the simulation is unreliable due to missing physics. The {\em short dashed black lines} indicate the power-law
${\cal P}_V\propto\Delta^\gamma$, for $\gamma=-2.5$ for comparison. This power law describes the scaling of \protect\cite{Mira00reion}'s model at high densities (see the main text). The low-mass \mini\ are photo-evaporated between the upper and lower panels.}
\label{fig:IGMdensity}
\end{figure*}

\begin{figure}
\includegraphics[width={0.48\textwidth}]{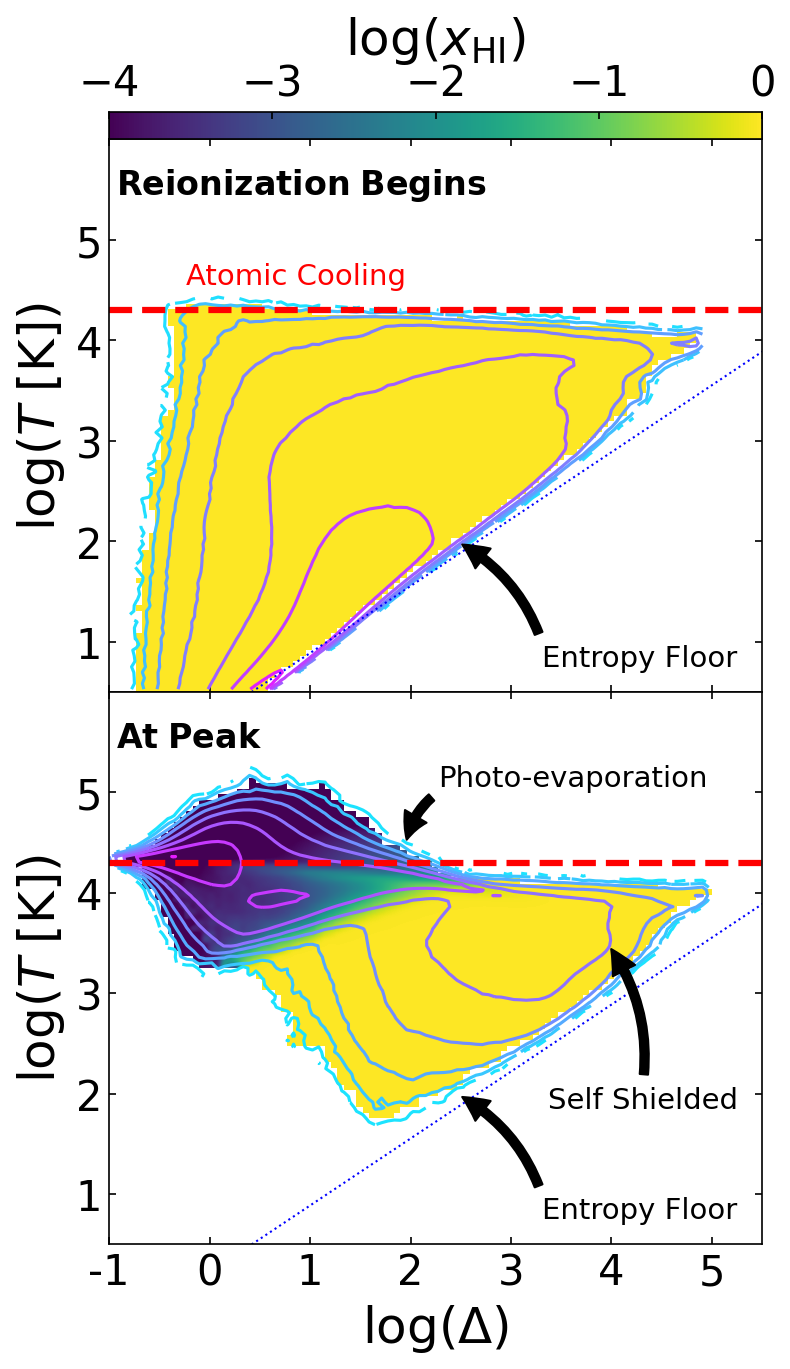}
\caption{Temperature vs over-density diagrams for the M512z8G03 run at the start of reionization, $z\lesssim z_i$,  ({\em upper panel}) and when the clumping factor $c_r$ is maximal ({\em lower panel}). The {\em colour scale} encodes the mass-weighted neutral fraction. The {\em red dashed line} is the temperature where the cooling due to H{\em I} is maximum. The {\em blue dotted line} is the initial entropy (Eq.~\ref{eq:entropyfloor}). {\em Contours} indicate the surface density
of gas particles in this plot, evenly spaced in log surface density.}
\label{fig:phase}
\end{figure}

\begin{figure}
\includegraphics[width={0.48\textwidth}]{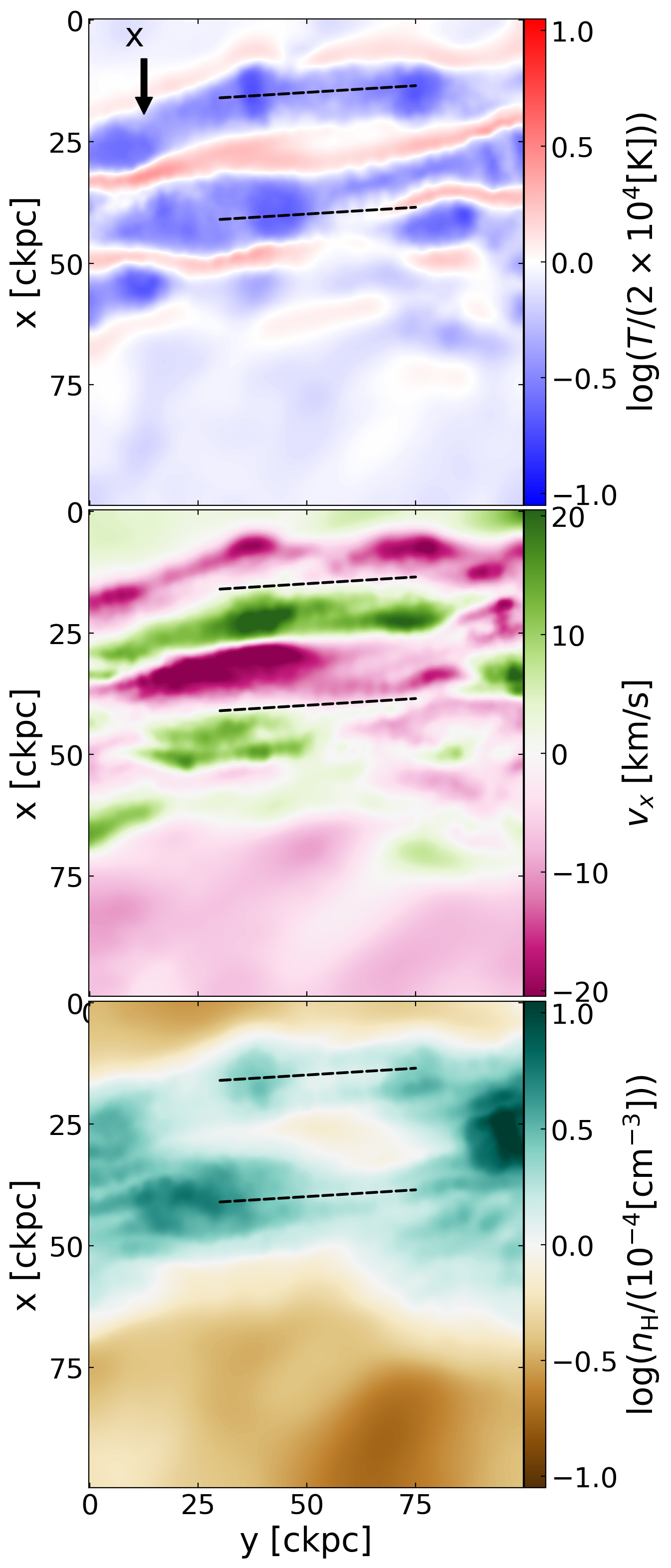}
\caption{Slices through the M512z8G03 run at $z=7.58$, after the \Ifront\ has crossed the simulation volume (see Fig.~\ref{fig:simoverview}). Panels show
temperature ({\em upper panel}), velocity in the $x$-direction ({\em central panel}), and hydrogen density ({\em lower panel}). These particular slices illustrate the photo-evaporation of two filaments, indicated by {\em dashed black lines} in each panel. At this moment in time, the density in the filaments is higher than in their surroundings (lower panel), yet the gas is cooler due to adiabatic expansion. The gas expands at a speed close to the sound speed (central panel) and compresses and heats the gas in the surrounding \ihm\, as seen in the upper panel. This type of compressional heating is the origin of the hotter gas, $T>10^{4.2}{\rm K}$, seen in Fig.~\ref{fig:phase}.}
\label{fig:filament}
\end{figure}

We examine the volume density distribution of our simulations in Fig. \ref{fig:IGMdensity}. Here we plot $\Delta^3\,{\cal P}_V(\Delta)$ as a function of $\log \Delta$. We show these curves for the total gas density (solid blue) and ionised gas density (dashed blue). First, we study the response of the \ihm\ to the passing of \Ifront\ for the simulation M512z8G03 (left panel) quantitatively. Upon the passage of \Ifront , low-density ($\Delta<5$) gas got highly ionised and heated to $T\sim 2\times 10^4{\rm K}$ (see also Fig.\ref{fig:phase}). But this gas changes relatively little in volume density (since it cannot expand further). On the other hand, self-shielded gas at high density ($\log\Delta>3$) remains mostly neutral and cold, whose $P_V$ also is not strongly affected by reionization due to self-shielding. Photo-heating significantly reduces $P_V$ of intermediate densities, $\log\Delta\sim 1-3$, since this gas is not self-shielded and expanded after photo-heating. More detailed analysis (including particle tracking) is presented in Appendix \ref{partracking}.

Other curves in Fig. \ref{fig:IGMdensity} show the total gas density \pdf's for all gas in halos (dashed orange), halos in the mass range $\log M_h[M_\odot]>8$ (green dashed), $7<\log M_h[M_\odot]<8$ (pink dotted), $6<\log M_h[M_\odot]<7$ (brown solid)
$5<\log M_h[M_\odot]<6$  (purple dashed) and $\log M_h[M_\odot]<5$ (grey dashed). The top panels correspond to the beginning of reionization when the \Ifront\ is still traversing the volume; the lower panels are at redshift $z=5$. Panels from left to right correspond to different values for $z_i$ and $\Gamma_{-12}$, as indicated. The straight black dashed line shows ${\cal P}_V\propto \Delta^{-2.5}$, which corresponds to the \pdf\ slope of an isothermal profile (see \citealt{Mira00reion} and discussions below).

In the left panel, we can see how halos with $\log M_h<7$ photo-evaporate, with their contribution to the high-density \pdf\, $\log\Delta>2$, decreasing dramatically between $z=6$ and $z=5$. 
The intermediate density gas, $1<\log\Delta<3$, is mostly associated with the more massive halos, $\log M_h>8$. 

The middle panels correspond to a case with $z_i=8$ and a lower photo-ionisation rate of $\Gamma_{-12}=0.03$
as compared to $z_i=6$ and $\Gamma_{-12}=0.3$. Nevertheless, the \pdf's at $z=5$ are quite similar, with the most striking difference being the location of the upturn in the \pdf, which occurs around $\log\Delta=2.6$ in the left panel and $\log\Delta=2$ in the middle panel. We note that this upturn is also the location where the gas turns from mostly ionised to mostly neutral.

The right panels correspond to a model with $z_i=10$ and $\Gamma_{-12}=0.3$. Although this model has the same value of the photo-ionisation rate as the model in the left panel, the \pdf's look strikingly different, with, in particular, very little gas at high densities. The reason for this becomes clear by comparing the top panels: the more massive halos have not formed yet by $z\sim 10$, and these halos do not accrete gas once it is photo-heated. This shows that reionization has a bigger impact on more massive halos by preventing them from accreting hot gas rather than by photo-evaporating their gas. 

\cite{Mira00reion} found that the gas \pdf\ is described well by the form ${\cal P}_V\propto \Delta^{\gamma}\,\exp(-\Delta^{4/3}/\sigma^2_\Delta)$, where $\gamma(=-2.5)$ and $\sigma_\Delta$ are fitting parameters. Their model is based on simulations with uniform UV background and the optically thin approximation \citep{Mira96Lyasim}.

With full radiation hydrodynamics here, our result (Fig. \ref{fig:IGMdensity}) does not agree with \cite{Mira00reion} (compared with the $\gamma=-2.5$ scaling in Fig. \ref{fig:IGMdensity}). Our \pdf\ has a stronger dip at $\log\Delta\sim 2$, followed by a stiff profile at higher density. With RHD, we capture the self-shielding of gas inside halos, which is less affected by the radiation, so the dense gas can maintain a stiff profile. The ionised gas, on the other hand, is photo-heated and evaporated, which explains the dip at $\log\Delta\sim 2$.

\cite{McQu11LLS} also showed a drop in gas density (at $\delta \sim 10^2$) compared to \cite{Mira00reion}. They turned off ionization background for $n_{\rm H}> 10^{-2}{\rm cm}^{-3}$ and consider self-shielding with pro-process radiative transfer. A dip in the \pdf\ around $\delta \sim 10^2$ was also found in other high-resolution RHD simulations, e.g. \cite{Park16clumping,DAlo20clumping}. These results support that the self-shielding of gas is responsible for the deviation from \cite{Mira00reion}.

The processes described here can be visualised
in the temperature-density diagrams of Fig.~\ref{fig:phase}, before (upper panel) and after (lower panel) the passage of the \Ifront.
The gas in the simulations has an initial entropy, $T/\rho^{\gamma-1}$, at the start of the calculations, set by the initial temperature of the gas (Eq. \ref{eq:entropyfloor}). This initial entropy is shown by the diagonal dotted line. But the gas can increase its entropy and move upwards from the dotted line through shocks during structure formation.

Before reionization, the gas temperature remains below $\sim 10^4\;{\rm K}$ due to Compton and atomic line cooling. Hence, pre-reionization gas exists in the triangular-shaped region of the upper panel. But we note that by mass, most of the gas remains close to the entropy floor, as shown by the contours.

After the \Ifront\ has crossed the computational volume, gas with $\log\Delta\le 2$ is ionized and photo-heated to $T\sim 2.2\times 10^4~{\rm K}$. Gas at higher densities self-shields, remains mostly neutral, and is at $\log T~[{\rm K}]\sim 2-4$.  The lower density gas with $\log\Delta\sim 0-1$ is mostly at $T\sim 2.2\times 10^4{\rm K}$, but some gas is hotter, and some gas is cooler. This results in a diamond-shaped region in Fig.~\ref{fig:phase}. As we explained previously, the origin of the hotter gas is due to adiabatic compression and shocking of under-dense gas by filaments\footnote{We tracked particles with $T>9\times 10^4 {\rm K}$ and $\log\Delta\sim 0-1$ back in time to investigate the evolution of their entropy. After being photo-ionized, their entropy changes slightly, whereas their density may change by one order of magnitude. This means that their temperature change is mostly due to adiabatic compression/expansion, with a small contribution from shock. This is consistent with the scatter in temperature being approximately symmetric around the $T\sim 2.2\times 10^4{\rm K}$. }. But clearly, some gas can also cool adiabatically, explaining the lower diamond region.

The connection with filaments is illustrated in more detail in Fig.~\ref{fig:filament}, which shows a close-up of two filaments that are nearly parallel to the $y$-axis (plotted horizontally in this figure). The middle panel shows that both filaments are expanding in the $x$ direction
with a speed $|v_x|$ of up to $20\;{\rm km}~{\rm s}^{-1}$ (which is higher than the sound speed). This adiabatic expansion cools the central region of the filament and heats the surroundings \ihm\ through compression (upper panel). Similar adiabatic changes are seen in the expansion of \mini\ and the resulting compression in their surroundings, e.g. Test~7 in \cite{Ilie09RTcom}.

\subsection{The inverse Str\"omgren layer}
\label{sec:shielding}
\begin{figure}
\includegraphics[width=0.45\textwidth]{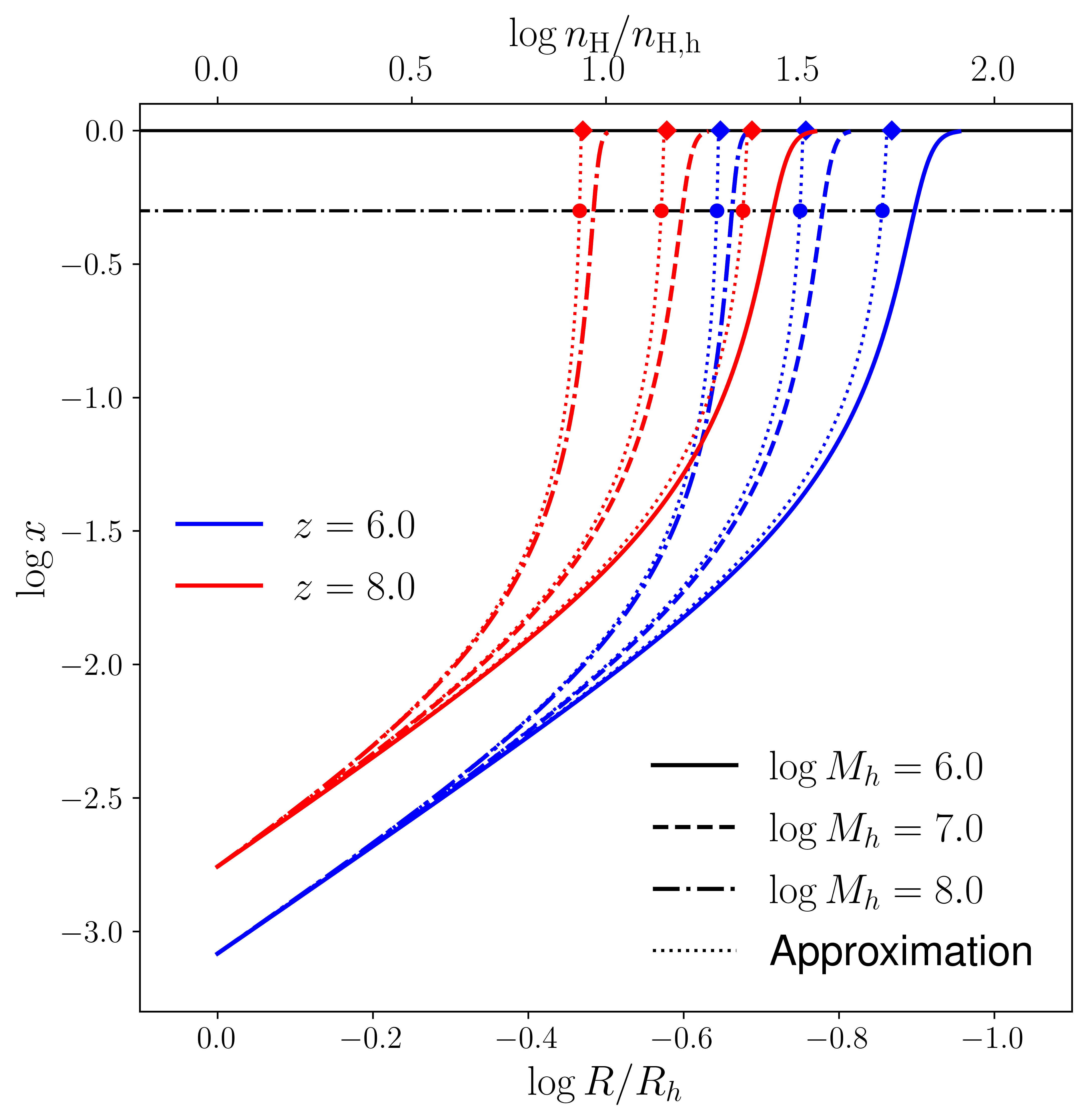}
\caption{Neutral fraction as a function of radius (lower horizontal axis) or density (upper horizontal axis) in halos with a spherical density profile of the form $n_{\rm H}=n_{\rm H,h}(R_h/R)^2$, where $R_h$ is the virial radius, for different halo masses (different line-styles) and redshifts (colours) as per the legend, obtained by integrating Eq.~(\ref{eq:x_R}). The {\em horizontal dot-dashed line}
and {\em solid line } indicates $x=1/2$ and $x=1$. The {\em dotted line} is the approximation for $x(R)$ from Eq.~(\ref{eq:x_R_approx}). {\em Small dots} and {\em small diamonds} indicate the value where $x=1/2$ and $x=1$ in the approximate model. The calculations take $\Gamma_h=10^{-12}{\rm s}$ for the photo-ionization rate at the virial radius, $\alpha_r=\alpha_{\rm B}(T=10^4{\rm K})$ for the recombination rate, and $\sigma_{\rm HI}=1.62\times 10^{-18}{\rm cm}^{-2}$ for the photo-ionization cross section. The approximate model is fairly accurate, even in predicting the location and density where $x=1/2$ and $x=1$. The value of $n_{\rm H,h}=8.6\times 10^{-3}$ and $4.1\times 10^{-3}{\rm cm}^{-3}$ at $z=8$ and $z=6$.}
\label{fig:xR}
\end{figure}

\begin{figure}
\includegraphics[width=0.45\textwidth]{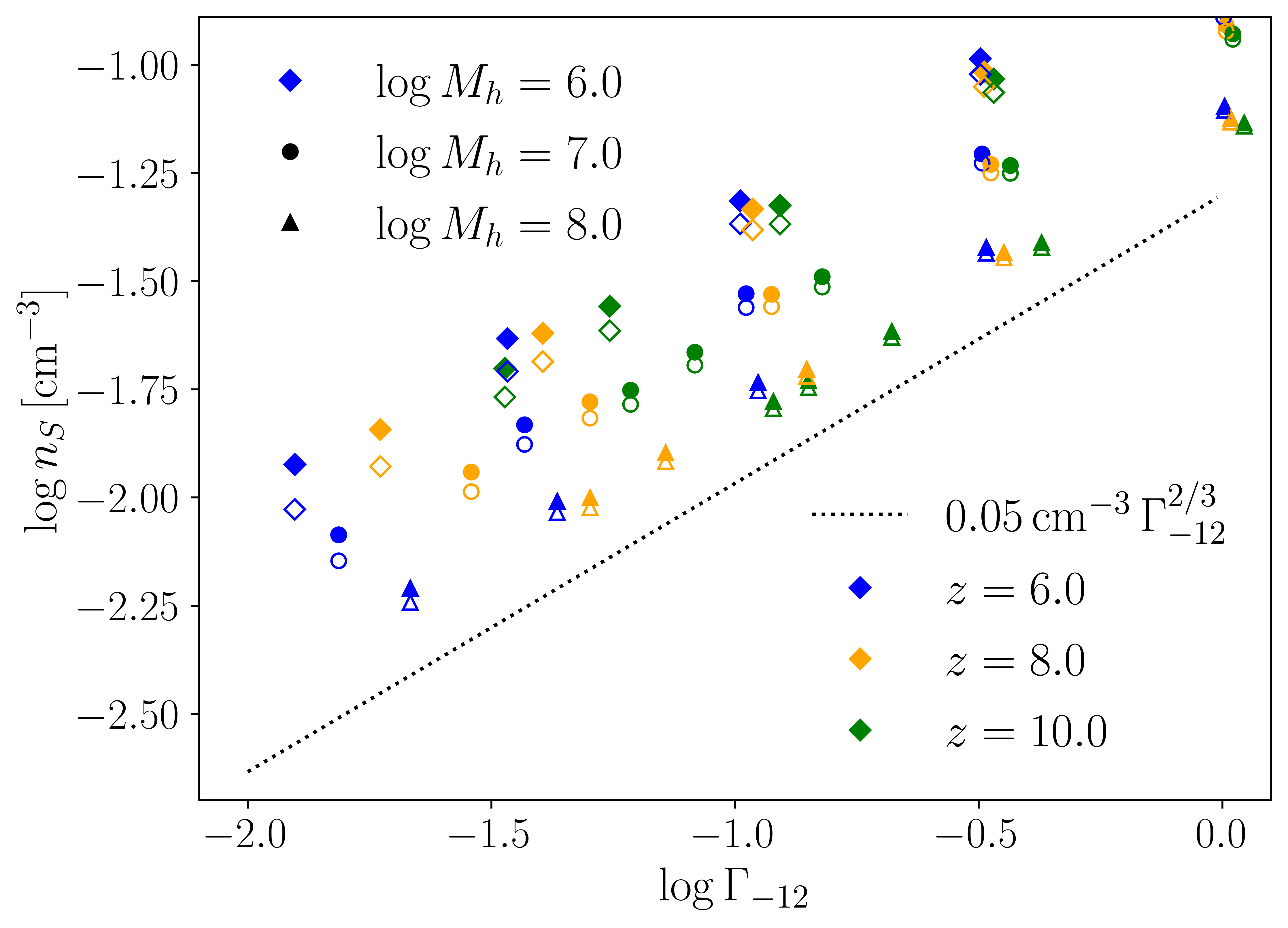}
\caption{The transition to neutral gas in the model
of \protect\cite{Theuns21}. {\em Filled symbols}
denote the density at the inverse Str\"omgren radius,
computed from Eq.~(\ref{eq:nSapprox}). {\em Open symbols} denote the minimum density where $x>1/2$, obtained from integrating Eq.~(\ref{eq:x_R}). In both cases we used Eq.~(\ref{eq:dGamma}) to relate the photo-ionization rate at the virial radius, $\Gamma_h$, to its volume-averaged value, $\Gamma_0\equiv 10^{-12}\Gamma_{-12}{\rm s}^{-1}$. The symbol type encodes virial mass, $M_h$ in units $\msun$, and its colour encodes redshift, as per the legends. The {\em black dotted line} shows the scaling $\propto\Gamma_0^{2/3}$. We have evaluated the recombination coefficient at $T=10^4~{\rm K}$ and set $\sigma_{\rm HI}=1.62\times 10^{-18}{\rm cm}^2$.}

\label{fig:nS}
\end{figure}
\begin{figure}
\includegraphics[width={0.48\textwidth}]{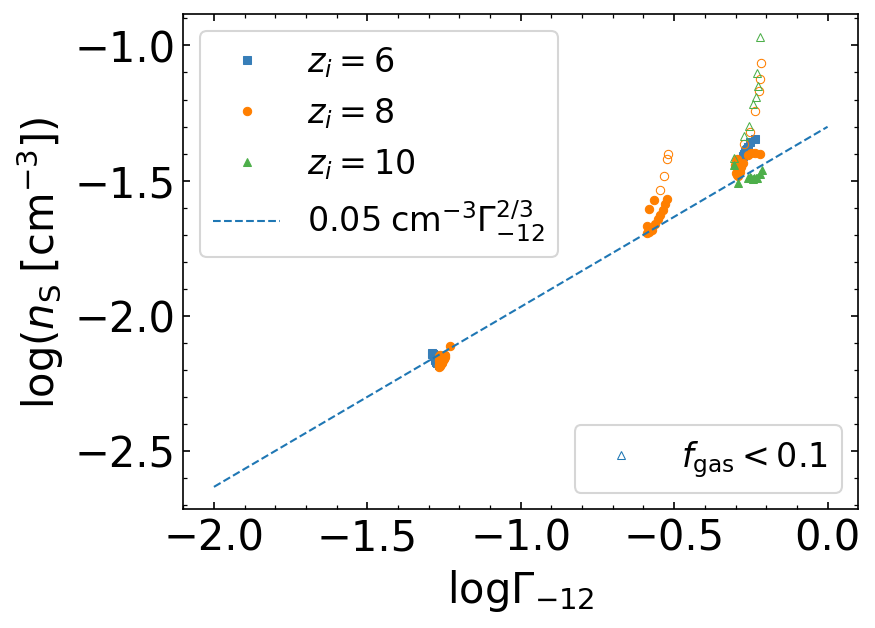}
\caption{Hydrogen density, $n_S$, where gas transitions to neutral, $x>1/2$, in the simulations plotted against the volume-averaged photo-ionization rate, $\Gamma_{-12}$, after the \Ifront\ has crossed the simulation volume. {\em Large markers} refer to gas in halos that contain more than 10~per cent of the cosmic baryon fraction, $f_b$; {\em small} markers are for halos containing less than 10~per cent of $f_b$. Colours and markers encode different values of $z_i$, as per the legend. The {\em dashed line} indicates the theoretical scaling ($n_S\propto \Gamma_{-12}^{2/3}$; Eq.~\protect\ref{eq:nSapprox}).}
\label{fig:nHtrans}
\end{figure}

When an \Ifront\ overruns neutral gas in a halo, its denser interior may be able to stall the front, provided the recombination rate is high enough. In \mini, the front will continue to move inwards as the gas in the outer parts photo-evaporates, but in more massive halos, the centre may continue to self-shield. In this section, we estimate the density, $n_S$, at the location of this \lq inverse Str\"omgren layer\rq, above which gas self-shields and remains mostly neutral \footnote{Note that $n_S$ is not the self-shielding density $n_{\rm H, SSh}$, used in \protect\cite{Rahm13HIshield}. The latter describes the density where the {\em optical depth due to self-shielding} reaches $\tau=1$, and $x$ can be much less than 1/2 even when $\tau=1$.}. In the simulations, we determine $n_S$ as the minimum gas density where $x>1/2$. The value of $n_S$ is of interest because it impacts the overall recombination rate and the clumping factor in a patch of Universe. A good understanding of what sets $n_S$ might also help to model self-shielding in simulations without performing computationally expensive \rt\ \cite[e.g.][]{Rahm14HIgalaxy,Ploe20COLIBREcooling}.

An analytical estimate for $n_S$ can be derived from the model of \cite{Theuns21} for damped Lyman-$\alpha$ systems. Assume that the density profile in a halo of virial radius $R_h$ is the power law
\begin{align}
    n_{\rm H}(r)&=n_{\rm H,h}\,\left(\frac{R_h}{R}\right)^2\,,
    \label{eq:prof}
\end{align}
where $n_{\rm H,h}(=200/3\langle n_{\rm H}\rangle)$ is the hydrogen density at the virial radius. Assuming that the gas is in photo-ionization equilibrium, the rate of change of the neutral density is zero,
\begin{align}
\frac{{\rm d}n_{\rm HI}}{{\rm d}t}=-\Gamma\,x\,n_{\rm H}+\alpha_{\rm B}(1-x)^2\,n^2_{\rm H}=0\,,
\end{align}
where $x$ is the neutral fraction and we have neglected collisional ionizations and any contribution from helium or other elements.
The photo-ionization rate $\Gamma$ at radius $R$ in the cloud is related to its value $\Gamma_h$ at $R_h$ by
\begin{align}
\Gamma = \Gamma_h\,\exp(-\tau)\,,
\end{align}
where the optical depth
\begin{align}
\tau=\int_R^{R_h}\,\sigma_{\rm HI}\,x\,n_{\rm H}\,dR\,.
\end{align}

Combining these yields
\begin{align}
\exp(-\tau) = \frac{\alpha_{\rm B}\,n_{\rm H, h}}{\Gamma_h}\frac{(1-x)^2}{x}\left(\frac{R_h}{R}\right)^2\,.
\end{align}
Taking the logarithm on both sides yields
\begin{align}
-\tau=-\int_R^{R_h}\,\sigma_{\rm HI}\,x\,n_{\rm H}\,dR = \ln\left[\frac{\alpha_{\rm B}\,n_{\rm H, h}}{\Gamma_h}\frac{(1-x)^2}{x}\left(\frac{R_h}{R}\right)^2\right]\,,
\end{align}
and taking the derivative with respect to $R$ 
\begin{align}
\sigma_{\rm HI}\,x\,n_{\rm H,h}\left(\frac{R_h}{R}\right)^2 = -\frac{1}{x}\,\frac{dx}{dR}-\frac{2}{1-x}\,\frac{dx}{dR}-\frac{2}{R}\,.
\end{align}

This is a differential equation for $x(r)$, which we write in dimensionless form as
\begin{align}
\left(-\frac{1}{x}-\frac{2}{1-x}\right)\,\frac{dx}{dr} = \frac{2}{r}+\tau_h\frac{x}{r^2}\,,
\label{eq:x_R}
\end{align}
where $\tau_h\equiv\sigma_{\rm HI}\,n_{\rm H,h}R_h$ 
is a characteristic optical depth for the halo, and $r\equiv R/R_h$. The boundary condition is the neutral fraction $x=x_h$ at $r=1$,
\begin{align}
\frac{x_h}{(1-x_h)^2}\approx x_h=\frac{\alpha_{\rm B}\,n_{\rm H,h}}{\Gamma_h}=\frac{t_i}{t_r}\,,
\end{align}
where the ionization and recombination time at $r=1$ are
$t_h\equiv {\Gamma_h}^{-1}$ and $t_h\equiv (\alpha_{\rm B}n_{\rm H,h})^{-1}$. This differential equation has no closed-form solution,
the numerical solution is plotted in Fig.~\ref{fig:xR} for a range of halo masses and two redshifts. In the ionized outskirts of the halo, we can take $x\ll 1$ so that $1-x\approx 1$, and the differential equation simplifies to
\begin{align}
-\frac{1}{x}\,\frac{dx}{dr} = \frac{2}{r}+\tau_h\frac{x}{r^2}\,,
\end{align}
with solution 
\begin{align}
x(r) = \frac{3rx_h}{-\tau_h\,x_h+(3+\tau_h\,x_h)r^3}\,.
\label{eq:x_R_approx}
\end{align}
This approximate solution is also plotted in Fig.~\ref{fig:xR}. Not surprisingly, it captures the increase in $x$ with increasing $n_{\rm H}$ very well when $x\ll 1$, but it also captures rather well the density and location where $x=1/2$ and even where $x=1$.
At a given redshift, $n_S$ (where $x=1$) is {\em higher} for lower halo masses. For a given halo mass, $n_S$ increases with increasing redshift.

We obtain the scaling of $n_S$ with $M_h$ and $z$ as follows.
We start by computing the value $R_S$ of the inverse Str\"omgren layer by expressing that the recombination rate along a ray, from $R_h$ to $R_S$, equals the impinging flux - this simply means that 
all photons impinging on the halo at $R_h$ have been used up by recombination between $R_h$ and $R_S$:
\begin{align}
\int_{R_S}^{R_h}\,\alpha_{\rm B}\,n^2_H(r)\,dr = \frac{\Gamma_h}{\sigma_{\rm HI}}\,,
\end{align}
so that for our $1/r^2$ density profile
\begin{align}
r_S \equiv\frac{R_S}{R_h}= \left(1+\frac{3\,t_r}{t_i\,\tau_h}\right)^{-1/3}\approx \left(\frac{3\,t_r}{t_i\,\tau_h}\right)^{-1/3}\,.
\label{eq:nS}
\end{align}
We note that $t_r$ depends on redshift but not on halo mass, whereas $\tau_h$ depends both on $M_h$ and redshift. The approximation of neglecting the \lq 1\rq\ in the round brackets applies to most cases of interest. In this approximation, we derive the following scaling relation for the hydrogen density $n_S$ at the Str\"omgren radius,
\begin{align}
    n_S&\approx \frac{1}{n_{\rm H,h}^{1/3}\,R_h^{2/3}}
    \,\left(\frac{3\Gamma_h}{\alpha_B\,\sigma_{\rm HI}}\right)^{2/3}\nonumber\\
    &\sim 0.09\,{\rm cm}^{-3}\left(\frac{1+z}{8}\right)^{-1/3}\left(\frac{M_h}{5\times 10^7\,M_\odot}\right)^{-2/9}
    \left(\frac{\Gamma_h}{10^{-12}{\rm s}^{-1}}\right)^{2/3}\,.
    \label{eq:nSapprox}
\end{align}
with the numerical value taking the gas temperature to be $T=10^4{\rm K}$ when evaluating the recombination coefficient, and
taking $\sigma_{\rm HI}=1.62\times 10^{-18}{\rm cm}^{-2}$. We note that the dependencies on halo mass and redshift are relatively weak. 

So far, we characterised the photo-ionization rate by its value $\Gamma_h$ at the virial radius. It is likely that $\Gamma_h<\Gamma_0$, where $\Gamma_0$ is the volume-averaged photo-ionization rate because the gas in the {\em surroundings} of the halo also causes absorption and hence suppresses the ionizing flux. We can estimate the importance of this effect by simply extrapolating the $1/r^2$ density profile to infinity:
\begin{align}
\int_{R_h}^\infty \alpha_B\,n^2_H\,dr = \frac{\Gamma_0-\Gamma_h}{\sigma_{\rm HI}}\,.
\end{align}
This yields the following relation,
\begin{align}
\Gamma_h &=\Gamma_0-\frac{n_{\rm H,h}\tau_h\,\alpha_r}{3}
\equiv \Gamma_0\,\left[1-\exp(-\tau_\infty)\right]\nonumber\\
\exp(-\tau_\infty) &= \frac{n_{\rm H,h}\tau_h\,\alpha_r}{3\Gamma_0}
\label{eq:dGamma}
\end{align}
We plot $n_S$ computed from Eq.~(\ref{eq:nSapprox}) and use the previous equation to relate $\Gamma_h$ to $\Gamma_0$ in Fig.~\ref{fig:nS}. 

The transition from neutral to ionised gas is relatively sharp in the simulations, as can be seen by comparing the \lq Ionised\rq\ and \lq Neutral\rq\ lines in Fig.~\ref{fig:IGMdensity}. This justifies the \cite{Mira00reion} assumption for a characteristic density dividing mostly neutral and mostly ionised gas, which was also shown in other numerical studies, e.g. \cite{McQu11LLS}, \cite{Park16clumping} and \cite{DAlo20clumping}. Therefore, the value of $n_S$ is relatively well defined, and operationally we determine it for each halo as the minimum density for which $x>1/2$. We plot this value in Fig.~\ref{fig:nHtrans} as a function of $\Gamma_{-12}$. The analytical relation of Eq.~(\ref{eq:nSapprox}) captures the $n_S\propto \Gamma^{2/3}_{-12}$
scaling, but the simulated value of $n_S$ is smaller. We suspect this is due to the density structure in the accreting gas, which the analytical model does not account for. 

We generalise the analytical calculation to the case of a halo overrun by a plane-parallel \Ifront\ in Appendix \ref{sec:minihalotest}. We use this to test the ability of the \rt\ scheme to capture an \Ifront\ at the typical numerical resolution with which we simulate \mini.

\subsection{Photo-evaporation of \mini}
\label{sec:mhevaporation}

\begin{figure}
\includegraphics[width={0.48\textwidth}]{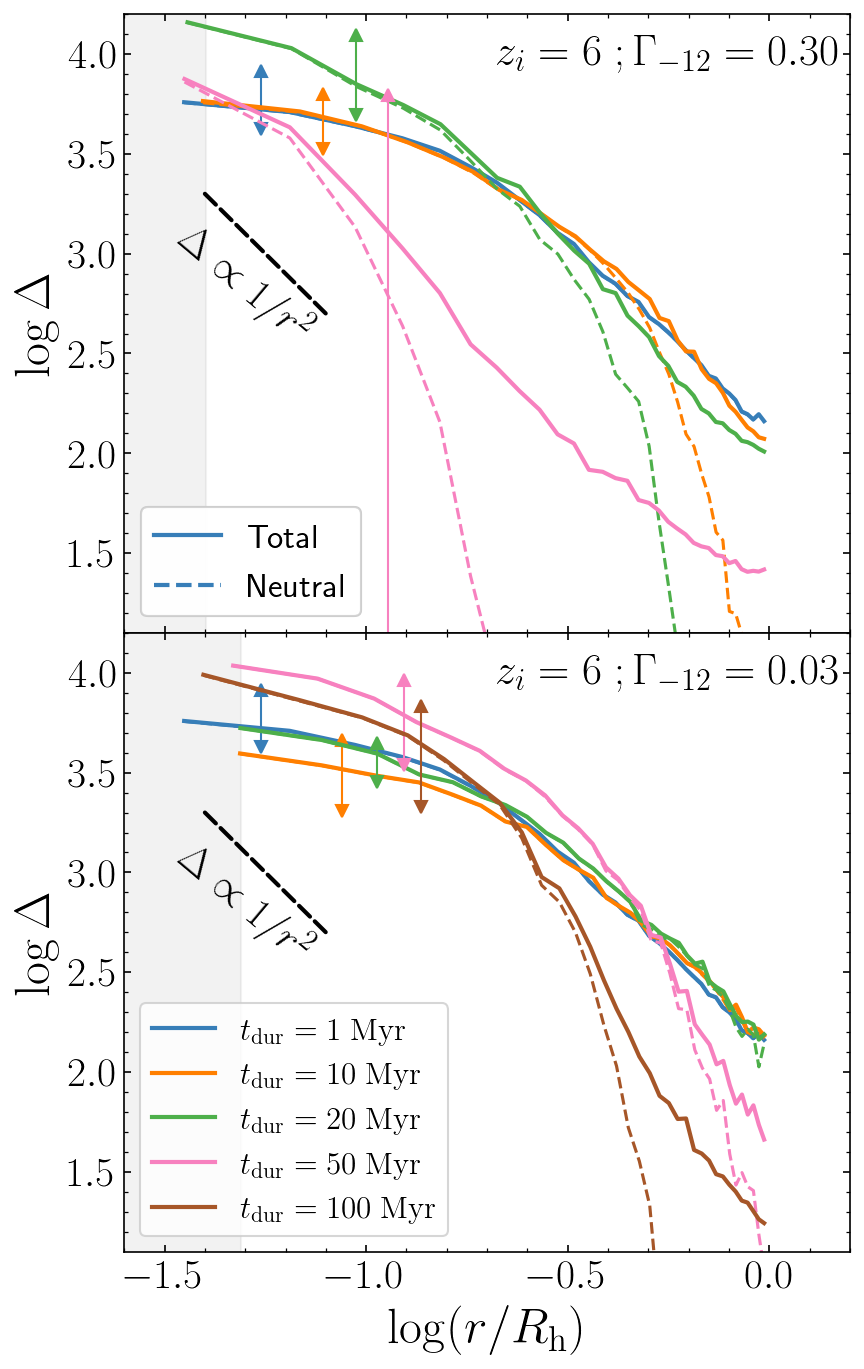}
\caption{Gas over-density profiles of halos with $M_{\rm h}\sim 10^6 \msun$ at different times, $t_{\rm dur}$, after they were over-run by an \Ifront.
The {\em solid lines} are total gas density, whereas the {\em dashed lines} are neutral gas density. The {\em arrows} indicate the 15$^{\rm th}$ and 85$^{\rm th}$ percentiles of halos with similar values of $t_{\rm dur}$. The {\em black dashed lines} show power-laws for isothermal profiles, $\rho(\Delta)\propto \Delta^{-2}$, that describe the run of the total density with radius in the range $0.1\le r/R_h\le 1$ relatively well ($R_h$ is the virial radius of the halos). Gas in the  {\em shaded region} is within a gravitational softening length from the centre.}
\label{fig:halodensityprofile}
\end{figure}

\begin{figure}
\includegraphics[width={0.45\textwidth}]{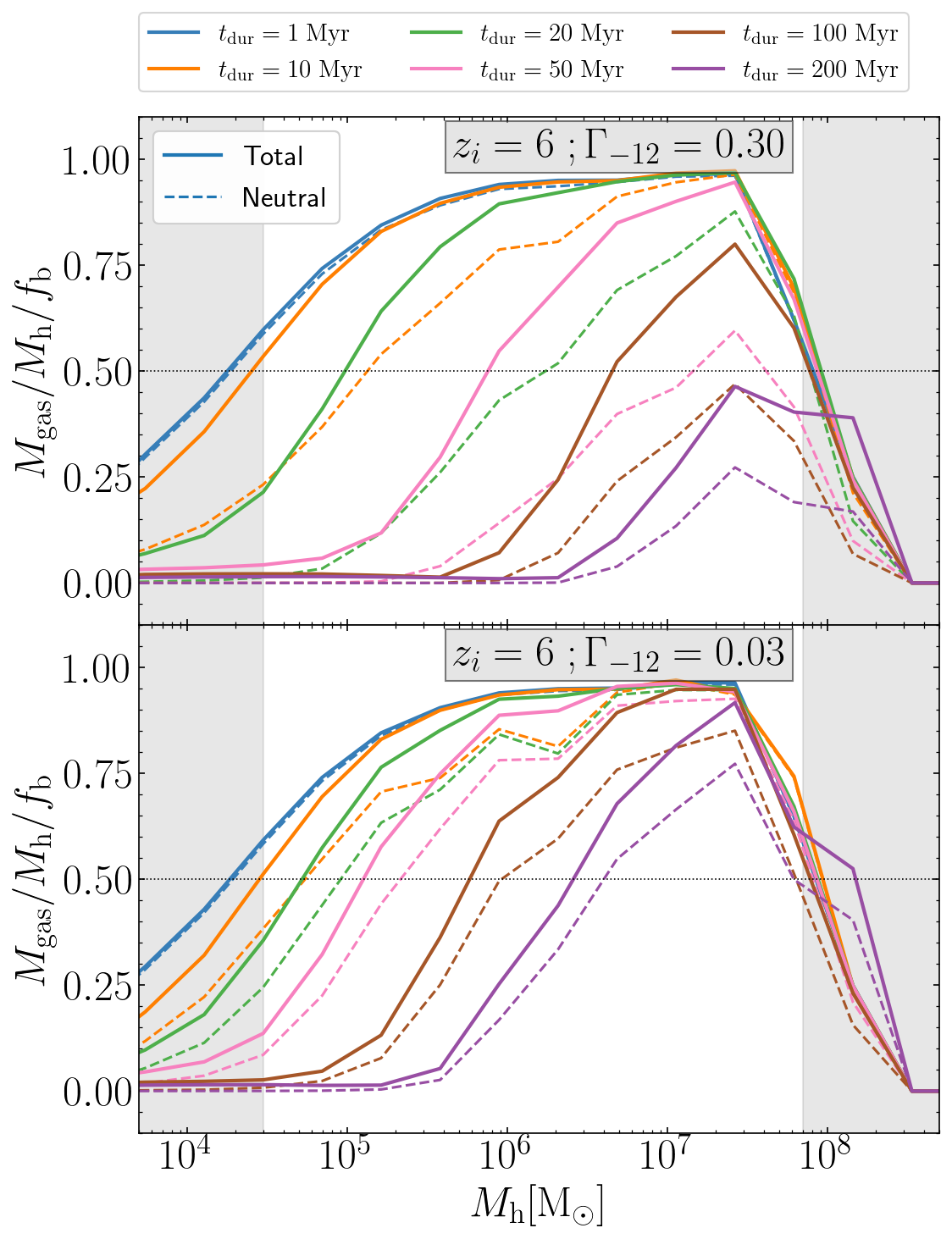}
\caption{Ratio between gas mass within the virial radius and the halo mass, in units of the cosmic baryon fraction, $f_b$. {\em Solid lines} refer to total gas mass, {\em dashed} lines to neutral gas. Different colours correspond to different times, $t_{\rm dur}$, since the halo was overrun by \Ifront, from {\em blue} (recently) to {\em purple} (long ago). The {\em upper panel} has $\Gamma_{-12}=0.3$, whereas the {\em lower panel} has the lower photo-ionization rate of $\Gamma_{-12}=0.03$. Halos in the {\em left shaded} regions are resolved with fewer than three hundred dark matter particles. Halos in the {\em right shaded region} are susceptible to atomic cooling. }
\label{fig:Htot_evap}
\end{figure}

\begin{figure*}
\includegraphics[width={1.0\textwidth}]{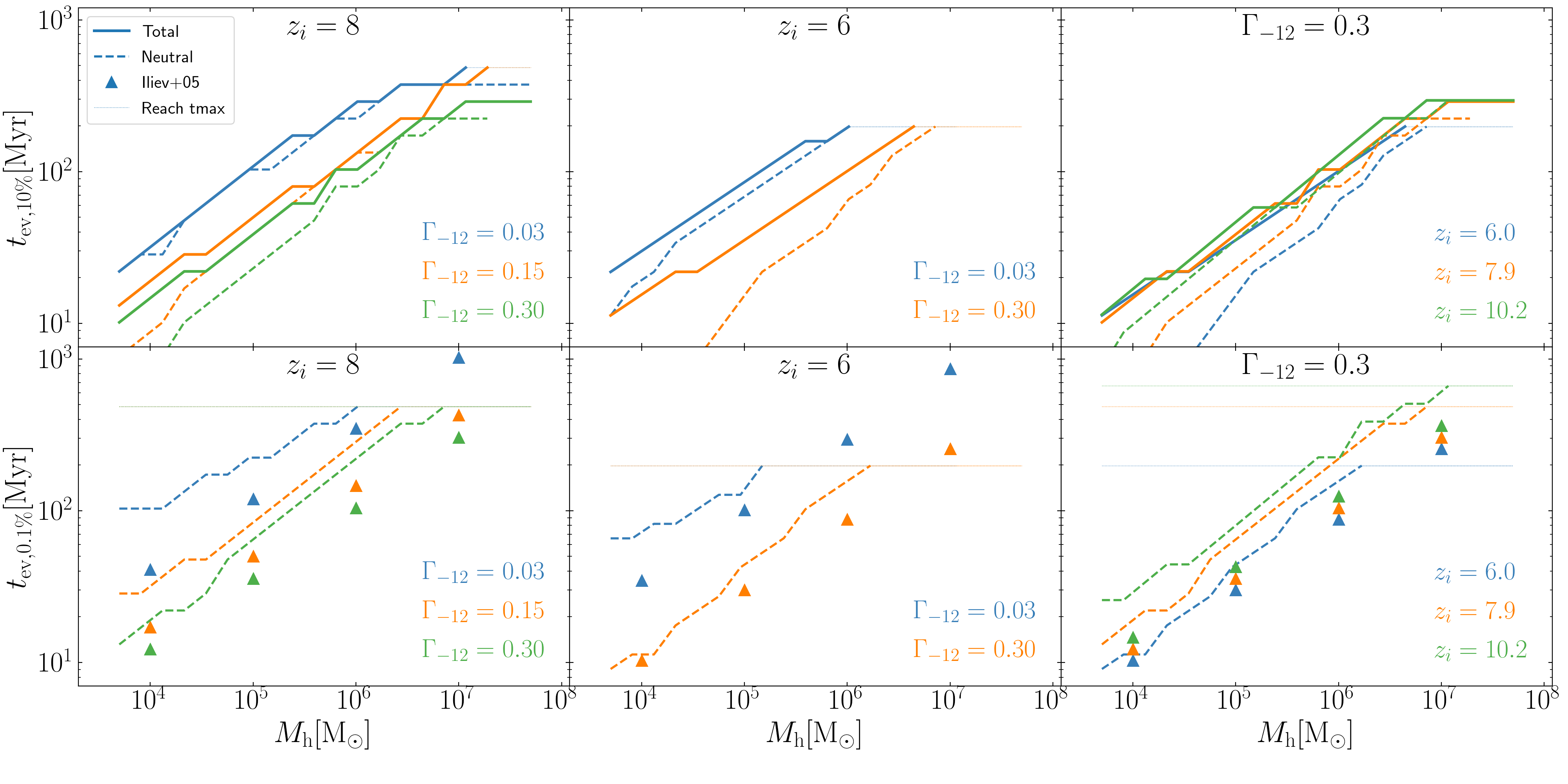}
\caption{Evaporation time, $t_{\rm ev}$, as a function of halo mass, $M_h$. The {\em upper} panels show the time to lose 90~per cent of the total gas mass ({\em solid lines}) or the neutral gas mass ({\em dashed lines}) (relative to the cosmic share of baryons; after the halo was overrun by \Ifront). {\em Thin dotted lines} indicate that evaporation is not yet complete when the simulation is stopped. The {\em lower panels} are similar, except that halos must lose 99.9~per cent of their gas. Triangles are the simulation results of \protect\cite{Ilie05minihalo}. Our evaporation times are slightly longer, but the trends with $M_h$ and $\Gamma_{-12}$ are very similar:
the evaporation time increases with $M_h$ and $z_i$, and decreases with increasing $\Gamma_{-12}$.
}
\label{fig:tevmulti}
\end{figure*}
Here we examine the time-dependent effects of the ionizing radiation on individual halos. To do so, we need to calculate how long \mini\ have been irradiated with ionizing radiation.

In our simulations, we inject ionizing photons from two opposing sides of the periodic volume. As a consequence, \mini\ close to the injection side are affected by the radiation earlier (and for longer) than those further away. To account for this time difference, we first calculate the \Ifront\ position, defined as the location where the volume-weighted neutral fraction is $\sim 0.5$ (we do this in cubic cells of extent ${\rm 10\;ckpc}$ instead of considering a plane-parallel \Ifront). For each halo, we then record the time since it encountered the \Ifront. Hence we can compute the duration $t_{\rm dur}$ between the current time and the time that the halo was first irradiated.

We select $\sim 20$ \mini\ with mass $M_{\rm h}\sim 10^6\msun$ and compute the median of their spherically averaged radial density profiles. We assume that the centre of the halo corresponds to the location where the density of neutral gas is highest\footnote{Taking a slightly different centre will affect the density profile mostly close to that centre. }.

The density profiles of these halos are plotted in Fig.~\ref{fig:halodensityprofile} for various values of $t_{\rm dur}$, and two values of the photo-ionization rate, $\Gamma_{-12}=0.3$ and 0.03. 

If radiative cooling is inefficient (as in mini-halos), the maximum gas density is limited by the entropy of the IGM after decoupling from radaiation \citep{Visb14gashalocenter}. Therefore the central density profiles are flat. Outside of the central regions, the profiles of the total density (solid lines) are reasonably well approximated by $n_{\rm H}(r)\propto r^{-2}$ (black dot-dashed line), at least before the \Ifront\ has propagated significantly into the halo's gas. 

Using the order-of-magnitude estimates derived in \S~\ref{sec:ana_eva}, we find for the sound crossing time $t_{\rm sc}\approx 20~{\rm Myr}$, the \Ifront\ crossing time $t_{\rm ic}=4\,\Gamma_{-12}^{-1}\,{\rm Myr}\approx13~{\rm Myr}$ for $\Gamma_{-12}=0.3$ (and 130~Myr for $\Gamma_{-12}=0.03)$, and the recombination time $t_{\rm r}=10~{\rm Myr}$.

Therefore halos are in the sound-speed limited
regime for $\Gamma_{-12}=0.3$ (upper panel, using the nomenclature of \S\ref{sec:ana_eva}). The \Ifront\ propagates rapidly into halos, reaching down to 5 per cent of the virial radius within 50~Myr. Gas starts photo-evaporating, but the time it takes to leave the halo is longer than $t_{\rm ic}$. Therefore the halo contains a large amount of highly ionized photo-heated gas. Given that the gas has density profile is $n_{\rm H}\propto r^{-2}$, the neutral gas has a density profile approximately $n_{\rm HI}(r)=x\,n_{\rm H}(r)\approx \alpha_B/\Gamma_{-12}n_{\rm H}^2\propto r^{-4}$.  

In contrast, halos are in the ionization limited regime when $\Gamma_{-12}=0.03$ (lower panel). The \Ifront\ propagates so slowly into the cloud that the photo-heated gas has time to photo-evaporate and leave the halo.
Consequently, the neutral and ionized gas profiles almost trace each other. For this low value of $\Gamma_{-12}$, even halos of this low mass can hold on to a large fraction of their gas for several $100{\rm Myr}$, and this affects the duration of the photo-evaporation phase of the \eor.

We demonstrate how fast a halo photo-evaporates
in Fig.~\ref{fig:Htot_evap}. We plot the halo baryon fraction in units of the cosmic mean, $M_{\rm gas}/(f_b\,M_h)$ (where $f_b\equiv \Omega_b/\Omega_m$), at various values of $t_{\rm dur}$. More massive halos can hold onto their gas for longer times. A $M_h=10^5~M_\odot$ halo loses half of its gas in $t_{\rm dur}=20~{\rm Myr}$ for $\Gamma_{-12}=0.3$. On the other hand, a halo with $M_h=10^7~M_\odot$ has only lost $\approx 20$~per cent of its gas at $t_{\rm dur}=50~{\rm Myr}$ at $\Gamma_{-12}=0.3$. A lower $\Gamma_{-12}$ can also slow down the photo-evaporation, e.g. the $M_h=10^5~M_\odot$ halo takes twice as long to photo-evaporate at $\Gamma_{-12}=0.03$ (than $\Gamma_{-12}=0.3$).

A more quantitative measure of the photo-evaporation timescale is
plotted in Fig.~\ref{fig:tevmulti}. We have computed the value of $t_{\rm dur}$ after which a halo contains only 10~per cent of the cosmic baryon fraction, which we refer to as $t_{\rm ev, 10\%}$ (upper panel).  For $\Gamma_{-12}=0.3$, we find the approximate scaling $t_{\rm ev, 10\%}\propto M_h^{1/2.3}$ (solid lines in the top right panel), which is somewhat shallower than the $t_{\rm ic}\propto M_h^{1/3}$ dependence of the \Ifront\ crossing time on halo mass. That panel also shows that the evaporation time depends surprisingly little on redshift. The dependence is even weaker than that of the sound crossing time, $t_{\rm sc}\propto (1+z)^{-1}$. $t_{\rm ev, 10\%}$ can be a few times 
$t_{\rm sc}$, because recombinations significantly delay ionization and hence photo-evaporation. 

In the left and central panels, we notice that the dependence of $t_{\rm ev, 10\%}$ on $\Gamma_{-12}$ is weaker than that of $t_{\rm ic}\propto \Gamma_{-12}^{-1}$. At the lower values of $\Gamma_{-12}=0.03$, the more massive halos have not yet reached $t_{\rm ev, 10\%}$ by the end of the simulation run: this is indicated by the thin dotted lines.

The dashed lines indicate when the halos contain less than 10~per cent of {\em neutral} gas. When $\Gamma_{-12}$ is low (e.g. $\Gamma_{-12}=0.03$), there is little difference between the total and neutral gas lines because once ionized, gas quickly leaves the halo: these halos are in the ionization-limited regime. However if $\Gamma_{-12}$ is larger
($\Gamma_{-12}=0.3$, say), some of the ionized gas is still inside the halo because it has not had time yet to photo-evaporate: these halos are in the sound-speed limited regime. The difference between the two regimes is more pronounced at lower $z$ and lower halo mass, as seen in the central panel.

The lower panel of Fig.~\ref{fig:tevmulti} shows $t_{\rm ev, 0.1\%}$, the time after which halos contain less than 0.1~per cent of the cosmic baryon fraction. The panels compare our results to those of \cite{Ilie05minihalo} (triangles), and we find an agreement within a factor of two.  Several differences between our simulations and theirs may explain the difference. Firstly, ours are 3D cosmological \rhd\ simulations, in which halos grow in mass through mergers and accretion during photo-evaporation. In contrast, theirs are 2D non-cosmological simulations. Secondly, our simulations include the effect of shadowing since we simulate a cosmological volume. Third, we consider radiation in one frequency bin, whereas their radiative transfer is multi-frequency. The multi-frequency treatment will allow high-energy radiation to penetrate further into halos and include effects of spectral hardening (which boost the photo-evaporation rates). Finally, our {\sc sph} simulation is particle-based, whereas they used a uniform grid. Our spatial resolution is comparable to theirs at high density but is lower in lower-density regions.

We do not plot the evaporation times of the total gas (solid) in the lower panels. It is because, in our simulations, the total gas fraction does not drop below 0.1~per cent: even \mini\ with $M_h\sim 10^4~{\rm M}_\odot$ can hold on to a small fraction of (highly ionized) gas.

\subsection{The Impact of Small-Scale Structure on Reionization}
\label{sec:clumping}

\begin{figure}
\includegraphics[width={0.45\textwidth}]{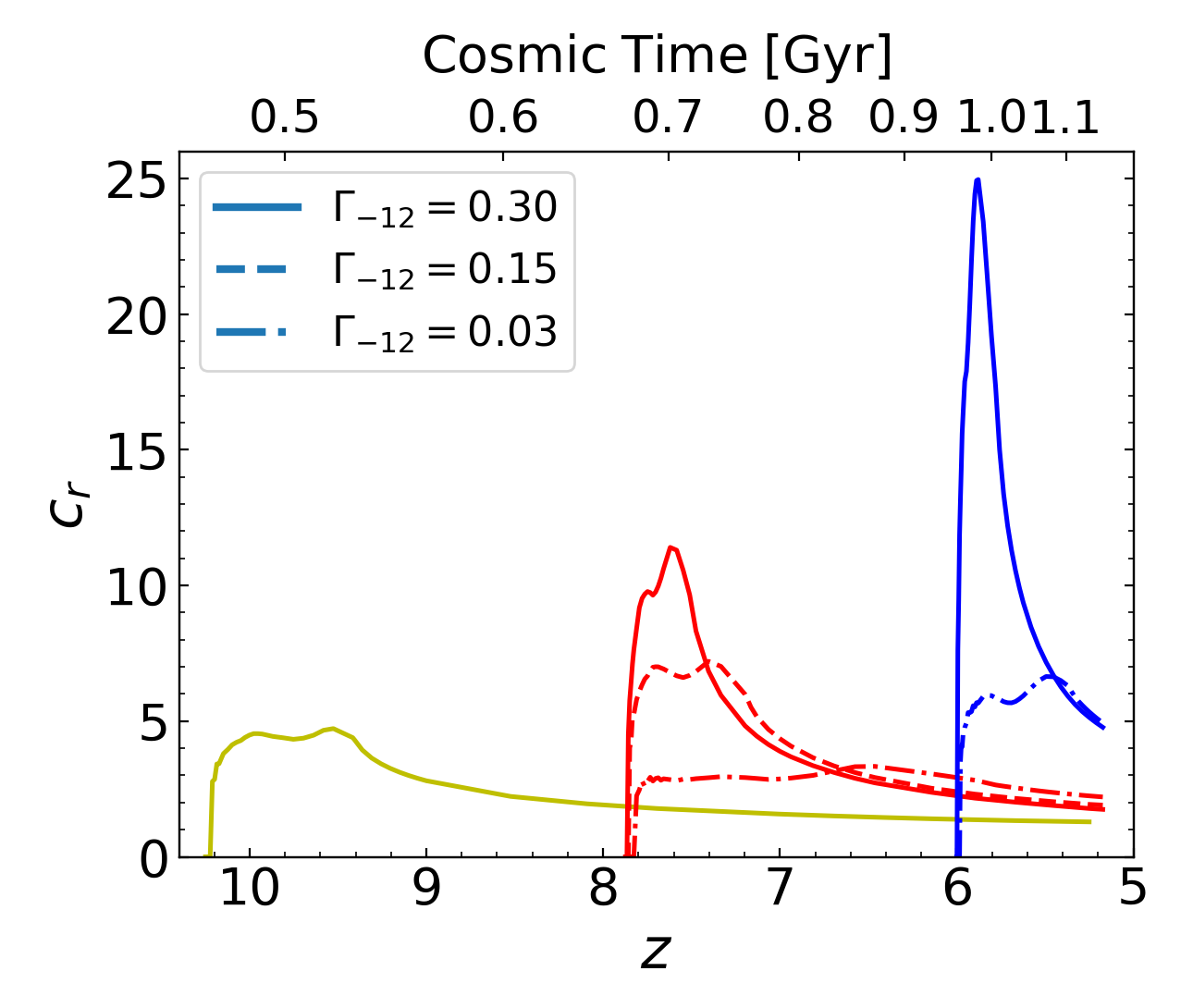}
\caption{Clumping factor, $c_r$ (Eq.~\ref{eq:crdef}), as a function of redshift, $z$, for simulations with different photo-ionization rates $\Gamma_{-12}$ ({\em solid}, {\em dashed}, {\em dot-dashed} lines correspond to $\Gamma_{-12}=$ 0.3, 0.15, and 0.03). The line colour indicates different redshifts of ionization ({\em yellow}, {\em red} and {\em blue} correspond to $z_i\sim$10, 8 and 6). When the \Ifront\ enters the simulation volume, $c_r$ rapidly increases to a peak value, $c_{r, {\rm peak}}$. Then, $c_r$ decreases on a time scale of order 100~Myrs before reaching asymptotically an approximately constant value. The value of $c_{r, {\rm peak}}$ increases with increasing $\Gamma_{-12}$ and decreasing $z_i$, but the asymptotic late-time value of $c_r$ is approximately independent of $\Gamma_{-12}$.
}
\label{fig:clumping}
\end{figure}

\begin{figure*}
\includegraphics[width={1.0\textwidth}]{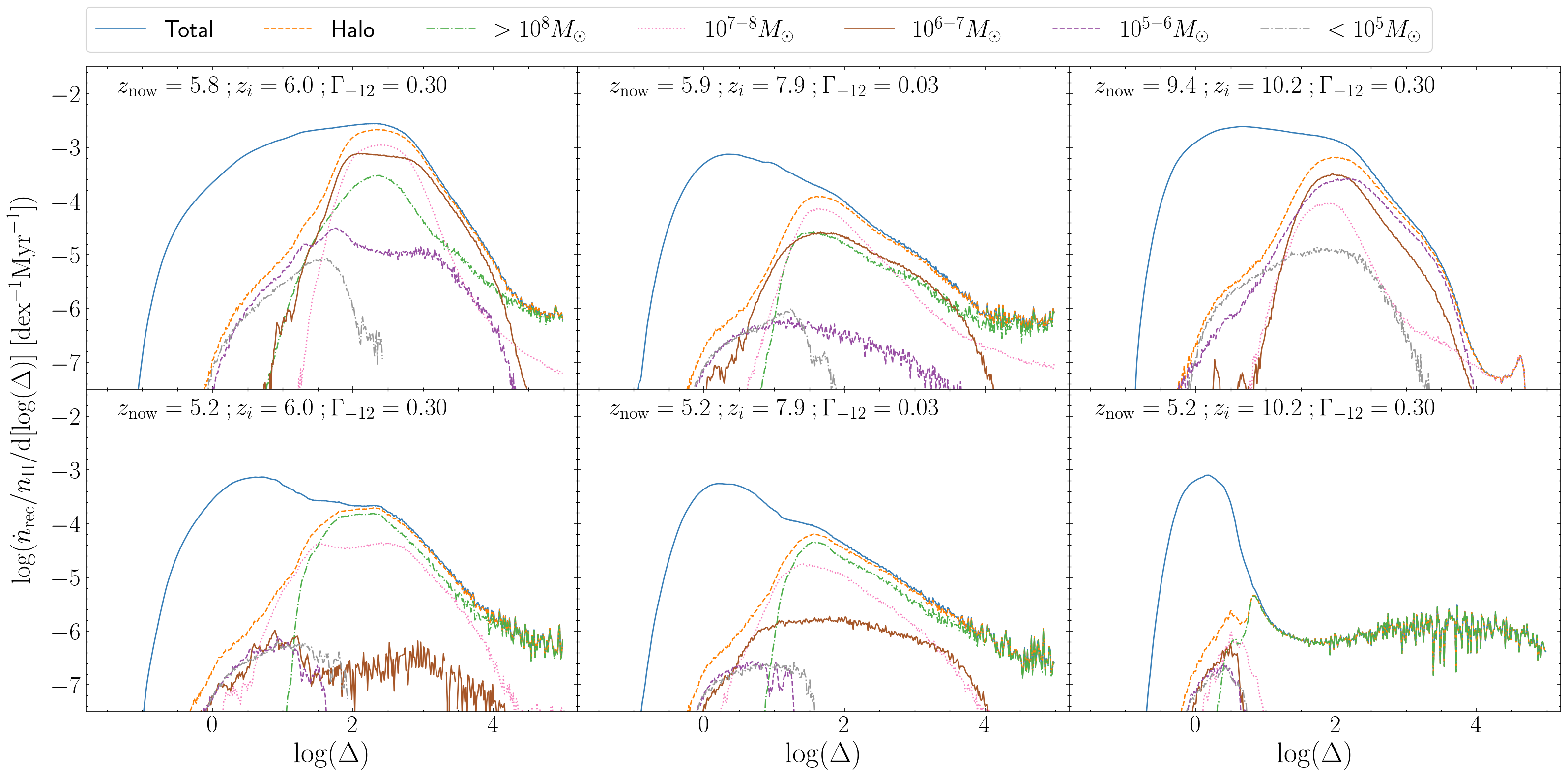}
\caption{Total recombination rate per hydrogen, $\dot n_{\rm rec}/n_{\rm H}$, as a function of gas over-density $\Delta$, per decade in $\Delta$. The area under the curve measures the contribution of gas at that over-density to the recombination rate in the simulation volume. Line styles and colours
are the same as in Fig.~\ref{fig:IGMdensity}. In a given column of panels, the {\em upper row} corresponds to $z\lessapprox z_i$, whereas the {\em lower row} corresponds to a lower redshift. Gas in halos contributes significantly to recombinations when $z\lessapprox z_i$ and $\Gamma_{-12}$ is high. As \mini\ photo-evaporate, the contribution of the \ihm\ to the overall recombination rate increases. At even lower redshifts (not shown), recombinations in more massive halos not contained in our small simulation volume dominate eventually.}
\label{fig:Recom_mbins}
\end{figure*}

\begin{figure}
\includegraphics[width={0.48\textwidth}]{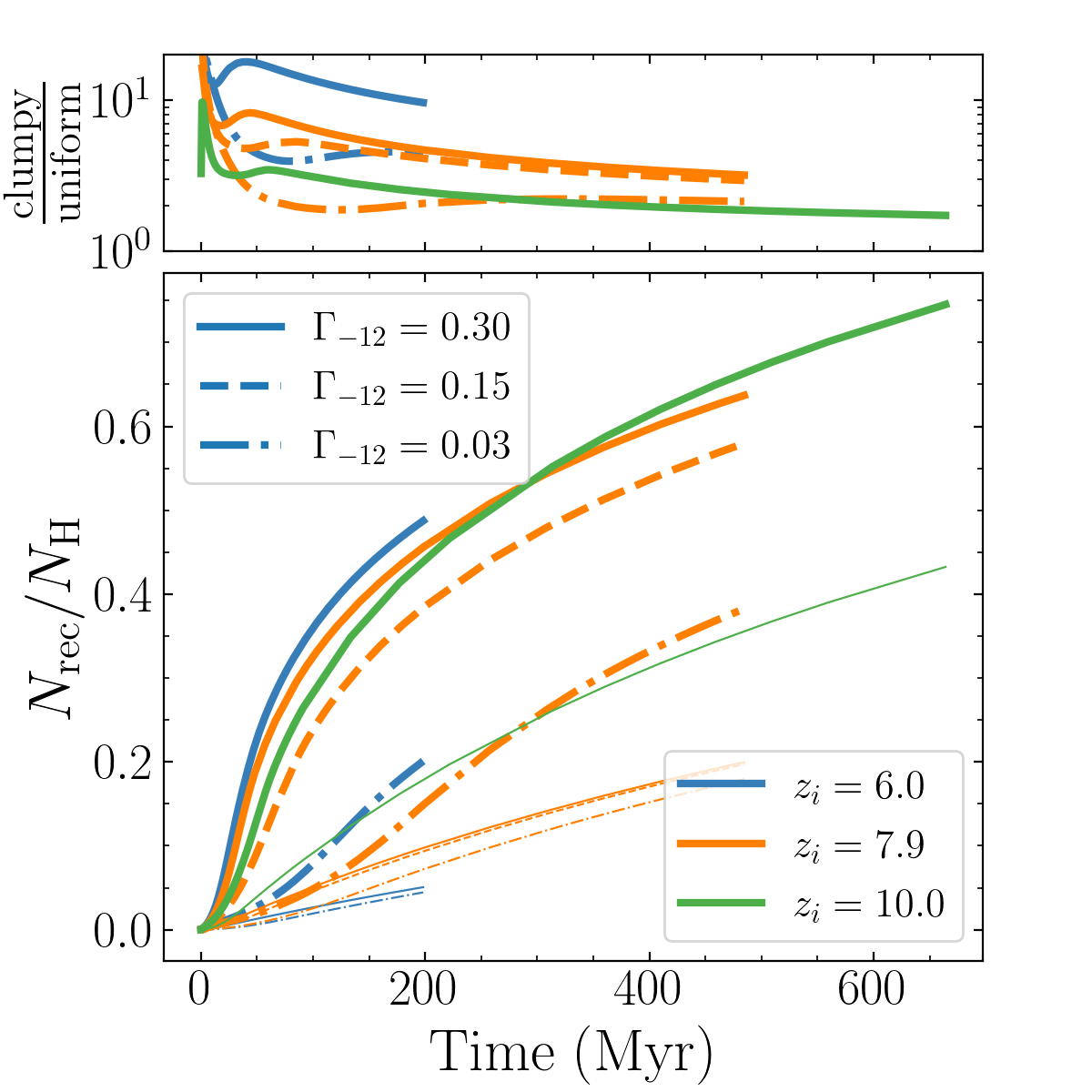}
\caption{ {\em Lower panel:} the cumulative number of recombinations per hydrogen atom, $N_{\rm rec}/N_{\rm H}$, as a function of time since ionizing photons were injected in the simulation volume. The {\em thick lines} represent the simulation results. The {\em thin lines} are computed for a homogeneous medium with the same median temperature. The ratio of these is plotted in the {\em upper panel}. Line colours and styles are as in Fig.~\ref{fig:clumping}.
Although the clumping factor can be quite high, $c_r=5-20$ (upper panel and Fig.~\ref{fig:clumping}), recombinations increase the required number of ionizing photons per hydrogen to complete reionization by a much smaller fraction, 1.2-1.7, because \mini\ photo-evaporate quickly. This fraction increases with $\Gamma_{-12}$ (different line styles) but is only weakly dependent on $z_i$
(different colours).}
\label{fig:recom}
\end{figure}

\subsubsection{The clumping factor, $c_r$}
The recombination rate in an inhomogeneous medium is enhanced compared with that of a uniform medium of the same mean density by the clumping factor, $c_r$\footnote{Note that $c_l$ as defined in the Introduction does not account for recombination rate but $c_r$ does.}, which we define here as
\begin{align}
\langle \frac{{\rm d}n_{\rm HI}}{{\rm d}t}|_{\rm rec}\rangle
= \langle\alpha_{\rm B}\,n_e\,n_{\rm HII}\rangle
\equiv c_{r, {\rm all}}\,\alpha_{\rm B}(\bar{T})\, \langle n_{\rm HII}\rangle^2\,,
\label{eq:crwholedef}
\end{align}
with the extra subscript \lq all\rq\ added for reasons explained below. Here, angular brackets, $\langle\rangle$, denote volume-averaging and $\bar{T}$ is the mass-averaged temperature. This definition captures the dependence of the recombination rate on temperature \citep[see also][]{Park16clumping}. Other definitions have appeared in the literature: these clumping factors can differ by tens of per cent, but the qualitative trends are similar (see Appendix \ref{sec:ressizecred} for details). 

Structures change the \Ifront\ speed and location; inserting Eq.~(\ref{eq:crwholedef}) into
Eq.~(\ref{eq:IFtrack}) yields the solution
\begin{align}
    Z &= \frac{\tau_r\,F}{c_r\,n_{\rm H}}\,\left[1-\exp(-c_rt/\tau_r)\right],\nonumber\\
    \frac{{\rm d}Z}{{\rm d}t} &=\frac{F}{n_{\rm H}}\,\exp(-c_rt/t_r)\,,
\end{align}
in the approximation that $n_{\rm H}$, $t_r$, and $F$ are all time independent. As expected, the \Ifront\ moves slower and stalls earlier when clumping is more pronounced (greater $c_r$).

We cannot apply Eq.~(\ref{eq:crwholedef}) directly to our simulations because ({\em i}) we inject radiation from the sides into the computational volume, and ({\em ii}) we include collisional ionization in the calculation. To account for ({\em i}), we only include gas downstream from the \Ifront\ in the calculation of $c_r$:
\begin{align}
c_r = \frac{\langle\alpha_{\rm B}\,n^2_{\rm HII}\rangle_{\rm IF}}{\alpha_{\rm B}(\bar{T}_{\rm IF})\, \langle n_{\rm HII}\rangle_{\rm IF}^2}.
\label{eq:crdef}
\end{align}
We approximate the \Ifront\ by a plane where the volume-weighted neutral fraction first reaches 50~per cent (and perpendicular to the injection direction). We also resolve (ii) this way since collisional ionization contributes relatively little to the recombination rate in the downstream region, where photo-ionization dominates. Furthermore, our small simulation volume does not contain many more massive halos where collisional ionizations {\em are} frequent.

\subsubsection{Evolution of the clumping factor}
\label{sec:clevo}
Fig.~\ref{fig:clumping} shows the evolution of the clumping factor as computed using Eq.~(\ref{eq:crdef}). In all runs, $c_r$ increases rapidly to a peak value\footnote{ There are two peaks in clumping factors. The second peak is due to the overlapping of I-fronts from opposite side.}, and after $\sim 100$~Myr, starts to decrease to an asymptotic value of $\lesssim 5$. The height of the peak, $c_{r, {\rm peak}}$, increases with increasing $\Gamma_{-12}$ and with decreasing $z_i$. We find that approximately
\begin{align}
c_{r,{\rm peak}} \sim 9 \times \left(\frac{1+z_i}{9}\right)^{-3}\left(\frac{\Gamma_{-12}}{0.3}\right)^\gamma,
\label{eq:crpeakz}
\end{align}
where the exponent $\gamma\sim 0.5-0.6$. The increase in $c_{r, {\rm peak}}$ with decreasing $z_i$ is simply due to structure formation: there are far fewer halos at higher $z$ to boost the recombination rate. The dependence on $\Gamma_{-12}$ is a result of the increase of $n_S$ (the density where the recombination rate is sufficiently high that the gas remains significantly neutral, Eq.~\ref{eq:nS}) with $\Gamma_{-12}$ and the dependence of the photo-evaporation time scale on $\Gamma_{-12}$. 

We caution the reader that our small simulation volume does not contain sufficient massive halos that dominate $c_r$ post-reionization. This may affect the simulation with $z_i\sim 10$ most because photo-heating of the \ihm\ at high $z$ prevents halos from accreting gas.

The contribution to the recombination rate of gas at different densities is plotted as the blue line, labelled as \lq Total\rq\ in Fig.~\ref{fig:Recom_mbins}. At low reionization redshift $z_i$, high $\Gamma_{-12}$, and close to the reionization redshift ($z\sim z_i$), the recombination rate is dominated by the denser gas with $\Delta>100$ that is mostly inside of halos (upper left panel). At slightly later times (lower left panel), lower-density gas with $\Delta$ of the order of a few starts to dominate because the denser gas has photo-evaporated (in fact, a considerable fraction of the gas with initial over-density $\Delta\sim 100$ decreases in density to values of a few following its photo-evaporation, see the discussion of Fig.~\ref{fig:simoverview_his}).

Our small simulation volume underestimates the contribution of halos to recombinations to some extent. Still, we find that similar trends (a rapid rise in $c_{r}$ to a peak value, followed by a decrease on a longer time scale) in a simulation performed in a larger computational volume (see Appendix \ref{sec:ressizecred}).

The central panels in Fig.~\ref{fig:Recom_mbins} also demonstrate how photo-evaporation decreases the contribution of halo gas to the recombination rate. Due to photo-evaporation, gas around the mean density, $\Delta\sim 1$, increasingly dominates the recombination rate at later times. This is even more striking in the right panels of the figure: early reionization prevents the accretion of gas onto more massive halos that form after $z_i$ so that halos now contribute much less to recombinations.

The other coloured lines in Fig.~\ref{fig:Recom_mbins} quantify the contribution to the recombination rate in halos of a given mass range, as indicated by the legend. For our choice of photo-ionization rate, $\Gamma_{-12}$, and reionization redshift, $z_i$, we find that \mini\ with mass below $10^6\msun$ (grey dashed line) do not contribute significantly to the recombination rate, and by extension to the clumping factor. This is partly because they photo-evaporate quickly, as seen by comparing the upper panel with the lower panel in any given column. Rather, the more massive \mini\ with masses $\gtrsim10^{6-7}\msun$ dominate recombinations during the \eor. 

Our findings agree with those of \cite{Ilie05minihalo}, who also concluded that \mini\ with $M_{\rm h}
\ll 10^6\msun$ contribute little to $c_r$. They find that halos with mass less than $\sim 10^6\msun$ increase\footnote{More precisely, they consider halo mass $<100\;M_{\rm J}$, where $M_{\rm J}$ is the pre-reionization Jeans mass.} the photon consumption during \eor\  by less than a 10-20~per cent. Since low-mass \mini\ do not affect $c_r$ significantly, we could relax the required resolution of our simulations to $m_{\rm DM}\lesssim10^3\msun$ (see Appendix ~\ref{sec:ressizecred}).

We now return to the late-time behaviour of the clumping factor, as plotted in Fig.~\ref{fig:clumping}. The value of $c_r$ drops from its peak to an
asymptotic value of $\sim 2-3$ after several 100~Myrs, in agreement with previous simulations of the post-\eor\ \ihm\ \citep[e.g.][]{Pawl09clumping,McQu11LLS}. The duration of the characteristic drop is set by the photo-evaporation time of \mini, which is of the order of 100~Myrs for the more massive \mini\ and $\lesssim$ 50 Myr for \mini\ with mass $M_h=10^5\msun$ (see Fig.~\ref{fig:tevmulti}).

Interestingly, we find that the value of the clumping factor at the end of the \eor\ is relatively insensitive to the photo-ionization rate. For example, in Fig.~\ref{fig:clumping}, changing $\Gamma_{-12}$ by over an order of magnitude does not affect the late-time clumping factors by more than $\gtrsim$ 10 per cents. This was also seen in the simulations of \cite{DAlo20clumping}.  This finding appears unexpected at first: larger values of $\Gamma_{-12}$ cause the ionized-neutral transition to occur at higher densities (Eq.\ref{eq:nS}) and this increases the recombination rate. Therefore 
if gas in halos were to dominate the recombination rate (and hence $c_r$), then we would expect that $c_r\propto \Gamma^{1/3}$ \cite[e.g.][]{McQu11LLS}. However, we find that recombinations mostly occur in gas with density around the mean at these early times, as seen in the lower panels of Fig.~\ref{fig:Recom_mbins}. Since this gas is highly ionized in any case, changing the value of $\Gamma_{-12}$ has little effect on $c_r$.

{ 
We now compare our results to those obtained by \cite{DAlo20clumping} and \cite{Park16clumping}. Note that the evolution of the clumpy factor depends strongly on the simulation setup and the reduced speed of light, so we only make qualitative comparisons here. Our values of $c_r$ are apparently $\sim$ 50 per cent higher than those plotted in Fig.~7 of \cite{DAlo20clumping}, when comparing the runs with $z_i=8,6$ and $\Gamma_{-12}=0.3$. However, \cite{DAlo20clumping} used the clumping factor definition of Eq.~(\ref{eq:cr4def}), which is tens of per cent lower than our fiducial definition Eq.~(\ref{eq:crdef}) (because Eq.~\ref{eq:cr4def} used a lower reference temperature for recombination rate; see Appendix \ref{sec:ressizecred}). Thus, our results are roughly consistent.
}

Our simulations are not directly comparable to those of \cite{Park16clumping}, since we used $\Gamma_{-12}=0.3$, roughly a factor of 3 lower than that used by them, $\Gamma_{-12}=0.92$. Applying the scaling of Eq.~(\ref{eq:crpeakz}) with $\gamma=0.5$, we would expect $c_{r, {\rm peak}}\approx 4.7\times (0.92/0.3)^{1/2} \approx 8.2$ for the case of $z_i=10$ and $\Gamma_{-12}=0.92$, which is reasonably close to their value of $c_{r, {\rm peak}}=7.5$ for that value of $\Gamma_{-12}$ and $z_i$ in their run M\_I-1\_z10. 
We also agree that the suppression of $c_r$ due to photo-evaporation is more significant at lower $\Gamma_{-12}$ because the gas in \mini\ starts to photo-evaporate before radiation ionizes the higher-density gas.

\subsubsection{Photon consumption in the clumpy medium}
\label{sec:photo-rec}
In Fig. \ref{fig:recom}, we plot the ratio ${\cal N}_r\equiv N_{\rm rec}/N_{\rm H}$, where $N_{\rm rec}$ is the cumulative number of recombination in the simulation volume up to a given time, and $N_{\rm H}$ is the total number of hydrogen atoms in that volume. We computed and recorded this ratio as the simulation progressed. We find that ${\cal N}_r$ increases with $\Gamma_{-12}$ (different line styles) but is rather 
strikingly only weakly dependent on $z_i$ (different colours).

Since $c_{r, {\rm peak}}\propto \Gamma_{-12}^\gamma$, 
a higher photo-ionization rate increases the clumping factor and hence also ${\cal N}_r$, as expected. The weak dependence of ${\cal N}_r$ on $z_i$ was also noted by \citealt{Park16clumping}. The mean instantaneous recombination rate per hydrogen atom is
\begin{equation}
    \frac{\int \left.\frac{dn_{\rm HI}}{dt}\right|_{\rm rec}\,dV}
    {\int n_{\rm H}\,dV}
    =\frac{c_r\,\alpha_B(\bar T)\,\langle n_{\rm H}\rangle^2}{\langle n_{\rm H}\rangle} = c_r\,\alpha_B(\bar T)\,\langle n_{\rm H}\rangle\,,
\end{equation}
where the integral is over the computational volume.
Since we found that approximately $c_r\propto (1+z)^{-3}$
(Eq.\ref{eq:crpeakz}), whereas $\langle n_{\rm H}\rangle\propto (1+z)^3$, the redshift dependence of the mean recombination rate per hydrogen atom is weak, which may explain the surprisingly weak dependence of ${\cal N}_r$ on $z_i$

Finally, we conclude that recombinations in the \ihm\ and \mini\ increase the number of ionizing photons per hydrogen atom required to ionize the Universe by a factor $1+{\cal N}_r\approx 1.2-1.7$. The lower value corresponds to $\Gamma_{-12}=0.03$, the upper value to $\Gamma_{-12}=0.3$. Hence the claim that recombinations increase $1+{\cal N}_r$ by 20-100 per cent, which we made in the abstract.

\section{Discussion}
\label{sec:discussion}

\label{sec:caveats}
Our simulation setup does not account for the formation of molecular hydrogen (${\rm H}_2$) and hence neglects cooling due to ${\rm H}_2$. The simulations of \cite{Shap04minihalo}, \cite{Embe13clumping}, \cite{Park16clumping}, and \cite{DAlo20clumping} made the same assumption. Molecular hydrogen cooling can trigger star formation in halos of masses as low as $10^6\msun$ at $z\sim 10$ \citep[e.g.][]{Tegm97firstobj}, orders of magnitude below the mass of halos in which gas can cool atomically ($M_{\rm ACH}$ discussed in \S~\ref{sec:analytic}). Feedback from the first Pop~{\sc iii} stars may remove gas from some \mini, thus reducing their impact on the propagation of the \Ifront\ and the value of $c_r$. However, molecular hydrogen can be photo-dissociated by Lyman-Werner (LW) photons \citep{Stec67photoH2}, which are emitted by those first stars \citep{Haim97desH2}. An LW radiation background may thus have already turned off the molecular cooling channel at the redshifts we considered ($z\lesssim 10$, e.g. \citealt{Tren09H2PopIII}), suppressing Pop.~{\sc iii} star formation and hence reducing its impact on \mini\footnote{The escape fraction of LW photons from the star-forming gas cloud and the impact of self-shielding in the LW band are both uncertain \citep{Skin20PopIIIshielding}. Therefore, the impact of Pop.~{\sc iii} stars on \mini\ during the \eor\ is unlikely to play a significant role, although we realise this is not currently well constrained.}.

We treat ionizing radiation using a frequency-averaged photo-ionization cross-section rather than following photons with a range of frequencies. This approximation does not capture the effects of spectral hardening or preheating. If included, these effects would most probably decrease the photo-evaporation time of \mini; hence, it may well be that our simulations overestimate the impact of such halos on the \eor. For example, \cite{Ilie05minihalo} found that a harder spectrum can reduce the photo-evaporation time by $\sim 50\%$.

We neglected sources of ionizing photons inside the computational volume. Instead, we injected photons from two opposing faces of the cubic simulation volume. Previous studies also inject radiation from boundaries and neglect the sources inside the volume \citep{Embe13clumping,Park16clumping,DAlo20clumping}. This approximation is reasonable for our purposes, given the small size of the computational volume. 

Our simulation volume contains halos with mass greater than $M_{\rm ACH}$ that would plausibly be able to form stars and contribute to the ionizing flux. Counter-intuitively, including stellar feedback from evolving stars forming in these halos could potentially {\em increase} the net recombination rate. Indeed, \cite{McQu11LLS} found that the recombination rate is higher in a model that included stellar winds compared to a pure \rhd\ run. Feedback can also avoid the over-cooling problem, where too much gas turns into stars \citep{Katz92overcooling}. Simulations with radiation sources within the simulation volume would be an important future improvement. These would be necessary for simulations of larger cosmological volumes where spatial correlations between photon sources and sinks are important.

Our simulations do not include pre-heating by X-rays emitted from an early generation of accreting black holes, as envisioned by \citet[e.g.][]{Rico04Xray}. Pre-heating would increase the minimum mass of halos that contain gas before reionization, reducing the impact of \mini\ on the \eor. \cite{DAlo20clumping} presented a simulation where the \ihm\ is pre-heated by X-rays below redshift $z=20$, and found a factor of two suppression of $c_{r, {\rm, peak}}$ for a simulation with $z_i=8$. They also concluded that X-ray pre-heating did not affect the value of $c_r$ at the end of the \eor. \cite{Park21xraystreaming} also studied the impact of X-ray preheating (starting at $z=20$) and found a much larger decrease in the value of $c_{r, {\rm peak}}$. However, their small simulation volume (200~$h^{-1}$~ckpc) does not sample the more massive \mini\ that may be less affected by X-ray pre-heating, which may lead them to overestimate the impact of pre-heating. \citet{Fial14latexray} suggested that X-ray pre-heating most likely occurs at lower redshifts, not long before $z_i$, which would reduce its impact on \mini.

We have neglected the relative velocity between dark matter and baryons post recombination (dark matter-baryon streaming), as discussed by \citet{Tsel10streaming}. The streaming 
velocity can be several times larger than the pre-reionization sound speed,
and this effect suppresses the early formation of low-mass halos and affects their baryon contents. Studying this effect, \cite{Park21xraystreaming} found a $c_r$ lower by a factor of two. In contrast, \cite{Cain20streamingclumping} concluded that baryon streaming decreases $c_r$ by only 5-10~per cent {(in regions with the root-mean-square streaming velocity)}.
They claimed that the small simulation volume used by \cite{Park21xraystreaming} exaggerates the streaming effects \footnote{ However, some of the difference between their results might be due to resolution \citep{Cain20streamingclumping}.}.

In summary, the previous discussion highlights some of the various ways in which our simulations could be improved in the future. Nonetheless, these are unlikely to change our main conclusions substantially.

\section{Conclusions}

\label{sec:final}
We presented cosmological radiation hydrodynamics simulations of the propagation of ionization fronts in a cosmological density field. Radiative transfer is performed with a two-moment method with local Eddington tensor closure, implemented in the \swift\ smoothed particle hydrodynamics code, as described by \cite{Chan21SPHM1RT}. Our simulations follow the ionization and heating of the gas, leading to the photo-evaporation of (gas in) low-mass halos (\mini; $T_{\rm vir}<10^4{\rm K};M_h\lesssim10^8\msun$), and self-shielding of gas in more massive halos. The simulation can resolve the smallest \mini\ that contain gas before reionization (gas mass resolution $m_{\rm gas}\sim 20 \;\msun$; spatial resolution  $\sim 0.1\;{\rm ckpc}$). We inject ionizing photons from two opposing sides of the computational volume at a given photo-ionization rate,$\Gamma_{-12}$, and for a given specified \lq reionization redshift\rq,  $z_i$. We assume that the ionizing sources have a black body spectrum with temperature $T_{\rm BB}=10^5{\rm K}$. 

These simulations improve upon the 2D \rhd\ simulations performed by \cite{Shap04minihalo} and \cite{Ilie05minihalo}, who studied a more idealised set-up of a single minihalo overrun by an \Ifront\ assuming cylindrical symmetry (see also \citealt{Naka20minihalo}). In contrast to previous cosmological \rhd\ simulations of the intergalactic medium \citep{Park16clumping,DAlo20clumping}, we have either a larger volume or higher (spatial) resolution. We also quantify the relative contribution to the clumping and recombination of \mini\ and the \ihm. The results are presented in \S\ref{sec:results} and summarised in \S\ref{sec:final}.

Our main results are as follows. In terms of the impact of the passage of the \Ifront\ on
the cosmological density distribution, we find that

\begin{itemize}
    \item  Upon the passage of \Ifront, the low density \ihm\ is rapidly heated to $T\sim2\times 10^4\;{\rm K}$. This value is expected when a black body spectrum of photons with temperature $T_{\rm BB}=10^5{\rm K}$ flash ionizes hydrogen when non-equilibrium effects are accounted for \citep{Chan21SPHM1RT}, but spectral hardening is neglected. Higher density filaments expand supersonically, heating the surrounding \ihm\ to $T\sim10^5\;{\rm K}$ through shocks and adiabatic compression. Gas also cools through adiabatic expansion. As a result, the temperature of ionized gas scatters around $T\sim 10^4{\rm K}$. Gas at higher densities in \mini\ slows the \Ifront\ as dense gas remains neutral due to self-shielding. 

    \item At later times, \mini\ photo-evaporate, which moves gas from an over-density of $\Delta\sim 100$ to $\Delta\sim 5$ (see \S \ref{sec:mhevaporation}). The photo-evaporation timescale becomes shorter with increasing $\Gamma_{-12}$, lowering $z_i$ and the mini-halo mass. The evaporation time is of order $50-100~{\rm Myrs}$, which can be several times longer than the sound crossing time.
    This is the case for halos of mass $M_{\rm h}\gtrsim 10^6\;\msun$, which trap the \Ifront\ for a long time ($\gg10\;{\rm Myr}$) due to the short recombination time of their gas.

    \item Our estimates of minihalo photo-evaporation times agree qualitatively with those of \cite{Shap04minihalo} and \cite{Ilie05minihalo}, who performed high-resolution idealized 2D radiation hydrodynamics simulations.

\end{itemize}

In terms of the impact of \mini\ on the overall recombination rate as quantified by the clumping factor, $c_r$, we found in \S~\ref{sec:clumping} that:

\begin{itemize}
    \item $c_r$ increases rapidly up to a peak value of $c_{r, {\rm peak}}\sim 20$ as the \Ifront\ overruns \mini. The value of  $c_{r, {\rm peak}}$ is higher when $\Gamma_{-12}$ is higher or $z_i$ is lower (\S~\ref{sec:clevo}). As \mini\ photo-evaporate over a time-scale of $\sim 100~{\rm Myrs}$, $c_r$ decreases to a value of $\approx 2-4$. At this stage, most recombinations occur in the \ihm\ rather than in halos. Consequently, at the final stages of the \eor, the value of $c_r$ depends only weakly on the value of $\Gamma_{-12}$.

    \item The inhomogeneous \ihm\ increases the reionization photon budget by 20-100~per cent, depending on the value of the photo-ionization rate during the \eor\ (higher values of $\Gamma_{-12}$ increase the budget)
    and to a lesser extent the value of $z_i$ (see Fig.~\ref{fig:recom}). 

    \item Low-mass \mini\ of mass $\ll 10^6{\rm M}_\odot$ do not contribute significantly to $c_r$ or the reionization photon budget because they photo-evaporate very quickly. The relative contribution to the recombination rate as a function of halo mass is plotted in Fig.~\ref{fig:Recom_mbins}.

\end{itemize}

We envision the following avenues for further research. Firstly, investigate the effect of the photo-evaporating \mini\ on the value and evolution of the mean free path of ionising photons during the \eor. What is the nature of absorbers that dominate the opacity (see, e.g. \citealt{Nasi21sinks})? Does this explain the rapid evolution claimed by \cite{Beck21mfp}? Do these absorbers leave a trace on Lyman $\alpha$ forest \citep{Park23MHLya}? Secondly, study the origin of the power-law-like gas density profile ($\rho\propto r^{-2}$) in \mini\ pre-reionization. Thirdly, predict the shape and evolution of the gas density probability distribution ({\sc pdf}), as well as the evolution of the {\sc pdf} of neutral gas before, during, and after the \eor. Finally, it would be worthwhile to incorporate the results presented here in a sub-grid physics model for recombinations (see e.g. \citealt{Cain21mfp,Cain23clump} for effects along these lines). Such a model could then be applied to a simulation at a much lower resolution but in a much larger volume. This would allow for the incorporation of \mini\ in reionization simulations that can connect with observations of the \eor\ in upcoming 21-cm surveys.

While this paper focuses on the reionization of a clumpy universe, it also serves as a benchmark of the two-moment method {\sc SPHM1RT} in cutting-edge cosmological simulations. We are coupling our radiation hydrodynamics method to interstellar medium and galaxy models. It will be promising in addressing a multitude of astrophysical problems, ranging from active galactic nuclei, HII regions, to self-consistent reionization.

\section*{ACKNOWLEDGEMENTS}
We thank John Helly for the help with nbodykit and the help from Matthieu Schaller and Mladen Ivkovic with {\small SWIFT}. We also thank the {\small SWIFT} collaboration for making the source code publicly available. We acknowledge the helpful discussions with Hyunbae Park, Joop Schaye, Nick Gnedin, Andrey Kravtsov, and Paul Shapiro.

We would like to pay our gratitude and respects to the late Prof. Richard Bower, who passed away in January of 2023. Richard was a Professor in the Physics department at the Durham University, with an expertise in understanding cosmology and galaxies with semi-analytic methods and simulations. He played a pioneering and leading role in many world-class astrophysics projects, including {\sc GALFORM}, {\sc EAGLE}, and {\sc SWIFT}. He was also an excellent supervisor and mentor for numerous students and postdocs, who continue his legacy of inspiration in academia and industry. We are greatly indebted to Richard for his innovative ideas and advice on this and related projects.

This work was supported by Science and Technology Facilities Council (STFC) astronomy consolidated grant ST/P000541/1
and ST/T000244/1. We acknowledge support from the European Research Council through ERC Advanced Investigator grant, DMIDAS [GA 786910] to CSF. 

TKC is supported by the `Improvement on Competitiveness in Hiring New Faculties' Funding Scheme from the Chinese University of Hong Kong (4937210, 4937211, 4937212). TKC was supported by the E. Margaret Burbidge Prize Postdoctoral Fellowship from the Brinson Foundation at the Departments of Astronomy and Astrophysics at the University of Chicago. 

ABL acknowledges support from the European Research Council (ERC) under the European Union's Horizon 2020 research and innovation program (GA 101026328).

This work used the DiRAC@Durham facility managed by the Institute for Computational Cosmology on behalf of the STFC DiRAC HPC Facility (www.dirac.ac.uk). The equipment was funded by BEIS capital funding via STFC capital grants ST/K00042X/1, ST/P002293/1, ST/R002371/1 and ST/S002502/1, Durham University and STFC operations grant ST/R000832/1. DiRAC is part of the National e-Infrastructure.

The research in this paper made use of the {\sc swift} open-source simulation code (http://www.swiftsim.com, \citealt{Scha18SWIFTascl}) version 0.9.0. This work also made use of matplotlib \citep{Hunt07matplotlib}, numpy \citep{vand11numpy}, scipy \citep{Jone01scipy}, swiftsimio \citep{Borr20swiftsimio}, nbodykit\citep{Hand18nbodykit}, {\sc colossus} \citep{Diemer18COLOSSUS}, and NASA’s Astrophysics Data System. We thank an anonymous referee for their careful reading of the paper and their valuable comments on its content.

\section*{DATA AVAILABILITY}
The data underlying this article will be shared on reasonable request to the corresponding author (TKC).

\bibliographystyle{mn2e}
\bibliography{mn-jour,mybib}

\begin{thebibliography}{}
\makeatletter
\relax
\def\mn@urlcharsother{\let\do\@makeother \do\$\do\&\do\#\do\^\do\_\do\%\do\~}
\def\mn@doi{\begingroup\mn@urlcharsother \@ifnextchar [ {\mn@doi@}
  {\mn@doi@[]}}
\def\mn@doi@[#1]#2{\def\@tempa{#1}\ifx\@tempa\@empty \href
  {http://dx.doi.org/#2} {doi:#2}\else \href {http://dx.doi.org/#2} {#1}\fi
  \endgroup}
\def\mn@eprint#1#2{\mn@eprint@#1:#2::\@nil}
\def\mn@eprint@arXiv#1{\href {http://arxiv.org/abs/#1} {{\tt arXiv:#1}}}
\def\mn@eprint@dblp#1{\href {http://dblp.uni-trier.de/rec/bibtex/#1.xml}
  {dblp:#1}}
\def\mn@eprint@#1:#2:#3:#4\@nil{\def\@tempa {#1}\def\@tempb {#2}\def\@tempc
  {#3}\ifx \@tempc \@empty \let \@tempc \@tempb \let \@tempb \@tempa \fi \ifx
  \@tempb \@empty \def\@tempb {arXiv}\fi \@ifundefined
  {mn@eprint@\@tempb}{\@tempb:\@tempc}{\expandafter \expandafter \csname
  mn@eprint@\@tempb\endcsname \expandafter{\@tempc}}}

\bibitem[\protect\citeauthoryear{{Abel}, {Bryan}  \& {Norman}}{{Abel}
  et~al.}{2002}]{Abel02firststar}
{Abel} T.,  {Bryan} G.~L.,   {Norman} M.~L.,  2002, \mn@doi [Science]
  {10.1126/science.295.5552.93}, \href
  {https://ui.adsabs.harvard.edu/abs/2002Sci...295...93A} {295, 93}

\bibitem[\protect\citeauthoryear{{Altay} \& {Theuns}}{{Altay} \&
  {Theuns}}{2013}]{Alta13revrt}
{Altay} G.,  {Theuns} T.,  2013, \mn@doi [\mnras] {10.1093/mnras/stt1067},
  \href {https://ui.adsabs.harvard.edu/abs/2013MNRAS.434..748A} {434, 748}

\bibitem[\protect\citeauthoryear{{Barkana} \& {Loeb}}{{Barkana} \&
  {Loeb}}{2004}]{Bark04ionbubble}
{Barkana} R.,  {Loeb} A.,  2004, \mn@doi [\apj] {10.1086/421079}, \href
  {https://ui.adsabs.harvard.edu/abs/2004ApJ...609..474B} {609, 474}

\bibitem[\protect\citeauthoryear{{Becker}, {D'Aloisio}, {Christenson}, {Zhu},
  {Worseck}  \& {Bolton}}{{Becker} et~al.}{2021}]{Beck21mfp}
{Becker} G.~D.,  {D'Aloisio} A.,  {Christenson} H.~M.,  {Zhu} Y.,  {Worseck}
  G.,   {Bolton} J.~S.,  2021, \mn@doi [\mnras] {10.1093/mnras/stab2696}, \href
  {https://ui.adsabs.harvard.edu/abs/2021MNRAS.508.1853B} {508, 1853}

\bibitem[\protect\citeauthoryear{{Benitez-Llambay} \&
  {Frenk}}{{Benitez-Llambay} \& {Frenk}}{2020}]{Beni20lmdwarf}
{Benitez-Llambay} A.,  {Frenk} C.,  2020, \mn@doi [\mnras]
  {10.1093/mnras/staa2698}, \href
  {https://ui.adsabs.harvard.edu/abs/2020MNRAS.498.4887B} {498, 4887}

\bibitem[\protect\citeauthoryear{{Ben{\'\i}tez-Llambay}
  et~al.,}{{Ben{\'\i}tez-Llambay} et~al.}{2017}]{Beni17RELHICs}
{Ben{\'\i}tez-Llambay} A.,  et~al., 2017, \mn@doi [\mnras]
  {10.1093/mnras/stw2982}, \href
  {https://ui.adsabs.harvard.edu/abs/2017MNRAS.465.3913B} {465, 3913}

\bibitem[\protect\citeauthoryear{{Bolton} \& {Haehnelt}}{{Bolton} \&
  {Haehnelt}}{2007}]{Bolt07extreion}
{Bolton} J.~S.,  {Haehnelt} M.~G.,  2007, \mn@doi [\mnras]
  {10.1111/j.1365-2966.2007.12372.x}, \href
  {https://ui.adsabs.harvard.edu/abs/2007MNRAS.382..325B} {382, 325}

\bibitem[\protect\citeauthoryear{Borrow \& Borrisov}{Borrow \&
  Borrisov}{2020}]{Borr20swiftsimio}
Borrow J.,  Borrisov A.,  2020, \mn@doi [Journal of Open Source Software]
  {10.21105/joss.02430}, 5, 2430

\bibitem[\protect\citeauthoryear{{Borrow}, {Schaller}, {Bower}  \&
  {Schaye}}{{Borrow} et~al.}{2022}]{Borr20SPHENIX}
{Borrow} J.,  {Schaller} M.,  {Bower} R.~G.,   {Schaye} J.,  2022, \mn@doi
  [\mnras] {10.1093/mnras/stab3166}, \href
  {https://ui.adsabs.harvard.edu/abs/2022MNRAS.511.2367B} {511, 2367}

\bibitem[\protect\citeauthoryear{{Bouwens}, {Illingworth}, {Oesch}, {Caruana},
  {Holwerda}, {Smit}  \& {Wilkins}}{{Bouwens} et~al.}{2015}]{Bouw15emiss}
{Bouwens} R.~J.,  {Illingworth} G.~D.,  {Oesch} P.~A.,  {Caruana} J.,
  {Holwerda} B.,  {Smit} R.,   {Wilkins} S.,  2015, \mn@doi [\apj]
  {10.1088/0004-637X/811/2/140}, \href
  {https://ui.adsabs.harvard.edu/abs/2015ApJ...811..140B} {811, 140}

\bibitem[\protect\citeauthoryear{{Bromm}, {Coppi}  \& {Larson}}{{Bromm}
  et~al.}{2002}]{Brom02firststar}
{Bromm} V.,  {Coppi} P.~S.,   {Larson} R.~B.,  2002, \mn@doi [\apj]
  {10.1086/323947}, \href
  {https://ui.adsabs.harvard.edu/abs/2002ApJ...564...23B} {564, 23}

\bibitem[\protect\citeauthoryear{{Bryan}, {Machacek}, {Anninos}  \&
  {Norman}}{{Bryan} et~al.}{1999}]{Bryan99}
{Bryan} G.~L.,  {Machacek} M.,  {Anninos} P.,   {Norman} M.~L.,  1999, \mn@doi
  [\apj] {10.1086/307173}, \href
  {https://ui.adsabs.harvard.edu/abs/1999ApJ...517...13B} {517, 13}

\bibitem[\protect\citeauthoryear{{Cain}, {D'Aloisio}, {Ir{\v{s}}i{\v{c}}},
  {McQuinn}  \& {Trac}}{{Cain} et~al.}{2020}]{Cain20streamingclumping}
{Cain} C.,  {D'Aloisio} A.,  {Ir{\v{s}}i{\v{c}}} V.,  {McQuinn} M.,   {Trac}
  H.,  2020, \mn@doi [\apj] {10.3847/1538-4357/aba26a}, \href
  {https://ui.adsabs.harvard.edu/abs/2020ApJ...898..168C} {898, 168}

\bibitem[\protect\citeauthoryear{{Cain}, {D'Aloisio}, {Gangolli}  \&
  {Becker}}{{Cain} et~al.}{2021}]{Cain21mfp}
{Cain} C.,  {D'Aloisio} A.,  {Gangolli} N.,   {Becker} G.~D.,  2021, \mn@doi
  [\apjl] {10.3847/2041-8213/ac1ace}, \href
  {https://ui.adsabs.harvard.edu/abs/2021ApJ...917L..37C} {917, L37}

\bibitem[\protect\citeauthoryear{{Cain}, {D'Aloisio}, {Gangolli}  \&
  {McQuinn}}{{Cain} et~al.}{2023}]{Cain23clump}
{Cain} C.,  {D'Aloisio} A.,  {Gangolli} N.,   {McQuinn} M.,  2023, \mn@doi
  [\mnras] {10.1093/mnras/stad1057}, \href
  {https://ui.adsabs.harvard.edu/abs/2023MNRAS.522.2047C} {522, 2047}

\bibitem[\protect\citeauthoryear{{Calverley}, {Becker}, {Haehnelt}  \&
  {Bolton}}{{Calverley} et~al.}{2011}]{Calv11UVbg}
{Calverley} A.~P.,  {Becker} G.~D.,  {Haehnelt} M.~G.,   {Bolton} J.~S.,  2011,
  \mn@doi [\mnras] {10.1111/j.1365-2966.2010.18072.x}, \href
  {https://ui.adsabs.harvard.edu/abs/2011MNRAS.412.2543C} {412, 2543}

\bibitem[\protect\citeauthoryear{{Chan}, {Theuns}, {Bower}  \& {Frenk}}{{Chan}
  et~al.}{2021}]{Chan21SPHM1RT}
{Chan} T.~K.,  {Theuns} T.,  {Bower} R.,   {Frenk} C.,  2021, \mn@doi [\mnras]
  {10.1093/mnras/stab1686}, \href
  {https://ui.adsabs.harvard.edu/abs/2021MNRAS.505.5784C} {505, 5784}

\bibitem[\protect\citeauthoryear{{Chan}, {Benitez-Llambay}, {Theuns}  \&
  {Frenk}}{{Chan} et~al.}{2023}]{Chan23IAUS}
{Chan} T.~K.,  {Benitez-Llambay} A.,  {Theuns} T.,   {Frenk} C.,  2023, \mn@doi
  [IAU Symposium] {10.1017/S1743921322001235}, \href
  {https://ui.adsabs.harvard.edu/abs/2023IAUS..362...15K} {362, 15}

\bibitem[\protect\citeauthoryear{{Chisholm} et~al.,}{{Chisholm}
  et~al.}{2018}]{Chrisholm18}
{Chisholm} J.,  et~al., 2018, \mn@doi [\aap] {10.1051/0004-6361/201832758},
  \href {https://ui.adsabs.harvard.edu/abs/2018A&A...616A..30C} {616, A30}

\bibitem[\protect\citeauthoryear{{Chisholm} et~al.,}{{Chisholm}
  et~al.}{2022}]{Chrisholm22}
{Chisholm} J.,  et~al., 2022, \mn@doi [\mnras] {10.1093/mnras/stac2874}, \href
  {https://ui.adsabs.harvard.edu/abs/2022MNRAS.517.5104C} {517, 5104}

\bibitem[\protect\citeauthoryear{{Ciardi}, {Scannapieco}, {Stoehr}, {Ferrara},
  {Iliev}  \& {Shapiro}}{{Ciardi} et~al.}{2006}]{Ciar06mhreion}
{Ciardi} B.,  {Scannapieco} E.,  {Stoehr} F.,  {Ferrara} A.,  {Iliev} I.~T.,
  {Shapiro} P.~R.,  2006, \mn@doi [\mnras] {10.1111/j.1365-2966.2005.09908.x},
  \href {https://ui.adsabs.harvard.edu/abs/2006MNRAS.366..689C} {366, 689}

\bibitem[\protect\citeauthoryear{{Courant}, {Friedrichs}  \& {Lewy}}{{Courant}
  et~al.}{1928}]{Courant28}
{Courant} R.,  {Friedrichs} K.,   {Lewy} H.,  1928, \mn@doi [Mathematische
  Annalen] {10.1007/BF01448839}, \href
  {https://ui.adsabs.harvard.edu/abs/1928MatAn.100...32C} {100, 32}

\bibitem[\protect\citeauthoryear{{D'Aloisio}, {McQuinn}, {Davies}  \&
  {Furlanetto}}{{D'Aloisio} et~al.}{2018}]{DAlo18UVbg}
{D'Aloisio} A.,  {McQuinn} M.,  {Davies} F.~B.,   {Furlanetto} S.~R.,  2018,
  \mn@doi [\mnras] {10.1093/mnras/stx2341}, \href
  {https://ui.adsabs.harvard.edu/abs/2018MNRAS.473..560D} {473, 560}

\bibitem[\protect\citeauthoryear{{D'Aloisio}, {McQuinn}, {Trac}, {Cain}  \&
  {Mesinger}}{{D'Aloisio} et~al.}{2020}]{DAlo20clumping}
{D'Aloisio} A.,  {McQuinn} M.,  {Trac} H.,  {Cain} C.,   {Mesinger} A.,  2020,
  \mn@doi [\apj] {10.3847/1538-4357/ab9f2f}, \href
  {https://ui.adsabs.harvard.edu/abs/2020ApJ...898..149D} {898, 149}

\bibitem[\protect\citeauthoryear{{Davis}, {Efstathiou}, {Frenk}  \&
  {White}}{{Davis} et~al.}{1985}]{Davi85fof}
{Davis} M.,  {Efstathiou} G.,  {Frenk} C.~S.,   {White} S.~D.~M.,  1985,
  \mn@doi [\apj] {10.1086/163168}, \href
  {https://ui.adsabs.harvard.edu/abs/1985ApJ...292..371D} {292, 371}

\bibitem[\protect\citeauthoryear{{Diemer}}{{Diemer}}{2018}]{Diemer18COLOSSUS}
{Diemer} B.,  2018, \mn@doi [\apjs] {10.3847/1538-4365/aaee8c}, \href
  {https://ui.adsabs.harvard.edu/abs/2018ApJS..239...35D} {239, 35}

\bibitem[\protect\citeauthoryear{{Doussot}, {Trac}  \& {Cen}}{{Doussot}
  et~al.}{2019}]{Dous19SCORCH}
{Doussot} A.,  {Trac} H.,   {Cen} R.,  2019, \mn@doi [\apj]
  {10.3847/1538-4357/aaef75}, \href
  {https://ui.adsabs.harvard.edu/abs/2019ApJ...870...18D} {870, 18}

\bibitem[\protect\citeauthoryear{{Draine}}{{Draine}}{2011}]{Draine11}
{Draine} B.~T.,  2011, {Physics of the Interstellar and Intergalactic Medium}.
Princeton University Press

\bibitem[\protect\citeauthoryear{Emberson, Thomas  \& Alvarez}{Emberson
  et~al.}{2013}]{Embe13clumping}
Emberson J.~D.,  Thomas R.~M.,   Alvarez M.~A.,  2013, \mn@doi [\apj]
  {10.1088/0004-637x/763/2/146}, 763, 146

\bibitem[\protect\citeauthoryear{{Fan} et~al.,}{{Fan} et~al.}{2006}]{Fan06EoR}
{Fan} X.,  et~al., 2006, \mn@doi [\aj] {10.1086/504836}, \href
  {https://ui.adsabs.harvard.edu/abs/2006AJ....132..117F} {132, 117}

\bibitem[\protect\citeauthoryear{{Faucher-Gigu{\`e}re}, {Lidz}, {Zaldarriaga}
  \& {Hernquist}}{{Faucher-Gigu{\`e}re} et~al.}{2009}]{Fauc09}
{Faucher-Gigu{\`e}re} C.-A.,  {Lidz} A.,  {Zaldarriaga} M.,   {Hernquist} L.,
  2009, \mn@doi [\apj] {10.1088/0004-637X/703/2/1416}, \href
  {http://adsabs.harvard.edu/abs/2009ApJ...703.1416F} {703, 1416}

\bibitem[\protect\citeauthoryear{{Fialkov}, {Barkana}  \& {Visbal}}{{Fialkov}
  et~al.}{2014}]{Fial14latexray}
{Fialkov} A.,  {Barkana} R.,   {Visbal} E.,  2014, \mn@doi [\nat]
  {10.1038/nature12999}, \href
  {https://ui.adsabs.harvard.edu/abs/2014Natur.506..197F} {506, 197}

\bibitem[\protect\citeauthoryear{{Finkelstein} et~al.,}{{Finkelstein}
  et~al.}{2019}]{Fink19lowescreion}
{Finkelstein} S.~L.,  et~al., 2019, \mn@doi [\apj] {10.3847/1538-4357/ab1ea8},
  \href {https://ui.adsabs.harvard.edu/abs/2019ApJ...879...36F} {879, 36}

\bibitem[\protect\citeauthoryear{{Finlator}, {Dav{\'e}}  \&
  {{\"O}zel}}{{Finlator} et~al.}{2011}]{Finl11VETgadget2}
{Finlator} K.,  {Dav{\'e}} R.,   {{\"O}zel} F.,  2011, \mn@doi [\apj]
  {10.1088/0004-637X/743/2/169}, \href
  {https://ui.adsabs.harvard.edu/abs/2011ApJ...743..169F} {743, 169}

\bibitem[\protect\citeauthoryear{Finlator, Oh, Özel  \& Davé}{Finlator
  et~al.}{2012}]{Finl12clumping}
Finlator K.,  Oh S.~P.,  Özel F.,   Davé R.,  2012, \mn@doi [\mnras]
  {10.1111/j.1365-2966.2012.22114.x}, 427, 2464

\bibitem[\protect\citeauthoryear{{Furlanetto}, {Zaldarriaga}  \&
  {Hernquist}}{{Furlanetto} et~al.}{2004}]{Furl04ionbubble}
{Furlanetto} S.~R.,  {Zaldarriaga} M.,   {Hernquist} L.,  2004, \mn@doi [\apj]
  {10.1086/423025}, \href
  {https://ui.adsabs.harvard.edu/abs/2004ApJ...613....1F} {613, 1}

\bibitem[\protect\citeauthoryear{{Furlanetto}, {Oh}  \& {Briggs}}{{Furlanetto}
  et~al.}{2006}]{Furlanetto06}
{Furlanetto} S.~R.,  {Oh} S.~P.,   {Briggs} F.~H.,  2006, \mn@doi [\physrep]
  {10.1016/j.physrep.2006.08.002}, \href
  {https://ui.adsabs.harvard.edu/abs/2006PhR...433..181F} {433, 181}

\bibitem[\protect\citeauthoryear{{George} et~al.,}{{George}
  et~al.}{2015}]{Geor15kSZreion}
{George} E.~M.,  et~al., 2015, \mn@doi [\apj] {10.1088/0004-637X/799/2/177},
  \href {https://ui.adsabs.harvard.edu/abs/2015ApJ...799..177G} {799, 177}

\bibitem[\protect\citeauthoryear{{Gingold} \& {Monaghan}}{{Gingold} \&
  {Monaghan}}{1977}]{Ging77SPH}
{Gingold} R.~A.,  {Monaghan} J.~J.,  1977, \mn@doi [\mnras]
  {10.1093/mnras/181.3.375}, \href
  {https://ui.adsabs.harvard.edu/abs/1977MNRAS.181..375G} {181, 375}

\bibitem[\protect\citeauthoryear{{Gnedin}}{{Gnedin}}{2000a}]{Gnedin00}
{Gnedin} N.~Y.,  2000a, \mn@doi [\apj] {10.1086/308876}, \href
  {https://ui.adsabs.harvard.edu/abs/2000ApJ...535..530G} {535, 530}

\bibitem[\protect\citeauthoryear{{Gnedin}}{{Gnedin}}{2000b}]{Gned00filtering}
{Gnedin} N.~Y.,  2000b, \mn@doi [\apj] {10.1086/317042}, \href
  {https://ui.adsabs.harvard.edu/abs/2000ApJ...542..535G} {542, 535}

\bibitem[\protect\citeauthoryear{{Gnedin}}{{Gnedin}}{2014}]{Gned14CROC}
{Gnedin} N.~Y.,  2014, \mn@doi [\apj] {10.1088/0004-637X/793/1/29}, \href
  {https://ui.adsabs.harvard.edu/abs/2014ApJ...793...29G} {793, 29}

\bibitem[\protect\citeauthoryear{{Gnedin} \& {Abel}}{{Gnedin} \&
  {Abel}}{2001}]{Gned01OTVET}
{Gnedin} N.~Y.,  {Abel} T.,  2001, \mn@doi [\na]
  {10.1016/S1384-1076(01)00068-9}, \href
  {https://ui.adsabs.harvard.edu/abs/2001NewA....6..437G} {6, 437}

\bibitem[\protect\citeauthoryear{{Gnedin} \& {Hui}}{{Gnedin} \&
  {Hui}}{1998}]{Gned98filtering}
{Gnedin} N.~Y.,  {Hui} L.,  1998, \mn@doi [\mnras]
  {10.1046/j.1365-8711.1998.01249.x}, \href
  {https://ui.adsabs.harvard.edu/abs/1998MNRAS.296...44G} {296, 44}

\bibitem[\protect\citeauthoryear{{Gnedin} \& {Ostriker}}{{Gnedin} \&
  {Ostriker}}{1997}]{Gned97reionclumping}
{Gnedin} N.~Y.,  {Ostriker} J.~P.,  1997, \mn@doi [\apj] {10.1086/304548},
  \href {https://ui.adsabs.harvard.edu/abs/1997ApJ...486..581G} {486, 581}

\bibitem[\protect\citeauthoryear{{Greig} et~al.,}{{Greig}
  et~al.}{2021}]{Greig21}
{Greig} B.,  et~al., 2021, \mn@doi [\mnras] {10.1093/mnras/staa3593}, \href
  {https://ui.adsabs.harvard.edu/abs/2021MNRAS.501....1G} {501, 1}

\bibitem[\protect\citeauthoryear{{HERA Collaboration} et~al.,}{{HERA
  Collaboration} et~al.}{2023}]{HERA23IGMheating}
{HERA Collaboration} et~al., 2023, \mn@doi [\apj] {10.3847/1538-4357/acaf50},
  \href {https://ui.adsabs.harvard.edu/abs/2023ApJ...945..124H} {945, 124}

\bibitem[\protect\citeauthoryear{{Haardt} \& {Madau}}{{Haardt} \&
  {Madau}}{2012}]{Haar12UVbackground}
{Haardt} F.,  {Madau} P.,  2012, \mn@doi [\apj] {10.1088/0004-637X/746/2/125},
  \href {https://ui.adsabs.harvard.edu/abs/2012ApJ...746..125H} {746, 125}

\bibitem[\protect\citeauthoryear{{Hahn} \& {Abel}}{{Hahn} \&
  {Abel}}{2011}]{Hahn11MUSIC}
{Hahn} O.,  {Abel} T.,  2011, \mn@doi [\mnras]
  {10.1111/j.1365-2966.2011.18820.x}, \href
  {http://adsabs.harvard.edu/abs/2011MNRAS.415.2101H} {415, 2101}

\bibitem[\protect\citeauthoryear{{Haiman}, {Rees}  \& {Loeb}}{{Haiman}
  et~al.}{1997}]{Haim97desH2}
{Haiman} Z.,  {Rees} M.~J.,   {Loeb} A.,  1997, \mn@doi [\apj]
  {10.1086/303647}, \href
  {https://ui.adsabs.harvard.edu/abs/1997ApJ...476..458H} {476, 458}

\bibitem[\protect\citeauthoryear{{Haiman}, {Abel}  \& {Madau}}{{Haiman}
  et~al.}{2001}]{Haim01minihalo}
{Haiman} Z.,  {Abel} T.,   {Madau} P.,  2001, \mn@doi [\apj] {10.1086/320232},
  \href {https://ui.adsabs.harvard.edu/abs/2001ApJ...551..599H} {551, 599}

\bibitem[\protect\citeauthoryear{{Hand}, {Feng}, {Beutler}, {Li}, {Modi},
  {Seljak}  \& {Slepian}}{{Hand} et~al.}{2018}]{Hand18nbodykit}
{Hand} N.,  {Feng} Y.,  {Beutler} F.,  {Li} Y.,  {Modi} C.,  {Seljak} U.,
  {Slepian} Z.,  2018, \mn@doi [\aj] {10.3847/1538-3881/aadae0}, \href
  {https://ui.adsabs.harvard.edu/abs/2018AJ....156..160H} {156, 160}

\bibitem[\protect\citeauthoryear{{Hunter}}{{Hunter}}{2007}]{Hunt07matplotlib}
{Hunter} J.~D.,  2007, \mn@doi [Computing in Science and Engineering]
  {10.1109/MCSE.2007.55}, \href
  {http://adsabs.harvard.edu/abs/2007CSE.....9...90H} {9, 90}

\bibitem[\protect\citeauthoryear{{Iliev}, {Shapiro}  \& {Raga}}{{Iliev}
  et~al.}{2005a}]{Ilie05minihalo}
{Iliev} I.~T.,  {Shapiro} P.~R.,   {Raga} A.~C.,  2005a, \mn@doi [\mnras]
  {10.1111/j.1365-2966.2005.09155.x}, \href
  {https://ui.adsabs.harvard.edu/abs/2005MNRAS.361..405I} {361, 405}

\bibitem[\protect\citeauthoryear{{Iliev}, {Scannapieco}  \& {Shapiro}}{{Iliev}
  et~al.}{2005b}]{Ilie05SSSIfront}
{Iliev} I.~T.,  {Scannapieco} E.,   {Shapiro} P.~R.,  2005b, \mn@doi [\apj]
  {10.1086/429083}, \href
  {https://ui.adsabs.harvard.edu/abs/2005ApJ...624..491I} {624, 491}

\bibitem[\protect\citeauthoryear{{Iliev}, {Mellema}, {Pen}, {Merz}, {Shapiro}
  \& {Alvarez}}{{Iliev} et~al.}{2006a}]{Ilie06ionbubble}
{Iliev} I.~T.,  {Mellema} G.,  {Pen} U.~L.,  {Merz} H.,  {Shapiro} P.~R.,
  {Alvarez} M.~A.,  2006a, \mn@doi [\mnras] {10.1111/j.1365-2966.2006.10502.x},
  \href {https://ui.adsabs.harvard.edu/abs/2006MNRAS.369.1625I} {369, 1625}

\bibitem[\protect\citeauthoryear{{Iliev} et~al.,}{{Iliev}
  et~al.}{2006b}]{Ilie06RTcom}
{Iliev} I.~T.,  et~al., 2006b, \mn@doi [\mnras]
  {10.1111/j.1365-2966.2006.10775.x}, \href
  {https://ui.adsabs.harvard.edu/abs/2006MNRAS.371.1057I} {371, 1057}

\bibitem[\protect\citeauthoryear{{Iliev}, {Mellema}, {Shapiro}  \&
  {Pen}}{{Iliev} et~al.}{2007}]{Ilie07srreion}
{Iliev} I.~T.,  {Mellema} G.,  {Shapiro} P.~R.,   {Pen} U.-L.,  2007, \mn@doi
  [\mnras] {10.1111/j.1365-2966.2007.11482.x}, \href
  {https://ui.adsabs.harvard.edu/abs/2007MNRAS.376..534I} {376, 534}

\bibitem[\protect\citeauthoryear{{Iliev} et~al.,}{{Iliev}
  et~al.}{2009}]{Ilie09RTcom}
{Iliev} I.~T.,  et~al., 2009, \mn@doi [\mnras]
  {10.1111/j.1365-2966.2009.15558.x}, \href
  {https://ui.adsabs.harvard.edu/abs/2009MNRAS.400.1283I} {400, 1283}

\bibitem[\protect\citeauthoryear{{Iliev}, {Mellema}, {Ahn}, {Shapiro}, {Mao}
  \& {Pen}}{{Iliev} et~al.}{2014}]{Ilie14reionvol}
{Iliev} I.~T.,  {Mellema} G.,  {Ahn} K.,  {Shapiro} P.~R.,  {Mao} Y.,   {Pen}
  U.-L.,  2014, \mn@doi [\mnras] {10.1093/mnras/stt2497}, \href
  {https://ui.adsabs.harvard.edu/abs/2014MNRAS.439..725I} {439, 725}

\bibitem[\protect\citeauthoryear{Jones, Oliphant, Peterson  et~al.}{Jones
  et~al.}{2001}]{Jone01scipy}
Jones E.,  Oliphant T.,  Peterson P.,   et~al., 2001, {SciPy}: Open source
  scientific tools for {Python}, \url {http://www.scipy.org/}

\bibitem[\protect\citeauthoryear{{Kahn}}{{Kahn}}{1954}]{Kahn54ifront}
{Kahn} F.~D.,  1954, \bain, \href
  {https://ui.adsabs.harvard.edu/abs/1954BAN....12..187K} {12, 187}

\bibitem[\protect\citeauthoryear{{Kannan}, {Smith}, {Garaldi}, {Shen},
  {Vogelsberger}, {Pakmor}, {Springel}  \& {Hernquist}}{{Kannan}
  et~al.}{2022}]{Kann21THESAN}
{Kannan} R.,  {Smith} A.,  {Garaldi} E.,  {Shen} X.,  {Vogelsberger} M.,
  {Pakmor} R.,  {Springel} V.,   {Hernquist} L.,  2022, \mn@doi [\mnras]
  {10.1093/mnras/stac1557}, \href
  {https://ui.adsabs.harvard.edu/abs/2022MNRAS.514.3857K} {514, 3857}

\bibitem[\protect\citeauthoryear{{Katz}}{{Katz}}{1992}]{Katz92overcooling}
{Katz} N.,  1992, \mn@doi [\apj] {10.1086/171366}, \href
  {https://ui.adsabs.harvard.edu/abs/1992ApJ...391..502K} {391, 502}

\bibitem[\protect\citeauthoryear{{Katz}, {Kimm}, {Sijacki}  \&
  {Haehnelt}}{{Katz} et~al.}{2017}]{Katz17reionVSL}
{Katz} H.,  {Kimm} T.,  {Sijacki} D.,   {Haehnelt} M.~G.,  2017, \mn@doi
  [\mnras] {10.1093/mnras/stx608}, \href
  {https://ui.adsabs.harvard.edu/abs/2017MNRAS.468.4831K} {468, 4831}

\bibitem[\protect\citeauthoryear{{Koopmans} et~al.,}{{Koopmans}
  et~al.}{2015}]{Koopman15}
{Koopmans} L.,  et~al., 2015, in Advancing Astrophysics with the Square
  Kilometre Array (AASKA14). p.~1 (\mn@eprint {arXiv} {1505.07568}),
  \mn@doi{10.22323/1.215.0001}

\bibitem[\protect\citeauthoryear{{Loeb} \& {Barkana}}{{Loeb} \&
  {Barkana}}{2001}]{Loeb01}
{Loeb} A.,  {Barkana} R.,  2001, \mn@doi [\araa]
  {10.1146/annurev.astro.39.1.19}, \href
  {https://ui.adsabs.harvard.edu/abs/2001ARA&A..39...19L} {39, 19}

\bibitem[\protect\citeauthoryear{{Lucy}}{{Lucy}}{1977}]{Lucy77SPH}
{Lucy} L.~B.,  1977, \mn@doi [\aj] {10.1086/112164}, \href
  {https://ui.adsabs.harvard.edu/abs/1977AJ.....82.1013L} {82, 1013}

\bibitem[\protect\citeauthoryear{{Madau}}{{Madau}}{2017}]{Madau17}
{Madau} P.,  2017, \mn@doi [\apj] {10.3847/1538-4357/aa9715}, \href
  {https://ui.adsabs.harvard.edu/abs/2017ApJ...851...50M} {851, 50}

\bibitem[\protect\citeauthoryear{{Mason}, {Treu}, {Dijkstra}, {Mesinger},
  {Trenti}, {Pentericci}, {de Barros}  \& {Vanzella}}{{Mason}
  et~al.}{2018}]{Maso18reion}
{Mason} C.~A.,  {Treu} T.,  {Dijkstra} M.,  {Mesinger} A.,  {Trenti} M.,
  {Pentericci} L.,  {de Barros} S.,   {Vanzella} E.,  2018, \mn@doi [\apj]
  {10.3847/1538-4357/aab0a7}, \href
  {https://ui.adsabs.harvard.edu/abs/2018ApJ...856....2M} {856, 2}

\bibitem[\protect\citeauthoryear{{McQuinn}, {Oh}  \&
  {Faucher-Gigu{\`e}re}}{{McQuinn} et~al.}{2011}]{McQu11LLS}
{McQuinn} M.,  {Oh} S.~P.,   {Faucher-Gigu{\`e}re} C.-A.,  2011, \mn@doi [\apj]
  {10.1088/0004-637X/743/1/82}, \href
  {https://ui.adsabs.harvard.edu/abs/2011ApJ...743...82M} {743, 82}

\bibitem[\protect\citeauthoryear{{Miralda-Escud{\'e}}, {Cen}, {Ostriker}  \&
  {Rauch}}{{Miralda-Escud{\'e}} et~al.}{1996}]{Mira96Lyasim}
{Miralda-Escud{\'e}} J.,  {Cen} R.,  {Ostriker} J.~P.,   {Rauch} M.,  1996,
  \mn@doi [\apj] {10.1086/177992}, \href
  {https://ui.adsabs.harvard.edu/abs/1996ApJ...471..582M} {471, 582}

\bibitem[\protect\citeauthoryear{{Miralda-Escud{\'e}}, {Haehnelt}  \&
  {Rees}}{{Miralda-Escud{\'e}} et~al.}{2000}]{Mira00reion}
{Miralda-Escud{\'e}} J.,  {Haehnelt} M.,   {Rees} M.~J.,  2000, \mn@doi [\apj]
  {10.1086/308330}, \href
  {https://ui.adsabs.harvard.edu/abs/2000ApJ...530....1M} {530, 1}

\bibitem[\protect\citeauthoryear{{Mo}, {van den Bosch}  \& {White}}{{Mo}
  et~al.}{2010}]{Mo10}
{Mo} H.,  {van den Bosch} F.~C.,   {White} S.,  2010, {Galaxy Formation and
  Evolution}.
UK: Cambridge University Press

\bibitem[\protect\citeauthoryear{{Mortlock} et~al.,}{{Mortlock}
  et~al.}{2011}]{Mortlock11}
{Mortlock} D.~J.,  et~al., 2011, \mn@doi [\nat] {10.1038/nature10159}, \href
  {https://ui.adsabs.harvard.edu/abs/2011Natur.474..616M} {474, 616}

\bibitem[\protect\citeauthoryear{{Naidu}, {Tacchella}, {Mason}, {Bose}, {Oesch}
   \& {Conroy}}{{Naidu} et~al.}{2020}]{Naidu20}
{Naidu} R.~P.,  {Tacchella} S.,  {Mason} C.~A.,  {Bose} S.,  {Oesch} P.~A.,
  {Conroy} C.,  2020, \mn@doi [\apj] {10.3847/1538-4357/ab7cc9}, \href
  {https://ui.adsabs.harvard.edu/abs/2020ApJ...892..109N} {892, 109}

\bibitem[\protect\citeauthoryear{{Nakatani}, {Fialkov}  \&
  {Yoshida}}{{Nakatani} et~al.}{2020}]{Naka20minihalo}
{Nakatani} R.,  {Fialkov} A.,   {Yoshida} N.,  2020, \mn@doi [\apj]
  {10.3847/1538-4357/abc5b4}, \href
  {https://ui.adsabs.harvard.edu/abs/2020ApJ...905..151N} {905, 151}

\bibitem[\protect\citeauthoryear{{Naoz} \& {Barkana}}{{Naoz} \&
  {Barkana}}{2007}]{Naoz07mhfg}
{Naoz} S.,  {Barkana} R.,  2007, \mn@doi [\mnras]
  {10.1111/j.1365-2966.2007.11636.x}, \href
  {https://ui.adsabs.harvard.edu/abs/2007MNRAS.377..667N} {377, 667}

\bibitem[\protect\citeauthoryear{{Naoz}, {Barkana}  \& {Mesinger}}{{Naoz}
  et~al.}{2009}]{Naoz09mhgf}
{Naoz} S.,  {Barkana} R.,   {Mesinger} A.,  2009, \mn@doi [\mnras]
  {10.1111/j.1365-2966.2009.15282.x}, \href
  {https://ui.adsabs.harvard.edu/abs/2009MNRAS.399..369N} {399, 369}

\bibitem[\protect\citeauthoryear{{Nasir}, {Cain}, {D'Aloisio}, {Gangolli}  \&
  {McQuinn}}{{Nasir} et~al.}{2021}]{Nasi21sinks}
{Nasir} F.,  {Cain} C.,  {D'Aloisio} A.,  {Gangolli} N.,   {McQuinn} M.,  2021,
  \mn@doi [\apj] {10.3847/1538-4357/ac2eb9}, \href
  {https://ui.adsabs.harvard.edu/abs/2021ApJ...923..161N} {923, 161}

\bibitem[\protect\citeauthoryear{Ocvirk et~al.,}{Ocvirk
  et~al.}{2016}]{Ocvi16CoDa}
Ocvirk P.,  et~al., 2016, \mn@doi [\mnras] {10.1093/mnras/stw2036}, 463, 1462

\bibitem[\protect\citeauthoryear{{Okamoto}, {Gao}  \& {Theuns}}{{Okamoto}
  et~al.}{2008}]{Okam08}
{Okamoto} T.,  {Gao} L.,   {Theuns} T.,  2008, \mn@doi [\mnras]
  {10.1111/j.1365-2966.2008.13830.x}, \href
  {http://adsabs.harvard.edu/abs/2008MNRAS.390..920O} {390, 920}

\bibitem[\protect\citeauthoryear{{Osterbrock} \& {Ferland}}{{Osterbrock} \&
  {Ferland}}{2006}]{Oste06nebulae}
{Osterbrock} D.~E.,  {Ferland} G.~J.,  2006, {Astrophysics of gaseous nebulae
  and active galactic nuclei}.
University Science Books

\bibitem[\protect\citeauthoryear{{Park}, {Shapiro}, {Choi}, {Yoshida}, {Hirano}
   \& {Ahn}}{{Park} et~al.}{2016}]{Park16clumping}
{Park} H.,  {Shapiro} P.~R.,  {Choi} J.-h.,  {Yoshida} N.,  {Hirano} S.,
  {Ahn} K.,  2016, \mn@doi [\apj] {10.3847/0004-637X/831/1/86}, \href
  {https://ui.adsabs.harvard.edu/abs/2016ApJ...831...86P} {831, 86}

\bibitem[\protect\citeauthoryear{{Park}, {Shapiro}, {Ahn}, {Yoshida}  \&
  {Hirano}}{{Park} et~al.}{2021}]{Park21xraystreaming}
{Park} H.,  {Shapiro} P.~R.,  {Ahn} K.,  {Yoshida} N.,   {Hirano} S.,  2021,
  \mn@doi [\apj] {10.3847/1538-4357/abd7f4}, \href
  {https://ui.adsabs.harvard.edu/abs/2021ApJ...908...96P} {908, 96}

\bibitem[\protect\citeauthoryear{{Park}, {Luki{\'c}}, {Sexton}  \&
  {Alvarez}}{{Park} et~al.}{2023}]{Park23MHLya}
{Park} H.,  {Luki{\'c}} Z.,  {Sexton} J.,   {Alvarez} M.,  2023, \mn@doi [arXiv
  e-prints] {10.48550/arXiv.2309.04129}, \href
  {https://ui.adsabs.harvard.edu/abs/2023arXiv230904129P} {p. arXiv:2309.04129}

\bibitem[\protect\citeauthoryear{{Pawlik}, {Schaye}  \& {van
  Scherpenzeel}}{{Pawlik} et~al.}{2009}]{Pawl09clumping}
{Pawlik} A.~H.,  {Schaye} J.,   {van Scherpenzeel} E.,  2009, \mn@doi [\mnras]
  {10.1111/j.1365-2966.2009.14486.x}, \href
  {https://ui.adsabs.harvard.edu/abs/2009MNRAS.394.1812P} {394, 1812}

\bibitem[\protect\citeauthoryear{{Pawlik}, {Rahmati}, {Schaye}, {Jeon}  \&
  {Dalla Vecchia}}{{Pawlik} et~al.}{2017}]{Pawl17Aurora}
{Pawlik} A.~H.,  {Rahmati} A.,  {Schaye} J.,  {Jeon} M.,   {Dalla Vecchia} C.,
  2017, \mn@doi [\mnras] {10.1093/mnras/stw2869}, \href
  {https://ui.adsabs.harvard.edu/abs/2017MNRAS.466..960P} {466, 960}

\bibitem[\protect\citeauthoryear{{Planck Collaboration} et~al.,}{{Planck
  Collaboration} et~al.}{2016}]{Planck16}
{Planck Collaboration} et~al., 2016, \mn@doi [\aap]
  {10.1051/0004-6361/201628897}, \href
  {https://ui.adsabs.harvard.edu/abs/2016A&A...596A.108P} {596, A108}

\bibitem[\protect\citeauthoryear{{Planck Collaboration} et~al.,}{{Planck
  Collaboration} et~al.}{2020}]{Planck2018CMB}
{Planck Collaboration} et~al., 2020, \mn@doi [\aap]
  {10.1051/0004-6361/201833910}, \href
  {https://ui.adsabs.harvard.edu/abs/2020A&A...641A...6P} {641, A6}

\bibitem[\protect\citeauthoryear{{Ploeckinger} \& {Schaye}}{{Ploeckinger} \&
  {Schaye}}{2020}]{Ploe20COLIBREcooling}
{Ploeckinger} S.,  {Schaye} J.,  2020, \mn@doi [\mnras]
  {10.1093/mnras/staa2172}, \href
  {https://ui.adsabs.harvard.edu/abs/2020MNRAS.497.4857P} {497, 4857}

\bibitem[\protect\citeauthoryear{{Pontzen}, {Rey}, {Cadiou}, {Agertz},
  {Teyssier}, {Read}  \& {Orkney}}{{Pontzen} et~al.}{2021}]{Pont20gridnoise}
{Pontzen} A.,  {Rey} M.~P.,  {Cadiou} C.,  {Agertz} O.,  {Teyssier} R.,  {Read}
  J.,   {Orkney} M. D.~A.,  2021, \mn@doi [\mnras] {10.1093/mnras/staa3645},
  \href {https://ui.adsabs.harvard.edu/abs/2021MNRAS.501.1755P} {501, 1755}

\bibitem[\protect\citeauthoryear{{Rahmati} \& {Schaye}}{{Rahmati} \&
  {Schaye}}{2014}]{Rahm14HIgalaxy}
{Rahmati} A.,  {Schaye} J.,  2014, \mn@doi [\mnras] {10.1093/mnras/stt2235},
  \href {https://ui.adsabs.harvard.edu/abs/2014MNRAS.438..529R} {438, 529}

\bibitem[\protect\citeauthoryear{{Rahmati}, {Pawlik}, {Rai{\v{c}}evi{\'c}}  \&
  {Schaye}}{{Rahmati} et~al.}{2013}]{Rahm13HIshield}
{Rahmati} A.,  {Pawlik} A.~H.,  {Rai{\v{c}}evi{\'c}} M.,   {Schaye} J.,  2013,
  \mn@doi [\mnras] {10.1093/mnras/stt066}, \href
  {https://ui.adsabs.harvard.edu/abs/2013MNRAS.430.2427R} {430, 2427}

\bibitem[\protect\citeauthoryear{{Rai{\v{c}}evi{\'c}} \&
  {Theuns}}{{Rai{\v{c}}evi{\'c}} \& {Theuns}}{2011}]{Raic11clumping}
{Rai{\v{c}}evi{\'c}} M.,  {Theuns} T.,  2011, \mn@doi [\mnras]
  {10.1111/j.1745-3933.2010.00993.x}, \href
  {https://ui.adsabs.harvard.edu/abs/2011MNRAS.412L..16R} {412, L16}

\bibitem[\protect\citeauthoryear{{Reed}, {Bower}, {Frenk}, {Jenkins}  \&
  {Theuns}}{{Reed} et~al.}{2007}]{Reed07}
{Reed} D.~S.,  {Bower} R.,  {Frenk} C.~S.,  {Jenkins} A.,   {Theuns} T.,  2007,
  \mn@doi [\mnras] {10.1111/j.1365-2966.2006.11204.x}, \href
  {https://ui.adsabs.harvard.edu/abs/2007MNRAS.374....2R} {374, 2}

\bibitem[\protect\citeauthoryear{{Ricotti} \& {Ostriker}}{{Ricotti} \&
  {Ostriker}}{2004}]{Rico04Xray}
{Ricotti} M.,  {Ostriker} J.~P.,  2004, \mn@doi [\mnras]
  {10.1111/j.1365-2966.2004.07942.x}, \href
  {https://ui.adsabs.harvard.edu/abs/2004MNRAS.352..547R} {352, 547}

\bibitem[\protect\citeauthoryear{{Robertson}}{{Robertson}}{2022}]{Robertson21}
{Robertson} B.~E.,  2022, \mn@doi [\araa]
  {10.1146/annurev-astro-120221-044656}, \href
  {https://ui.adsabs.harvard.edu/abs/2022ARA&A..60..121R} {60, 121}

\bibitem[\protect\citeauthoryear{Robertson, Kravtsov, Gnedin, Abel  \&
  Rudd}{Robertson et~al.}{2010}]{Robe10gridcodeerror}
Robertson B.~E.,  Kravtsov A.~V.,  Gnedin N.~Y.,  Abel T.,   Rudd D.~H.,  2010,
  \mn@doi [\mnras] {10.1111/j.1365-2966.2009.15823.x}, 401, 2463

\bibitem[\protect\citeauthoryear{{Robertson} et~al.,}{{Robertson}
  et~al.}{2013}]{Robe13faintreion}
{Robertson} B.~E.,  et~al., 2013, \mn@doi [\apj] {10.1088/0004-637X/768/1/71},
  \href {https://ui.adsabs.harvard.edu/abs/2013ApJ...768...71R} {768, 71}

\bibitem[\protect\citeauthoryear{{Rosdahl}, {Blaizot}, {Aubert}, {Stranex}  \&
  {Teyssier}}{{Rosdahl} et~al.}{2013}]{Rosd13ramsert}
{Rosdahl} J.,  {Blaizot} J.,  {Aubert} D.,  {Stranex} T.,   {Teyssier} R.,
  2013, \mn@doi [\mnras] {10.1093/mnras/stt1722}, \href
  {https://ui.adsabs.harvard.edu/abs/2013MNRAS.436.2188R} {436, 2188}

\bibitem[\protect\citeauthoryear{{Rosdahl} et~al.,}{{Rosdahl}
  et~al.}{2018}]{Rosd18SPHINX}
{Rosdahl} J.,  et~al., 2018, \mn@doi [\mnras] {10.1093/mnras/sty1655}, \href
  {https://ui.adsabs.harvard.edu/abs/2018MNRAS.479..994R} {479, 994}

\bibitem[\protect\citeauthoryear{{Schaller}, {Gonnet}, {Chalk}  \&
  {Draper}}{{Schaller} et~al.}{2016}]{Scha16SWIFT}
{Schaller} M.,  {Gonnet} P.,  {Chalk} A. B.~G.,   {Draper} P.~W.,  2016, in
  Proceedings of the Platform for Advanced Scientific Computing Conference.
  p.~2 (\mn@eprint {arXiv} {1606.02738}), \mn@doi{10.1145/2929908.2929916}

\bibitem[\protect\citeauthoryear{{Schaller} et~al.}{{Schaller}
  et~al.}{2018}]{Scha18SWIFTascl}
{Schaller} M.,  et~al., 2018, {SWIFT: SPH With Inter-dependent Fine-grained
  Tasking}, Astrophysics Source Code Library (\mn@eprint {ascl} {1805.020})

\bibitem[\protect\citeauthoryear{{Schaller} et~al.,}{{Schaller}
  et~al.}{2023}]{Scha23SWIFT}
{Schaller} M.,  et~al., 2023, \mn@doi [arXiv e-prints]
  {10.48550/arXiv.2305.13380}, \href
  {https://ui.adsabs.harvard.edu/abs/2023arXiv230513380S} {p. arXiv:2305.13380}

\bibitem[\protect\citeauthoryear{{Shapiro} \& {Giroux}}{{Shapiro} \&
  {Giroux}}{1987}]{Shap87cosHII}
{Shapiro} P.~R.,  {Giroux} M.~L.,  1987, \mn@doi [\apjl] {10.1086/185015},
  \href {https://ui.adsabs.harvard.edu/abs/1987ApJ...321L.107S} {321, L107}

\bibitem[\protect\citeauthoryear{{Shapiro}, {Iliev}  \& {Raga}}{{Shapiro}
  et~al.}{2004}]{Shap04minihalo}
{Shapiro} P.~R.,  {Iliev} I.~T.,   {Raga} A.~C.,  2004, \mn@doi [\mnras]
  {10.1111/j.1365-2966.2004.07364.x}, \href
  {https://ui.adsabs.harvard.edu/abs/2004MNRAS.348..753S} {348, 753}

\bibitem[\protect\citeauthoryear{{Sharma}, {Theuns}, {Frenk}, {Bower}, {Crain},
  {Schaller}  \& {Schaye}}{{Sharma} et~al.}{2016}]{Sharma16}
{Sharma} M.,  {Theuns} T.,  {Frenk} C.,  {Bower} R.,  {Crain} R.,  {Schaller}
  M.,   {Schaye} J.,  2016, \mn@doi [\mnras] {10.1093/mnrasl/slw021}, \href
  {https://ui.adsabs.harvard.edu/abs/2016MNRAS.458L..94S} {458, L94}

\bibitem[\protect\citeauthoryear{{Sharma}, {Theuns}, {Frenk}, {Bower}, {Crain},
  {Schaller}  \& {Schaye}}{{Sharma} et~al.}{2017}]{Shar17starburstreion}
{Sharma} M.,  {Theuns} T.,  {Frenk} C.,  {Bower} R.~G.,  {Crain} R.~A.,
  {Schaller} M.,   {Schaye} J.,  2017, \mn@doi [\mnras] {10.1093/mnras/stx578},
  \href {https://ui.adsabs.harvard.edu/abs/2017MNRAS.468.2176S} {468, 2176}

\bibitem[\protect\citeauthoryear{{Shull}, {Harness}, {Trenti}  \&
  {Smith}}{{Shull} et~al.}{2012}]{Shul12clump}
{Shull} J.~M.,  {Harness} A.,  {Trenti} M.,   {Smith} B.~D.,  2012, \mn@doi
  [\apj] {10.1088/0004-637X/747/2/100}, \href
  {https://ui.adsabs.harvard.edu/abs/2012ApJ...747..100S} {747, 100}

\bibitem[\protect\citeauthoryear{{Skinner} \& {Wise}}{{Skinner} \&
  {Wise}}{2020}]{Skin20PopIIIshielding}
{Skinner} D.,  {Wise} J.~H.,  2020, \mn@doi [\mnras] {10.1093/mnras/staa139},
  \href {https://ui.adsabs.harvard.edu/abs/2020MNRAS.492.4386S} {492, 4386}

\bibitem[\protect\citeauthoryear{{Springel}}{{Springel}}{2005}]{Spri05Gadget2}
{Springel} V.,  2005, \mn@doi [\mnras] {10.1111/j.1365-2966.2005.09655.x},
  \href {http://adsabs.harvard.edu/abs/2005MNRAS.364.1105S} {364, 1105}

\bibitem[\protect\citeauthoryear{{Springel} \& {Hernquist}}{{Springel} \&
  {Hernquist}}{2002}]{Spri02esph}
{Springel} V.,  {Hernquist} L.,  2002, \mn@doi [\mnras]
  {10.1046/j.1365-8711.2002.05445.x}, \href
  {https://ui.adsabs.harvard.edu/abs/2002MNRAS.333..649S} {333, 649}

\bibitem[\protect\citeauthoryear{{Stecher} \& {Williams}}{{Stecher} \&
  {Williams}}{1967}]{Stec67photoH2}
{Stecher} T.~P.,  {Williams} D.~A.,  1967, \mn@doi [\apjl] {10.1086/180047},
  \href {https://ui.adsabs.harvard.edu/abs/1967ApJ...149L..29S} {149, L29}

\bibitem[\protect\citeauthoryear{{Tegmark}, {Silk}, {Rees}, {Blanchard}, {Abel}
   \& {Palla}}{{Tegmark} et~al.}{1997}]{Tegm97firstobj}
{Tegmark} M.,  {Silk} J.,  {Rees} M.~J.,  {Blanchard} A.,  {Abel} T.,   {Palla}
  F.,  1997, \mn@doi [\apj] {10.1086/303434}, \href
  {https://ui.adsabs.harvard.edu/abs/1997ApJ...474....1T} {474, 1}

\bibitem[\protect\citeauthoryear{{Theuns}}{{Theuns}}{2021}]{Theuns21}
{Theuns} T.,  2021, \mn@doi [\mnras] {10.1093/mnras/staa3412}, \href
  {https://ui.adsabs.harvard.edu/abs/2021MNRAS.500.2741T} {500, 2741}

\bibitem[\protect\citeauthoryear{{Trac} \& {Cen}}{{Trac} \&
  {Cen}}{2007}]{Trac07rtreion}
{Trac} H.,  {Cen} R.,  2007, \mn@doi [\apj] {10.1086/522566}, \href
  {https://ui.adsabs.harvard.edu/abs/2007ApJ...671....1T} {671, 1}

\bibitem[\protect\citeauthoryear{{Trenti} \& {Stiavelli}}{{Trenti} \&
  {Stiavelli}}{2009}]{Tren09H2PopIII}
{Trenti} M.,  {Stiavelli} M.,  2009, \mn@doi [\apj]
  {10.1088/0004-637X/694/2/879}, \href
  {https://ui.adsabs.harvard.edu/abs/2009ApJ...694..879T} {694, 879}

\bibitem[\protect\citeauthoryear{{Tricco} \& {Price}}{{Tricco} \&
  {Price}}{2012}]{Tric12divBclean}
{Tricco} T.~S.,  {Price} D.~J.,  2012, \mn@doi [Journal of Computational
  Physics] {10.1016/j.jcp.2012.06.039}, \href
  {https://ui.adsabs.harvard.edu/abs/2012JCoPh.231.7214T} {231, 7214}

\bibitem[\protect\citeauthoryear{{Tseliakhovich} \& {Hirata}}{{Tseliakhovich}
  \& {Hirata}}{2010}]{Tsel10streaming}
{Tseliakhovich} D.,  {Hirata} C.,  2010, \mn@doi [\prd]
  {10.1103/PhysRevD.82.083520}, \href
  {https://ui.adsabs.harvard.edu/abs/2010PhRvD..82h3520T} {82, 083520}

\bibitem[\protect\citeauthoryear{{Viel}, {Haehnelt}  \& {Springel}}{{Viel}
  et~al.}{2004}]{Viel04Lyalpha}
{Viel} M.,  {Haehnelt} M.~G.,   {Springel} V.,  2004, \mn@doi [\mnras]
  {10.1111/j.1365-2966.2004.08224.x}, \href
  {https://ui.adsabs.harvard.edu/abs/2004MNRAS.354..684V} {354, 684}

\bibitem[\protect\citeauthoryear{{Visbal}, {Haiman}  \& {Bryan}}{{Visbal}
  et~al.}{2014}]{Visb14gashalocenter}
{Visbal} E.,  {Haiman} Z.,   {Bryan} G.~L.,  2014, \mn@doi [\mnras]
  {10.1093/mnrasl/slu063}, \href
  {https://ui.adsabs.harvard.edu/abs/2014MNRAS.442L.100V} {442, L100}

\bibitem[\protect\citeauthoryear{{Wise}}{{Wise}}{2019}]{Wise19}
{Wise} J.~H.,  2019, arXiv e-prints, \href
  {https://ui.adsabs.harvard.edu/abs/2019arXiv190706653W} {p. arXiv:1907.06653}

\bibitem[\protect\citeauthoryear{{Wu}, {McQuinn}  \& {Eisenstein}}{{Wu}
  et~al.}{2021}]{Wu21M1accuracy}
{Wu} X.,  {McQuinn} M.,   {Eisenstein} D.,  2021, \mn@doi [\jcap]
  {10.1088/1475-7516/2021/02/042}, \href
  {https://ui.adsabs.harvard.edu/abs/2021JCAP...02..042W} {2021, 042}

\bibitem[\protect\citeauthoryear{{Wyithe} \& {Bolton}}{{Wyithe} \&
  {Bolton}}{2011}]{Wyit11UVbg}
{Wyithe} J. S.~B.,  {Bolton} J.~S.,  2011, \mn@doi [\mnras]
  {10.1111/j.1365-2966.2010.18030.x}, \href
  {https://ui.adsabs.harvard.edu/abs/2011MNRAS.412.1926W} {412, 1926}

\bibitem[\protect\citeauthoryear{{Zahn} et~al.,}{{Zahn}
  et~al.}{2012}]{Zahn12kSZreion}
{Zahn} O.,  et~al., 2012, \mn@doi [\apj] {10.1088/0004-637X/756/1/65}, \href
  {https://ui.adsabs.harvard.edu/abs/2012ApJ...756...65Z} {756, 65}

\bibitem[\protect\citeauthoryear{van~der Walt, Colbert  \& Varoquaux}{van~der
  Walt et~al.}{2011}]{vand11numpy}
van~der Walt S.,  Colbert S.~C.,   Varoquaux G.,  2011, \mn@doi [Computing in
  Science Engineering] {10.1109/MCSE.2011.37}, 13, 22

\makeatother
\end{thebibliography}

\appendix
\section{Filtering and minimum minihalo mass}
\label{sec:filtering}
Because the gas temperature in the \ihm\ is finite, the gas density distribution is smoother than that of the dark matter. However, the gas smoothing depends on the entire temperature evolution. 

The {\it filtering} scale is where the baryonic perturbation are smoothed, given the temperature evolution. The filtering scale can be understood in terms of an evolving Jeans length, as described by \cite{Gned98filtering}. First define the (co-moving) Jeans wave-number, $k_J$, at a given scale factor ($a$) as, 
\begin{align}
k_J \equiv \frac{a^{-1/2}}{c_S}\sqrt{4\pi G \overline{\rho}_{m,0}}\,,
\end{align}
where $\overline{\rho}_{m,0}=\Omega_m\rho_c$ is the average matter density at $z=0$, and $c_S$ is the sound speed.
The filtering wave-number, $k_F$, is obtained by integrating Jeans wave-number over time:
\begin{align}
\frac{1}{k^2_{F}(t)} = \frac{1}{D_+(t)}\int^t_0dt'a^2(t')\frac{\ddot{D}_+(t')+2H(t')\dot{D}_+(t')}{k_J^2(t')}\int_{t'}^t\frac{dt''}{a^2(t'')}
\label{eq:filering}
\end{align}
where $D_+$ is the linear growth factor.

In a flat Universe, $D_+$ is equal to the scale factor, $a$, so $D_+\propto a\propto t^{2/3}$ at redshifts
$z\gtrsim 2$. Therefore
\begin{align}
\frac{1}{k^2_{F}(a)} = \frac{3}{a}\int^a_0\frac{da'}{k_J^2(a')}\left[1-\left(\frac{a'}{a}\right)^{1/2}\right]\,,
\label{eq:fileringflat}
\end{align}
where $z_{\rm dec}=1/a_{\rm dec}-1\approx 130$ is the redshift where the gas temperature decouples from the \cmb\ temperature.

Before reionization and the onset of any other heating sources, the \ihm\ temperature is:
\begin{align}
T=  \left \{\begin{matrix}
 2.73 {\rm K}/a& a<a_{\rm dec}\\ 
 2.73 {\rm K} a_{\rm dec}/a^2& {\rm otherwise}, 
\end{matrix} \right.
\end{align}
\begin{align}
c_S^2=  \left \{\begin{matrix}
 \frac{\gamma R}{\mu}2.73 {\rm K}/a& a<a_{\rm dec}\\ 
  \frac{\gamma R}{\mu}2.73 {\rm K} a_{\rm dec}/a^2& {\rm otherwise},
\end{matrix} \right.
\end{align}
\begin{align}
\frac{1}{k^2_{J}}=  \left \{\begin{matrix}
 4\pi G\overline{\rho}_c \frac{\gamma R}{\mu}2.73 {\rm K}& a<a_{\rm dec}\\ 
  4\pi G\overline{\rho}_c \frac{\gamma R}{\mu}2.73 {\rm K}a_{\rm dec}/a& {\rm otherwise}, 
\end{matrix} \right.
\label{eq:adeceq}
\end{align}
Therefore the (co-moving) Jeans scale is independent of redshift before decoupling.

We combine Eqs. \ref{eq:fileringflat} and \ref{eq:adeceq} to calculate the filtering wave-number:
\begin{align}
\frac{1}{k^2_{F}(a)} &= \frac{3a_{\rm dec}}{a}\int^a_{a_{\rm dec}}\frac{da'}{a'}4\pi G\overline{\rho}_c \frac{\gamma R}{\mu}2.73 {\rm K}\left[1-\left(\frac{a'}{a}\right)^{1/2}\right]\nonumber\\
&+ \frac{3}{a}\int^{a_{\rm dec}}_0 da'4\pi G\overline{\rho}_c \frac{\gamma R}{\mu}2.73 {\rm K}\left[1-\left(\frac{a'}{a}\right)^{1/2}\right],\nonumber\\
&= \frac{3}{k^2_J(a)}\left\{\int^a_{a_{\rm dec}}\frac{da'}{a'}\left[1-\left(\frac{a'}{a}\right)^{1/2}\right]\right.\nonumber\\
&\left.+ \frac{1}{a_{\rm dec}}\int^{a_{\rm dec}}_0 da'\left[1-\left(\frac{a'}{a}\right)^{1/2}\right]\right\},
\end{align}

In the simulations, we have neglected the difference between the baryon and dark matter power spectra at the starting redshift. Thus to enable a fair comparison with the simulations, we only integrate Eq.~(\ref{eq:fileringflat}) from $a_{\rm dec}$ to find,
\begin{align}
\frac{1}{k^2_{F}(a)} = \frac{1}{k^2_{J}(a)}\left[3\ln (a/a_{\rm dec})-6+6\left(\frac{a_{\rm dec}}{a}\right)^{1/2}\right].
\end{align}
The filtering mass (eq\ref{eq:MF}) can then be inferred from the resultant $k_{F}$ and $M_J$.

When generating initial conditions for the simulations, we assume that gas traces the dark matter. For consistency, we therefore
replace $z_{\rm dec}$ by the starting redshift
of the simulations, $z_{\rm ic}=127$, to estimate
$M_F$ in the simulations, with the result
plotted in Fig.~\ref{fig:difm_z}. The resulting value of $M_{\rm F}$ is lower than that quoted by \cite{Gned98filtering}, who start the integration in the limit of $z\to\infty$. 
Our estimate of $M_F$ is close to the more accurate calculation performed by \cite{Naoz07mhfg}, who take account of the fact that the gas and \cmb\ temperatures do not decouple completely at $z_{\rm dec}$. Presumably this better agreement is mostly coincidental. 

\section{Variable speed of light}
\label{sec:vsl}

\begin{figure}
 \includegraphics[width=0.48\textwidth]{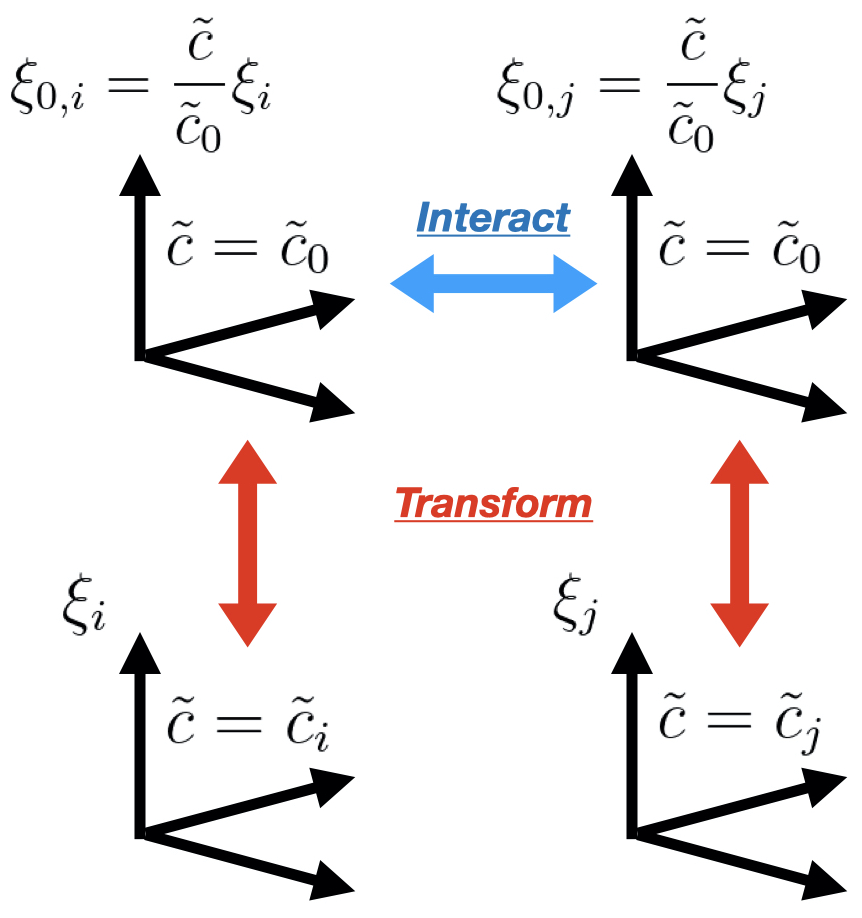}
\caption{Schematic diagram illustrating our variable-speed-of-light implementation. The left and right lower panels represent particles $i$ and $j$ with different specific radiation energies $\xi$ and reduced speeds of light $\tilde{c}$ respectively. To evaluate the energy and flux exchanges, particles $i$ and $j$ are transformed to a frame with a reduced speed of light $\tilde{c}_0$ (upper panels). After evaluating the particle-particle interaction, radiation variables are transformed back to the original frame. }
\label{fig:VSLschematic}
\end{figure}

\begin{figure}
 \includegraphics[width=0.48\textwidth]{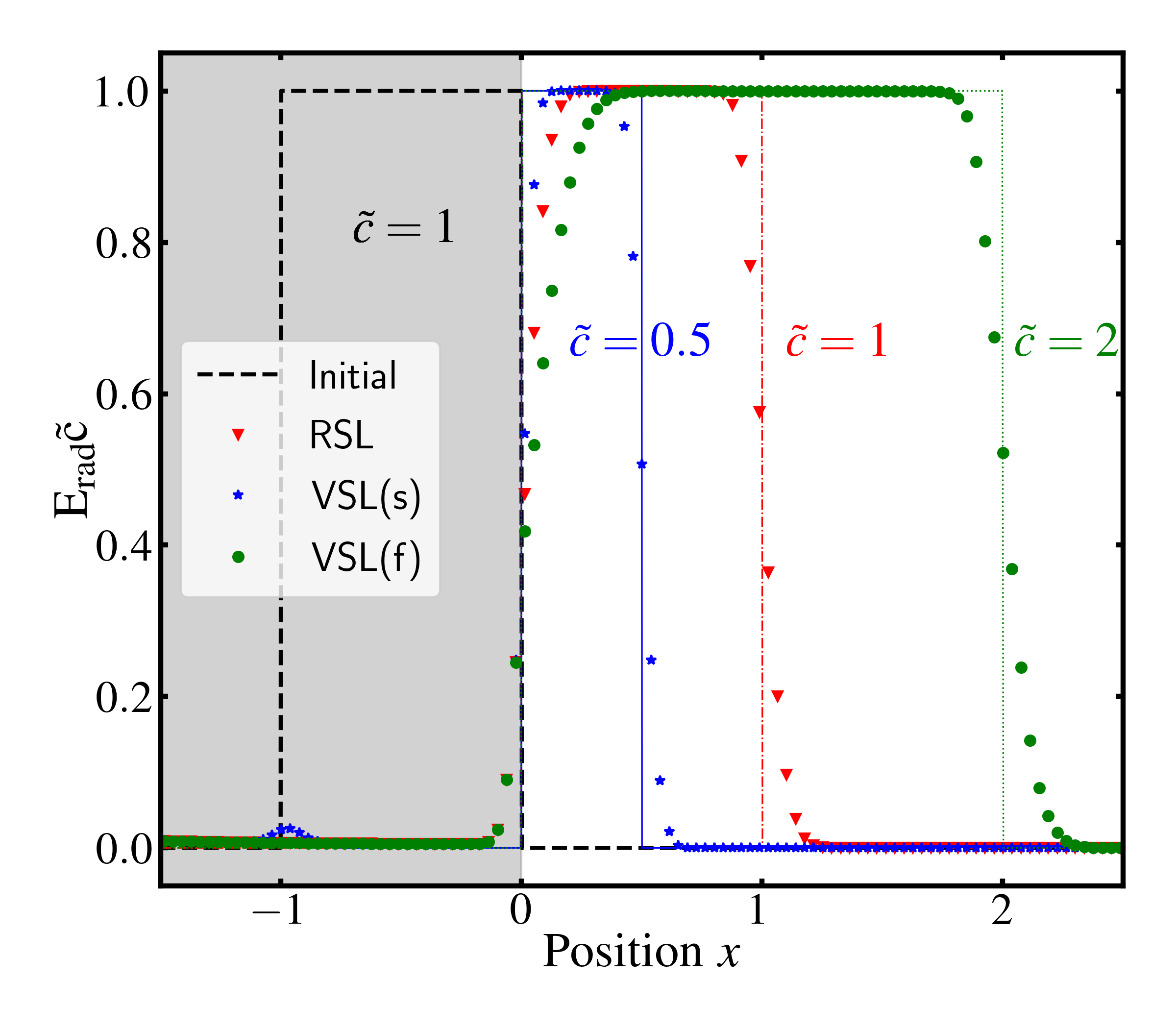}
\caption{Tests of the \rsl\ and \vsl\ implementation, in which a packet of radiation streams in the optically thin limit from left to right. Shown in all cases is the product of the energy density times $\tilde{c}$, which is independent of the choice of $\tilde{c}$.  The {\em black dashed line} is the initial shape of the packet, with {\em thin coloured lines} the correct solution at later times. The corresponding numerical solution in the \rsl\ case with $\tilde{c}=1$ is shown as {\em red downwards triangles}. The {\em blue stars} and {\em green circles} correspond to the numerical solution in the \vsl\ approximation, which uses $\tilde{c}=1$ for $x<0$ and
$\tilde{c}=0.5$ and $\tilde{c}=2.0$) for $x>0$, respectively. The numerical calculation uses $10^3$ particles in 1 dimension, and in all cases is shown at time $t=1$. The simulation reproduces the correct solution, apart from the inevitable diffusion associated with the SPH scheme.}
\label{fig:radfront1dvsl}
\end{figure}

\begin{figure}
 \includegraphics[width=0.48\textwidth]{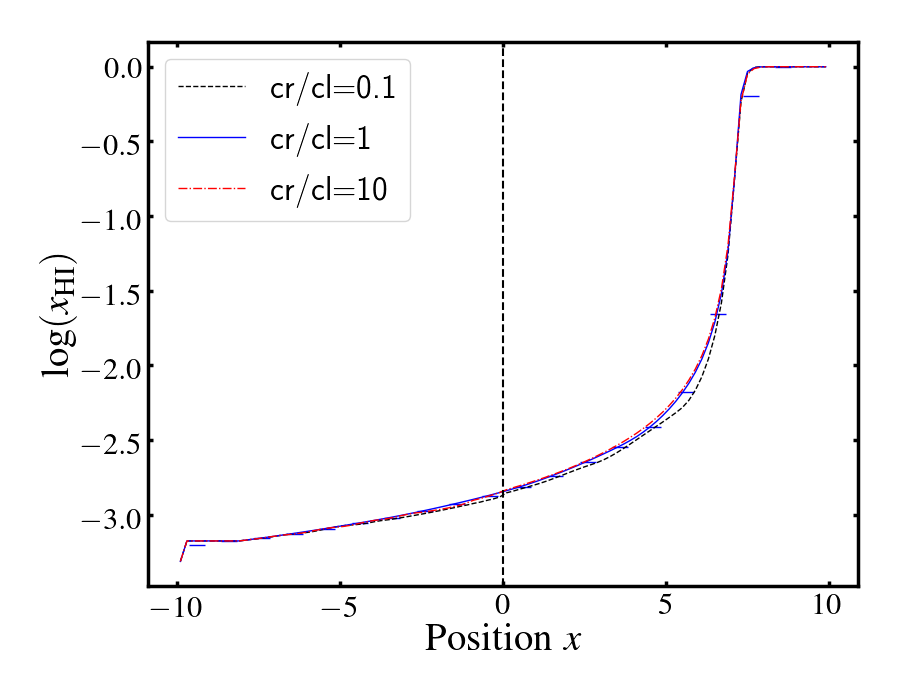}
\caption{Neutral gas fraction, $x_{\rm HI}=n_{\rm HI}/n_{\rm H}$, in a 1D photo-ionization test of the \vsl\ approximation. We use dimensionless variables, in which
the simulation volume contains 100 gas particles with hydrogen density $n_{\rm H}=1.0$, photo-ionization cross-section $\sigma_{\rm HI}=10^2$ and recombination coefficient $\alpha_{\rm B}=4.0$. Radiation is injected at
the location $x=-8$ with a constant radiation flux $F_\gamma=60$. Different curves show the run of $x_{\rm HI}$ with $x$ for simulations with different values of \rsl\, with $cr$ the value of \rsl\ at $x>0$ in units of the \rsl\ at $x<0$. Small horizontal lines indicate the local value of the smoothing length. The profile is shown at time $t=10^2$ when $x_{\rm HI}$ is in equilibrium. The test shows that the equilibrium value does not depend on the choice of the \rsl. }
\label{fig:radbound1dvsl}
\end{figure}

\begin{figure}
 \includegraphics[width=0.48\textwidth]{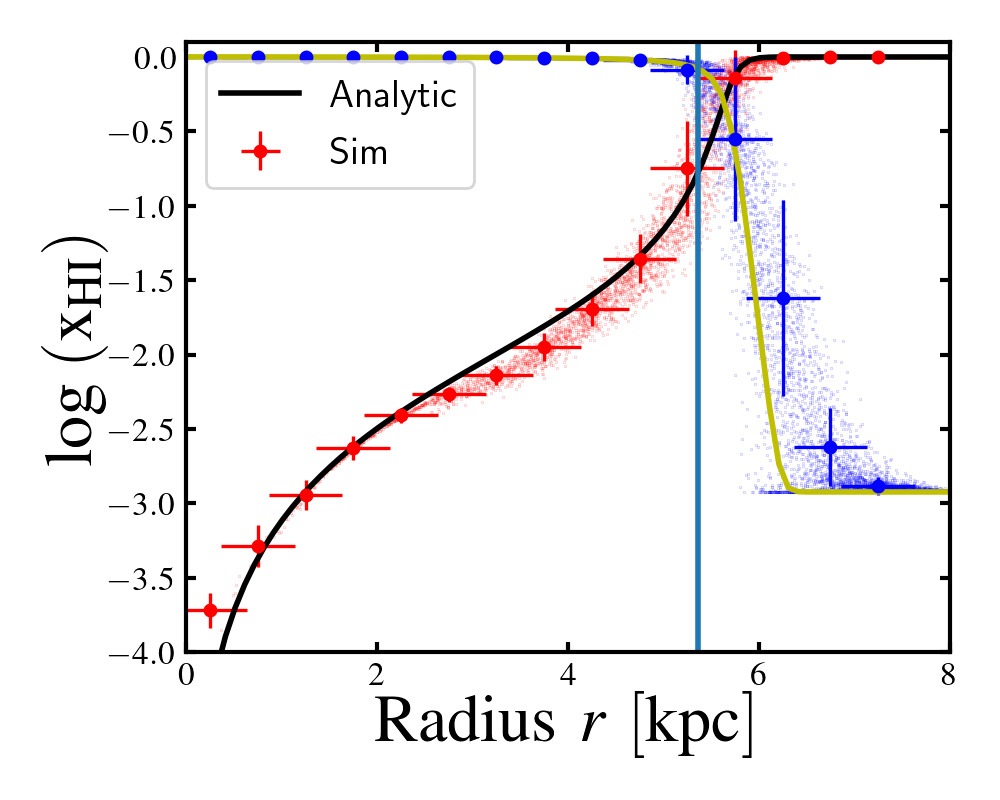}
\caption{Isothermal Str\"omgren sphere from \protect\cite{Ilie06RTcom} Test 1: a source of radiation, located at radius $r=0$, photo-ionizes hydrogen gas, kept at constant density and constant temperature. The system is shown a time $t=500~{\rm Myrs}$ after the source is switched on. {\em Red points} are the neutral hydrogen fraction, $x_{\rm HI}=n_{\rm HI}/n_{\rm H}$, of individual gas particles, with the {\em thick red points} showing binned values, with horizontal error bars indicating the bin width, and vertical error bars the standard deviation; the {\em black line} is the analytical solution. {\em Small blue points}, and {\em thick blue points} with error bars show the corresponding ionized fraction, $1-x_{\rm HI}$, with the {\em yellow line} the analytical solution. The {\em vertical light blue line} is the approximate analytic location of the Str\"omgren radius. In this \vsl\ test, $\tilde{c}$ increases by a factor two at $r=4~{\rm kpc}$. This has little effect on the numerical solution, which is almost identical to the case where $\tilde{c}$ is constant (Fig.~B4 in \protect\citealt{Chan21SPHM1RT}).}
\label{fig:stromgren3dvsl}
\end{figure}

The time-step in simulations is limited by the Courant–Friedrichs–Lewy condition (CFL, \citealt{Courant28}), $\Delta t<{\cal C}h/v_{\rm sign}$, where $h$ is the spatial resolution, $v_{\rm sign}$ the signal speed in the simulated fluid, and ${\cal C}$ is a constant of order unity. For radiation, the signal speed is the speed of light ($v_{\rm sign}=c$), so this time step is often many orders of magnitude smaller than that for the fluid itself, where the signal speed is of the order of the sound speed. Therefore, the inclusion of \rt\ can dramatically reduce the time step and, in turn, increases the compute time. 

However, even though radiation travels with speed $c$, ionization fronts typically travel {\em much} slower, see for example Eq.~(\ref{eq:IFspeed}). \cite{Gned01OTVET} first suggested 
that reducing the speed of light, $c\to\tilde{c}\ll c$, would improve the computational speed of the code without sacrificing its accuracy: this is the \lq reduced speed of light\rq\ approximation (hereafter \rsl). The essential idea is that the state of the gas, when it is in equilibrium with the radiation (e.g. photo-ionization equilibrium), is independent of the propagation speed. In fact, the \rsl\ approximation may give accurate results even in some non-equilibrium situations, provided the light crossing time is much shorter than other time-scales \citep[see e.g. ][]{Rosd13ramsert, Chan21SPHM1RT}.

The required $\tilde{c}$ varies across the simulation volume. Indeed, \Ifront\ moves fast in low-density regions, and hence $\tilde{c}$ should not be much smaller than $c$. However, such a large $\tilde{c}$ results in very short time-steps in high-density regions\footnote{The required time-step can be less than one thousand years in particles with sizes smaller than 0.1~ckpc, one million times smaller than the duration of simulations.}, where $h$ is small. Fortunately, the \Ifront\ speed is actually very low in dense regions, so we could apply a much lower value of $\tilde{c}$ without loss of accuracy. Therefore, it is very tempting
to allow $\tilde{c}$ to {\em vary} in space and time. This is the variable speed of light approximation (hereafter \vsl, e.g. one \vsl\ version in \citealt{Katz17reionVSL}).

The first requirement (1) of a \vsl\ scheme is to keep the {\it equilibrium} photo-ionization rate independent of $\tilde{c}$, so does the {\it equilibrium} neutral gas fraction.
The photo-ionization rate is proportional to the number density of photons, the speed of light, and the photo-ionization cross-section, i.e. $n_\gamma\,c\,\sigma_{\rm HI}$. This implies that the product of the radiation density times the reduced speed of light, $n_{\gamma}\tilde{c}$, and the radiation flux should not change when $\tilde{c}$ is used\footnote{We use the notation of \cite{Chan21SPHM1RT}, in which $\rho\xi$ is the radiation density, and $\rho{\bf f}$ is the radiative flux, where $\rho$ is the gas density.}.

The second requirement (2) is that the scheme should reduce to the \rsl\ approximation when ${\tilde c}$ is uniform.

For reference, the terms entering the rate of change of radiation variables 
that involve spatial derivatives of these variables (and hence are computed by summing over {\sc sph} neighbours that may have a different value of $\tilde{c})$,
are given by (see \citealt{Chan21SPHM1RT}, their Eqs.~5-7),
\begin{align}
&\left.\left(\frac{D\xi}{Dt}\right)\right|_{\rm prop}=-\frac{1}{\rho}\nabla\cdot(\rho{\bf f}),
\end{align}
\begin{align}
&\left.\frac{D {\bf f}}{Dt}\right|_{\rm prop}=-\tilde{c}^2\frac{\nabla\cdot (\xi\mathbb{F})}{\rho}\,,
\end{align}
where the subscript \lq prop\rq\ means that we only write terms that involve the propagation of radiation across particles. We are going to modify these equations for \vsl, so that they can be applied consistently even if two particles that {\em exchange} radiation have different values of $\tilde{c}$. To be definite, let these particles be $i$ and $j$, with \rsl\ values $\tilde{c}_i$ and $\tilde{c}_j$.

To satisfy requirements (1) and (2), we devise a new three-step \vsl\ scheme, as illustrated in Fig.~\ref{fig:VSLschematic}. In short, (A) we transform 
the equations to a system with a common \rsl, $\tilde{c}_0$. Next, (B) we solve the propagation equation with $\tilde{c}_0$. Finally, (C) we transform the equations back to the $\tilde{c}_i$ and $\tilde{c}_j$.

In the first step (A), we construct a reference frame with $\tilde{c}_0$ and then transform the quantities from the local (particle i) frame (with the \rsl\ $\tilde{c}_i$) to the reference frame through:
\begin{align}
\xi_{i} \rightarrow \frac{\tilde{c}_0}{\tilde{c}}\xi_{0,i}; \;\;\;\;{\bf f}_{i} \rightarrow {\bf f}_{0,i},
\end{align}
where we label the $\tilde{c}_0$ reference frame with the subscript ``0''. We apply the same transform for particles j, the neighbourhood particles of i. In this transform, $\rho {\bf f}$ and $\rho \tilde{c}\xi$ are invariant, so the radiation energy density in the reference frame is equal to that in the original frame, i.e $\rho\tilde{c}_0\tilde{\xi_0}=\rho\tilde{c}_i\xi_i$, so this satisfies the requirement (1).

Hence, the radiation equations in the $\tilde{c}_0$ frame become:
\begin{align}
&\left.\left(\frac{D\xi_{0}}{Dt}\right)\right|_{\rm prop}
=-\frac{1}{\rho}\nabla\cdot(\rho{\bf f}_{0}),
\label{eq:sphxi}
\end{align}
\begin{align}
&\left.\frac{D {\bf f}_{0}}{Dt}\right|_{\rm prop}=-\tilde{c}_0^2\frac{\nabla\cdot (\xi_{0}\mathbb{F}_{0})}{\rho}.
\label{eq:SPHdfrad}
\end{align}

In the second step (B), we solve Eq. \ref{eq:sphxi} and Eq. \ref{eq:SPHdfrad} in one neighbourhood by applying the SPH algorithm to evaluate the derivatives,
\begin{align}
\left.\left(\frac{D\xi_{0,i}}{Dt}\right)\right|_{\rm prop}=-\sum_j \frac{m_j}{\Omega_i\rho_i^2}(\rho_i{\bf f}_{0,i}-\rho_j{\bf f}_{0,j})\cdot\nabla_i W_{ij}(h_i),
\end{align}
and
\begin{align}
\left.\left(\frac{D {\bf f}_i}{Dt}\right)\right|_{\rm prop}=-\sum_j\frac{m_j\tilde{c}_0^2}{\Omega_i\rho_i^2}\left (\rho_i \xi_{0,i}\mathbb{F}_{0,i} -\rho_j\xi_{0,j}\mathbb{F}_{0,j}\right )\cdot\nabla_i W_{ij}(h_i).
\label{eq:fradVSL}
\end{align}
Note that we have to pick the ``difference'' SPH differential form (e.g. see \citealt{Tric12divBclean} and \citealt{Chan21SPHM1RT}) if the speed of light changes discontinuously. Otherwise, the estimated gradient will be inaccurate, which would lead to numerical reflections at discontinuities.

In the final step (C), we convert back to the local ($\tilde{c}_i$) frames,
\begin{align}
\xi_{0,i}\rightarrow  \frac{\tilde{c}}{\tilde{c}_0}\xi_{i}; \;\;\;\;{\bf f}_{0,i} \rightarrow {\bf f}_{i},
\end{align}
and similarly for particle $j$.

The solution will depend on our choice for $\tilde{c}_0$.
We take $\tilde{c}_0={\rm max}(\tilde{c}_i,\tilde{c}_j)$ for each pairwise
interaction term to ensure symmetry and recover the \rsl\ limit (i.e. satisfying the requirement (2)). In experiments, we found that the choice of $\tilde{c}_0$ makes little difference\footnote{The RSL solution will also depend on $\tilde{c}$, e.g. in the optically thin limit. But the equilibrium solution, e.g. Str\"omgren sphere, will not.} as long as (I) $\tilde{c}_0$ is between $\tilde{c}_i$ and $\tilde{c}_j$ and (II) $\tilde{c}_0$ is symmetric with respect to $i$ and $j$.

We also apply these changes to the artificial diffusion
(see \S~2.5 in \citealt{Chan21SPHM1RT}).
\begin{align}
\left.\frac{{\rm D}\tilde{\xi}_{0,i}}{{\rm D} t}\right|_{\rm diss}=\sum^N_{j=1}D_{\xi,ij}\frac{m_j}{\rho_i\rho_j}(\rho_i\xi_{0,i}-\rho_j\xi_{0,j})\frac{{\bf \hat{r}}_{ij}\cdot\overline{\nabla_i W_{ij}}}{r_{ij}},
\label{eq:xidiss}
\end{align}
and artificial viscosity:
\begin{align}
\left.\frac{{\rm D}{\bf f}}{{\rm D} t}\right|_{\rm diss} = \frac{1}{\rho} \nabla\cdot\left[ \mathbb{D}^{\bf f}\nabla\cdot(\rho{\bf f})\right],
\label{eq:fdissaniso}
\end{align}
where $\mathbb{D}^{\bf f}$ is given by
\begin{align}
\mathbb{D}^{\bf f} =\alpha_{\bf f} v_{\rm sig} h \hat{{\bf n}}\hat{{\bf n}}.
\end{align}
Eq.~(\ref{eq:fdissaniso}) has to be solved with the ``difference'' SPH form too. 

We demonstrate the robustness of our scheme in Fig.~\ref{fig:radfront1dvsl}, which shows a case of radiation propagation in the optically thin limit. We consider a rapid change of $\tilde{c}$ at $x=0$ and investigate whether our scheme can work in the following aspects. 

First, both \rsl(s) and \vsl(f) keep the product $E_{\rm rad}\tilde{c}$ approximately constant within the plateau region, so they have the same ionizing power. Second, both the \vsl(s) and \vsl(f) profiles follow $\tilde{c}$ specified in different regions. For example, the \vsl(s) line reaches $x=0.5$ at $t=1$ (with $\tilde{c}=0.5$). To achieve this speed, the \vsl(s) profile is compressed by a factor of two. Third, we have explicitly checked that photon conservation is better than 1\% in all runs even with sudden changes in $\tilde{c}$. Finally, our scheme is also robust during the transition. Even with a rapid change in $\tilde{c}$ at $x=0$, the radiation profile at $x=0$ is remarkably smooth. 

The second test is the 1D photo-ionization test in Fig.~\ref{fig:radbound1dvsl}. A constant photon flux is injected from the left and recombinations balance photo-ionizations. In the \vsl\ runs, the reduced speed of light increases or decreases by a factor of ten at $x=0$. We find that the equilibrium neutral fraction profiles computed with the \vsl\ approximation match well that computed with the \rsl\ approximation, demonstrating the accuracy of the  \vsl\ implementation.

The final test is a 3D isothermal Str\"omgren sphere (\protect\citealt{Ilie06RTcom} Test 1) in Fig. \ref{fig:stromgren3dvsl}. A source at the centre is emitting ionizing radiation at a constant rate of $\dot N_\gamma=5\times10^{48} {\rm photons~s}^{-1}$. The volume has a linear extent of 20~kpc, filled with $32^3$ gas particles in a glass-like distribution. The gas is pure hydrogen with density $n_{\rm H}=10^{-3}{\rm cm^{-3}}$. The collisional ionization coefficient is $\beta=3.1\times10^{-16}{\rm cm^3s^{-1}}$, whereas the recombination coefficient is $\alpha_{\rm B}=2.59\times10^{-13}{\rm cm^3s^{-1}}$ (with the on-the-spot approximation). The photo-ionization cross section is $\sigma_{\rm \gamma HI}=8.13\times 10^{-18}{\rm cm^2}$. We apply the variable speed of light approximation such that $\tilde{c}=0.05c$ for $r<4\;{\rm pkpc}$ and $\tilde{c}=0.1c$ for $r>4\;{\rm pkpc}$.  Even with a sudden change of $\tilde{c}$ by a factor of two, the equilibrium neutral fraction profile still agrees well with the analytic solution (from \citealt{Chan21SPHM1RT}).

Our \vsl\ scheme improves upon the original scheme described by \cite{Katz17reionVSL} as follows. Our scheme allows for adaptive time stepping without the need for sub-cycling. Our scheme can also be applied to any moment-based radiative transfer scheme, e.g. adaptive mesh refinement, moving mesh, or any meshless method.

\section{Cosmological Radiative Transfer}
\label{sec:comoving}

\begin{figure}
 \includegraphics[width=0.45\textwidth]{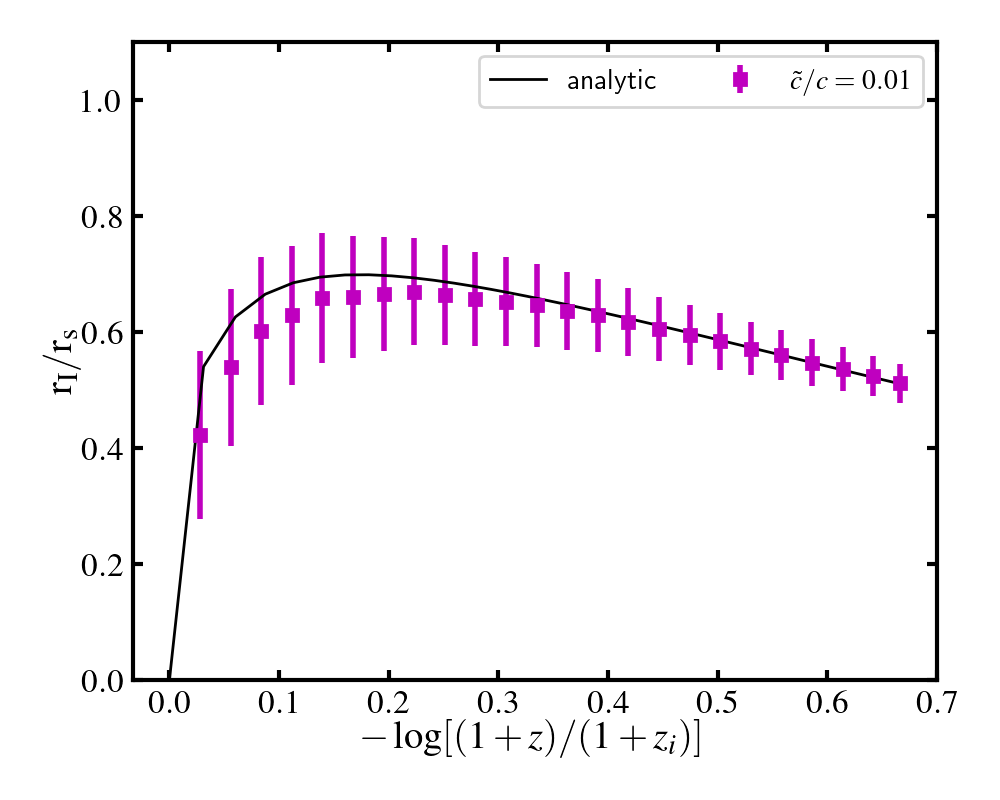}
\caption{\Ifront\ location, $r_I$, divided by the Str\"omgren radius, $r_S$, as a function of redshift; $r_I$ is operationally defined as the location where 50~per cent of the gas is ionized. The source switches on at redshift $z_i=9$. {\em Magenta squares} show the simulation result, with error-bars depicting the gas smoothing lengths divided by $r_S$. The {\em black line} is the analytic solution from \protect\cite{Shap87cosHII}.}
\label{fig:ifrontcosmo}
\end{figure}

To account for cosmological expansion, we convert from proper to co-moving variables in the usual way,

\begin{align}
{\bf r}=a{\bf r}';\;\chi=a^2\chi';\;\rho = \frac{1}{a^3}\rho';\;W_{ij}(h)=\frac{1}{a^3}W'_{ij}(h');\;\nonumber\\
\nabla_i = \frac{1}{a}\nabla_i';\;\nabla_iW_{ij}(h)=\frac{1}{a^4}\nabla' W'_{ij}(h');\;\dot{r}'=\frac{1}{a^2}v';\;\frac{\dot a}{a}=H,
\end{align}
where primed variables are co-moving and $a$ is the expansion factor. Furthermore, ${\bf r}$ is position, $\chi$ opacity, $\rho$ gas density, $W$ is the kernel and $\nabla$ the gradient operator. 

Furthermore, we transform the radiation quantities as follows,
\begin{align}
\xi =\xi';\;{\bf f}=a{\bf f'};\;\tilde{c}=a\tilde{c}';\;\mathbb{P}=\frac{1}{a^3}\mathbb{P}'.
\label{eq:comovingxi}
\end{align}
Here, $\xi(=E_{\rm rad}/\rho)$ is the ratio of the radiation energy density over the gas density
${\bf f}$ is the ratio of the radiative flux over the gas density and $\mathbb{P}$ is the radiation stress tensor (see table~1 in \citealt{Chan21SPHM1RT}). Our choice accounts only for the cosmological dilution of photons but not the redshifting of the wavelengths of photons. As a result, both gas and photon densities have the same dependency on scale factor, so $\xi=E_{\rm rad}/\rho =E'_{\rm rad}/\rho' =\xi'$. 

The primed radiation equations are then unchanged 
from the original proper equations, 
\begin{align}
\frac{D\xi'}{Dt}=-\frac{1}{\rho'}\nabla'\cdot(\rho'{\bf f}'),
\label{eq:duradcos}
\end{align}
\begin{align}
\frac{1}{\tilde{c}'^2}\frac{D}{Dt}{\bf f}'=-\frac{{\bf \nabla'\cdot \mathbb{P}'}}{\rho'}-\frac{\chi'\rho'}{\tilde{c}'}{\bf f}',
\label{eq:dfradcos}
\end{align}
if we neglect small cosmological terms when the scale factor changes slowly. 
These are small corrections for the current application\footnote{One of the neglected terms is the cosmological redshifting of the wavelength of photons. This is relevant for those photons that can travel over distances comparable to the Hubble distance, but not for the ionizing photons that have a mean-free path of only a few Mpc in our set-up. } and ignored in many previous works, e.g. in  \cite{Shap87cosHII,Rosd13ramsert}.

We verify the implementation of these equations by
simulating the evolution of a cosmological H{\sc ii} region
in a homogeneous, constant-temperature, matter-dominated flat universe, with 10~per cent of its mass density in hydrogen
(and neglecting helium). The hydrogen is initially fully neutral and is ionized by a source that 
switches on at $z_i=9$, emitting ionizing radiation at a constant rate of $\dot{N}_\gamma=5\times10^{48} {\rm photons~s}^{-1}$ at $h\nu=13.6{\rm eV}$ (and keeping the temperature constant). The collisional ionization coefficient is $\beta=3.1\times10^{-16}{\rm cm^3s^{-1}}$, whereas the recombination coefficient is $\alpha_{\rm B}=2.59\times10^{-13}{\rm cm^3s^{-1}}$ (at $T=10^4{\rm K}$); making the on-the-spot approximation). The photo-ionization cross section is $\sigma_{\rm \gamma HI}=8.13\times 10^{-18}{\rm cm^2}$. 

In the non-cosmological case, provided the density and temperature are uniform and constant, the \Ifront\ initially moves at the speed of light, before slowing down to the speed 
derived in Eq.~(\ref{eq:IFspeed}) which eventually approaches zero due to recombinations.
Its position $r_I$ as a function of time is given by
\begin{align}
    y(t) \equiv \left(\frac{r_I(t)}{r_S}\right)^3\to 1-\exp\left(-\frac{t-t_i}{t_r}\right)\,.
\end{align}
Here, $t_i$ is the time that the sources switches on, $t_r$ is the recombination time, $t_r^{-1}=\alpha_{\rm B}\,n_{\rm H}$, and $r_S=\left(3t_r\dot N_\gamma/(4\pi n_{\rm H})\right)^{1/3}$ is the Str\"omgren radius.

In the cosmological case, the \Ifront\ initially also moves at the speed of light before slowing down as described by Eq.~(\ref{eq:IFspeed}). However, the Str\"omgren radius increases as the density decreases due to the expansion of the Universe, 
$R_s\propto a^2$ (provided the temperature is taken to be constant, rather than decreasing with the gas density). As a consequence,
the ratio $r_I/r_S$ {\em decreases} again at late times. The analytical solution is given by \citep{Shap87cosHII}
\begin{align}
y(t) \to \lambda\,\exp(\lambda\,\frac{t_i}{t})\,\left[\frac{t}{t_i}\,E_2(\lambda \frac{t_i}{t})-E_2(\lambda)\right]\,,
\label{eq:cosIF}
\end{align}
where $E_2$ is the exponential integral of order 2, and the constant $\lambda$ is the ratio of the initial time over the initial recombination time, $\lambda\equiv t_i/t_r(t_I)$.

We simulate the set-up using $64^3$ gas particles in a computational volume with a linear extent of $40~{\rm ckpc}$, using $\tilde{c}=0.01\,c$, and compare the outcome to the analytical solution in Fig.~\ref{fig:ifrontcosmo}. 
The H{\sc ii} region is not well resolved initially, but at later times the agreement between the simulation and the analytical result is excellent. The good agreement at late times demonstrates that our choice of co-moving radiation variables works correctly.

\section{Numerical convergence}
\label{sec:ressizecred}
\subsection{Choice of clumping factor}
\label{sec:clumptest}
\begin{figure}
\includegraphics[width=0.45\textwidth]{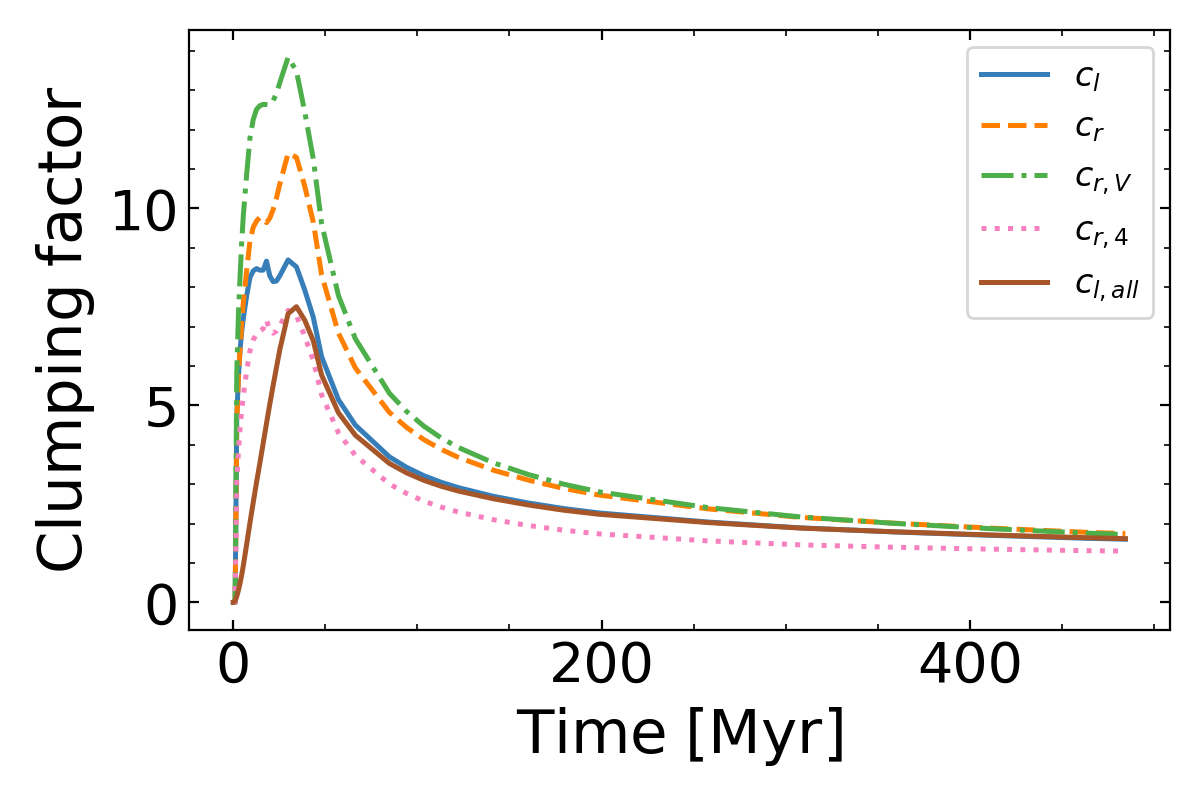}
\caption{Evolution of the clumping factor in simulation M512z8G03, where $t=0$ is the instance the \Ifront\ enters the simulation volume. Different curves correspond to different definition of the clumping factor, as described in section~\ref{sec:clumptest}.}
\label{fig:clumptest}
\end{figure}

In this section we contrast various definitions of the clumping factor, $c_r$, that appear in the literature. For the first three of those, we only consider gas that is downstream of the \Ifront\ (as we did in the main text).

The first definition is $c_l$ from \cite{Embe13clumping}:
\begin{align}
c_l = \frac{\langle n^2_{\rm HII}\rangle_{\rm IF}}{\langle n_{\rm HII}\rangle_{\rm IF}^2}\,.
\label{eq:cl}
\end{align}
This expression does not account for any spatial variations in temperature in the ionized regions.
Such variations affect the value of the recombination rate.

The second definition is $c_r$ from \cite{Park16clumping}, as used in the main text (Eq.~\ref{eq:crdef}). This expression does account for spatial variations in the temperature: the denominator evaluates the recombination rate at the mass-averaged temperature, $\bar{T}_{\rm IF}$.

The third definition is $c_{r,V}$, which is similar to the second definition, except that the recombination rate is evaluated at the {\em volume} averaged temperature, 
$\langle T\rangle_{\rm IF}$, yielding
\begin{align}
c_{r,V} \equiv \frac{\langle\alpha_{\rm B}\,n^2_{\rm HII}\rangle_{\rm IF}}{\alpha_{\rm B}(\langle T\rangle_{\rm IF})\, \langle n_{\rm HII}\rangle_{\rm IF}^2},
\label{eq:crVdef}
\end{align}

The fourth definition is similar to \cite{DAlo20clumping}. We evaluated the recombination rate at $T=10^4{\rm K}$:
\begin{align}
c_{r,V} \equiv \frac{\langle\alpha_{\rm B}\,n^2_{\rm HII}\rangle_{\rm IF}}{\alpha_{\rm B}(10^4{\rm K})\, \langle n_{\rm HII}\rangle_{\rm IF}^2}.
\label{eq:cr4def}
\end{align}

In the fifth definition, we consider all gas in the computational volume:
\begin{align}
c_{l,all} = \frac{\langle n^2_{\rm HII}\rangle}{\langle n_{\rm H}\rangle^2}.
\label{eq:clall}
\end{align}
Unlike the previous definitions, the recombination rate can be estimated from $c_{l,all}$ without knowing the ionization fraction of the volume.

We have computed these clumping factors in simulation M512z8G03 and they are plotted as a function of time since the \Ifront\ entered the computational volume in Fig.~\ref{fig:clumptest}. All clumping factors exhibit a similar evolution: a fast rise to a peak value, followed by a decline to an asymptotic value. 

The clumping factor $c_r$ that we use in the main text ({\em i.e.} computed following the second definition, Eq.~\ref{eq:crdef}) is larger than $c_l$ by $\sim 30$~per cent, which is due to spatial variation in the temperature which  $c_{r}$ accounts for. In the simulations, dense gas is shielded and remains cooler than lower-density gas. The lower temperature increases the recombination rate in the numerator, and hence the clumping factor (see also Fig.~2 in \citealt{Park16clumping}). On the other hand, $c_r$ is {\em smaller} than $c_{r,V}$ by $\sim 30$~per cent. This is because the volume-averaged temperature (used in $c_{r,V}$) is higher than the mass-averaged temperature (used in $c_r$), hence $\alpha_{\rm B}(\langle T\rangle)<\alpha_{\rm B}(\bar T)$ since $\alpha_{\rm B}(T)\propto T^{-0.7}$. However, both $c_{r,V}$ and $c_{r}$ can be accurate measures of recombination rate, as long as the corresponding $\alpha_{\rm B}$ is accounted for
when computing the net recombination rate in Eq.~(\ref{eq:crdef}). $c_{r,4}$ is lower than both $c_{r,V}$ and $c_{r}$, because the reference temperature $T=10^4{\rm K}$ is lower than the mean temperature of the ionized gas. Finally, $c_{l,all}$ is the lowest since it considers all of the gas in the denominator. It quickly converges to $c_l$ when all of the volume is ionized, around 50~Myr in this simulation.

All definitions here converge to the same asymptotic value 
(times $t\geq 400$~Myr in this simulations), because once the volume is ionized, temperature variations are relatively small. However, at earlier times, it is important to use the same definition of the clumping factor when comparing numerical simulations, given the 30-60~per cent differences that we find. We think that $c_r$ or $c_{r, V}$ are preferred since they capture the initial spatial temperature fluctuations better.

\subsection{Convergence with the value of $\tilde{c}$}
\begin{figure}
\includegraphics[width=0.45\textwidth]{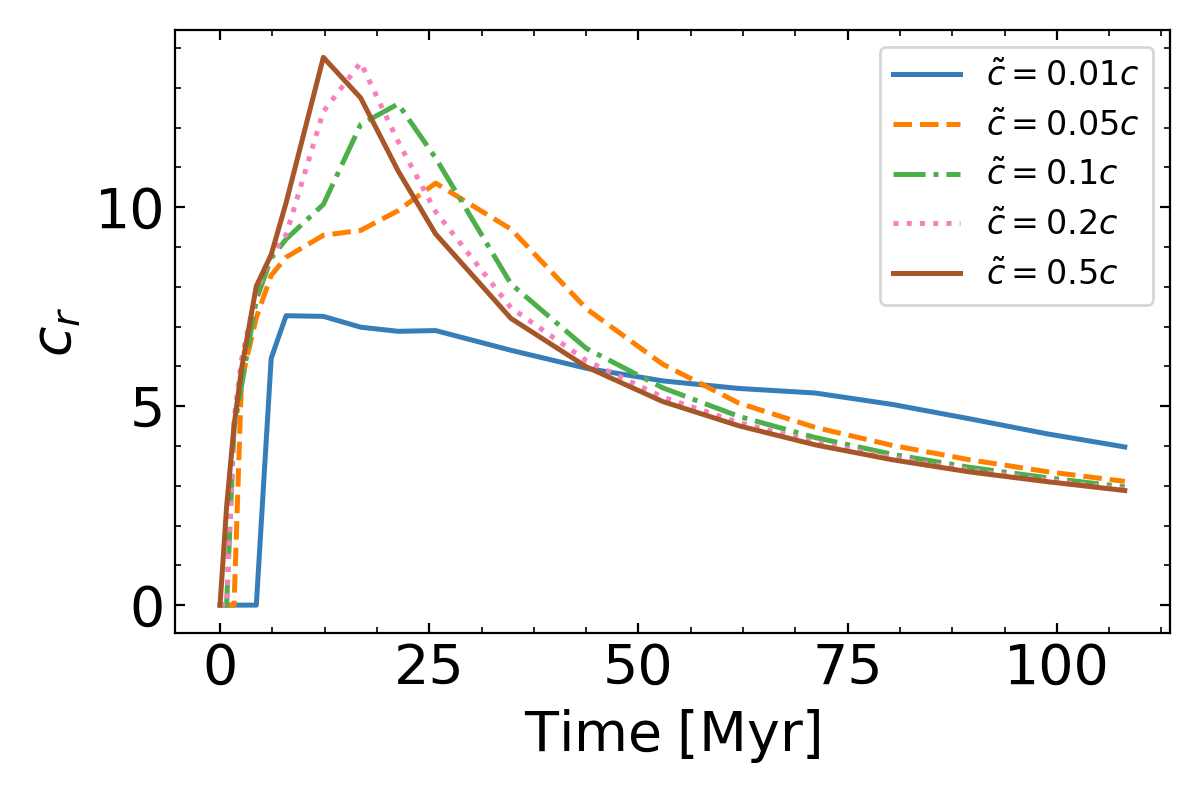}
\caption{Convergence of the clumping factor for different reduced speeds of light $\tilde{c}$ (at the mean density of the simulation volume). We consider the small box (400 ckpc), $\Gamma_{-12}=0.3$, and $m_{\rm gas}=20\;\msun$ (i.e. the S256G03 runs in Table \ref{table:sim}). The peak clumping factor converges at $\tilde{c}\sim 0.2c$, which is around the value we adopt in the production runs.}
\label{fig:credtest}
\end{figure}
\begin{figure}
\includegraphics[width=0.45\textwidth]{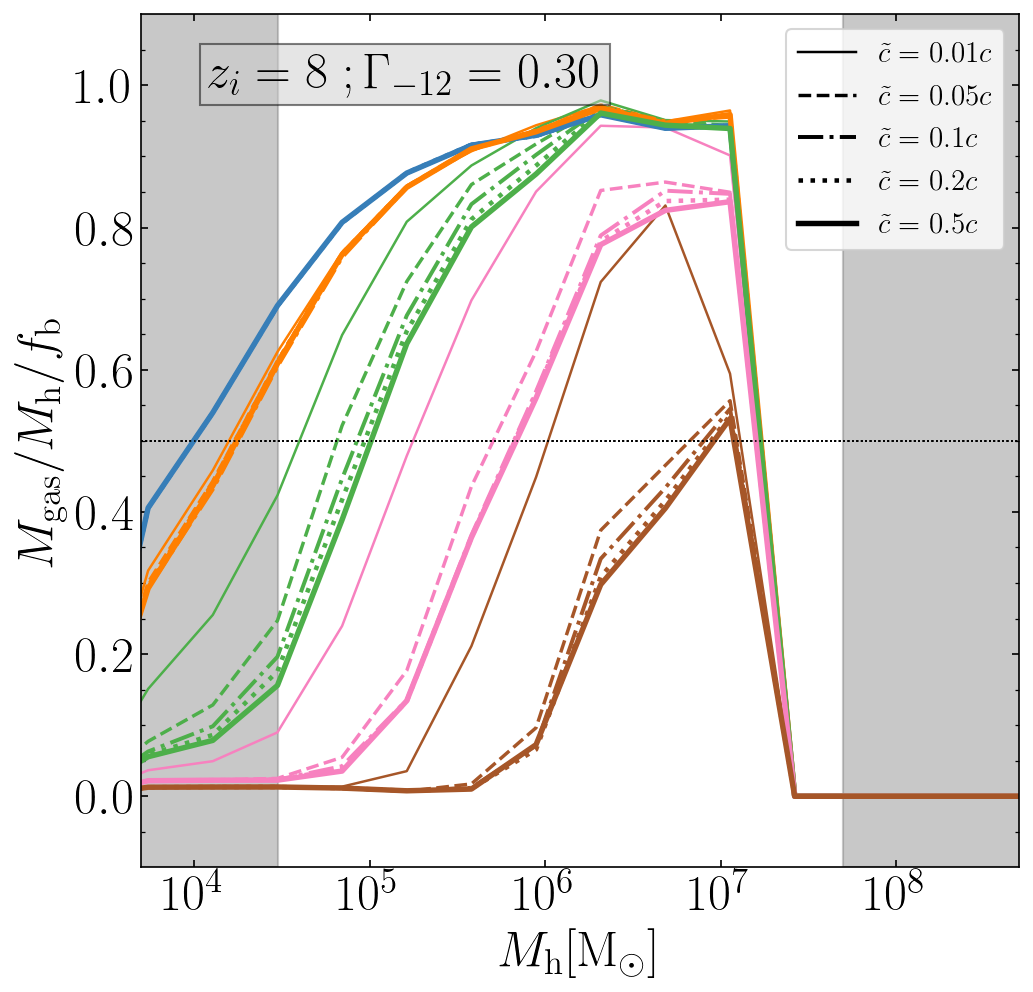}
\caption{Ratio between gas mass within the virial radius and the halo mass, in units of the cosmic baryon fraction, $f_b$. It is similar to Fig.~\ref{fig:halocatunheated} but for simulations performed with different values of the reduced speed of light $\tilde{c}$, as per the legend. Different colours correspond to different times, $t_{\rm dur}$, since the halo was overrun by \Ifront, from {\em blue} (recently) to {\em purple} (long ago) (see Fig.~\ref{fig:halocatunheated} legend).  The baryon fractions converge to better than 10~per cent when $\tilde{c}>0.05c$, and significantly better when $\tilde{c}>0.1$.}
\label{fig:credtest_Hfrac}
\end{figure}

We use a reduced speed of light (\rsl) approximation, $c\to\tilde{c}\ll c$, in the simulations: this improves computational efficiency. The value of $\tilde{c}$ is further allowed to vary spatially: this is the variable speed of light (\vsl) approximation, described in Appendix~\ref{sec:vsl}. 

We test by how much we can reduce $\tilde{c}$ in the \vsl\ approximation and still capture accurately the evolution of the clumping factor $c_r$ in Fig. \ref{fig:credtest}. The value of $c_r$ at its peak
is constant to within 10-15~per cent provided $\tilde{c}\geq 0.1~c$. This justifies our
choice of $\tilde{c}=0.15~c$ in the main simulations. The asymptotic value of $c_r$ does not dependent on the choice of $\tilde{c}$, provided $\tilde{c}\geq 0.05$.

\subsection{Convergence with particle mass}
\label{sec:restest}
\begin{figure}
\includegraphics[width=0.45\textwidth]{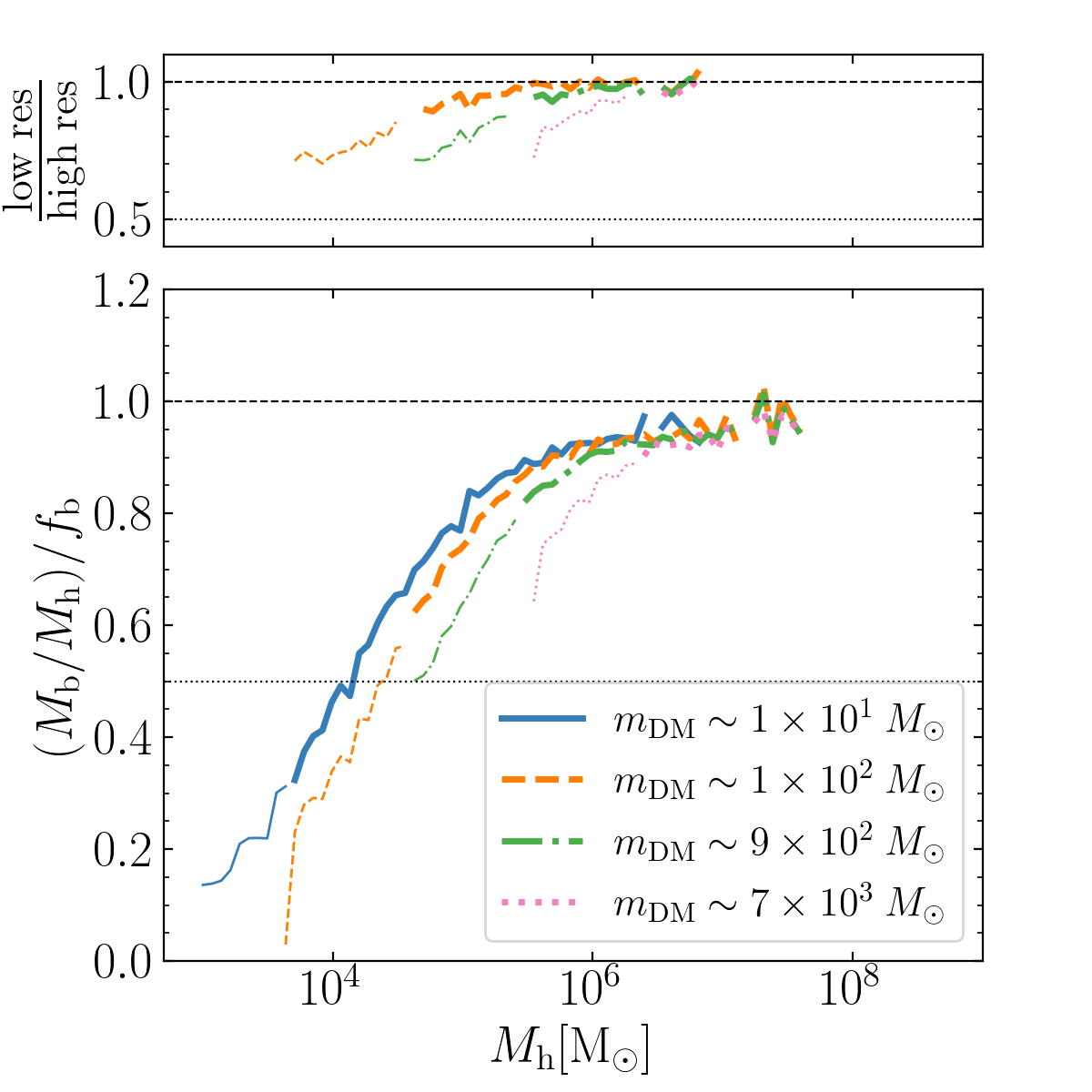}
\caption{{\em Lower panel:} Baryon fraction of halos in units of the cosmic mean, $f_b$, 
for simulations with different particles masses (colours). {\em Upper panel:} the baryon mass divided by the baryon mass of the highest resolution simulation. In both panels, {\em thick} ({\em thin}) lines refer to halos resolved with more than (less than) 300 DM particles. Results are shown at redshift $z=8$, before radiation was injected in the simulations. The baryon fraction is converged to within 20-30~per cent, provided the halo is resolved with more than $\sim 300$ particles.}
\label{fig:restest_barfrac}
\end{figure}

\begin{figure}
\includegraphics[width=0.45\textwidth]{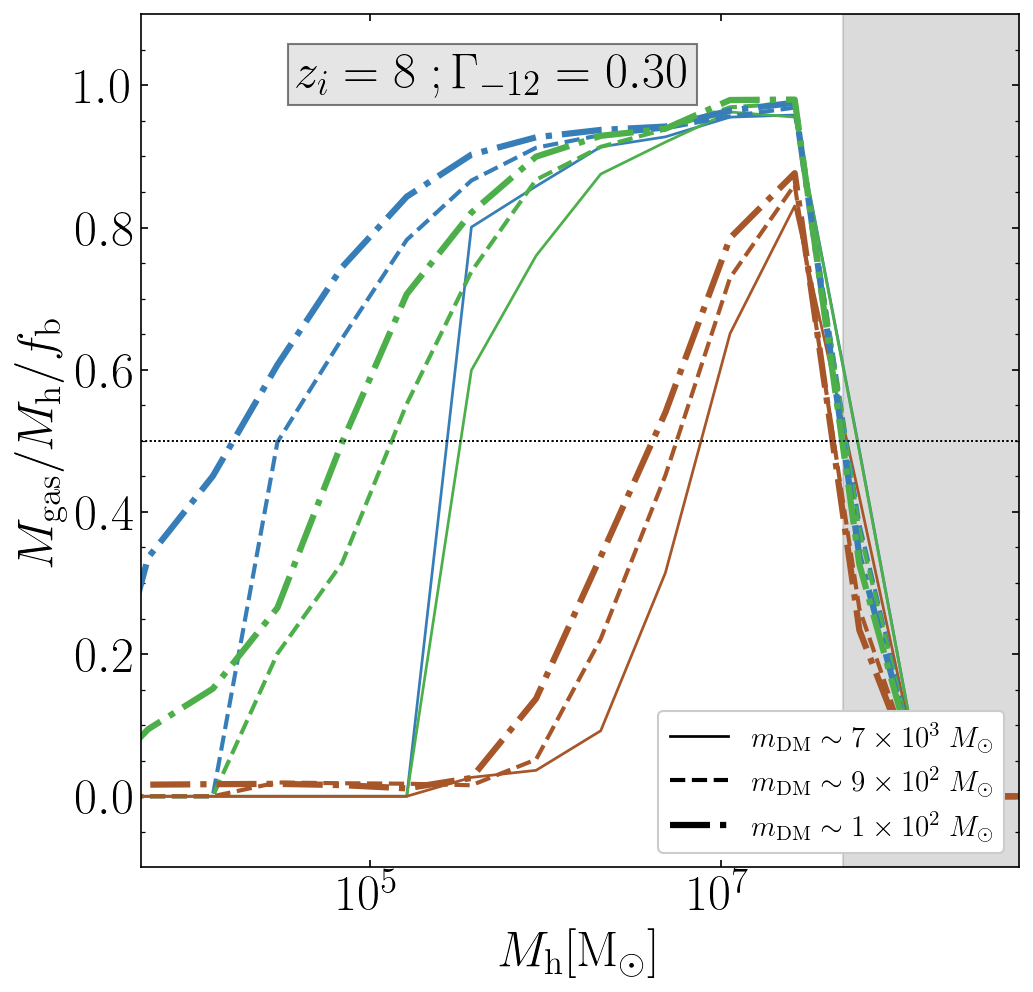}
\caption{Ratio between gas and halo mass, in units of the cosmic baryon fraction, $f_b$. It is similar to Fig.~\ref{fig:halocatunheated}, but for simulations with different particles masses
(plotted using different line styles). Different colours correspond to different times, $t_{\rm dur}$, since the halo was over run by \Ifront:  {\em blue}: 1 Myr; {\em green}: 20 Myr; {\em brown}: 100 Myr. The simulation volume is 800~ckpc on a side. Higher resolution simulations yield longer photo-evaporation time scales at a given halo mass.}
\label{fig:restest_Mfrac}
\end{figure}

\begin{figure}
\includegraphics[width=0.45\textwidth]{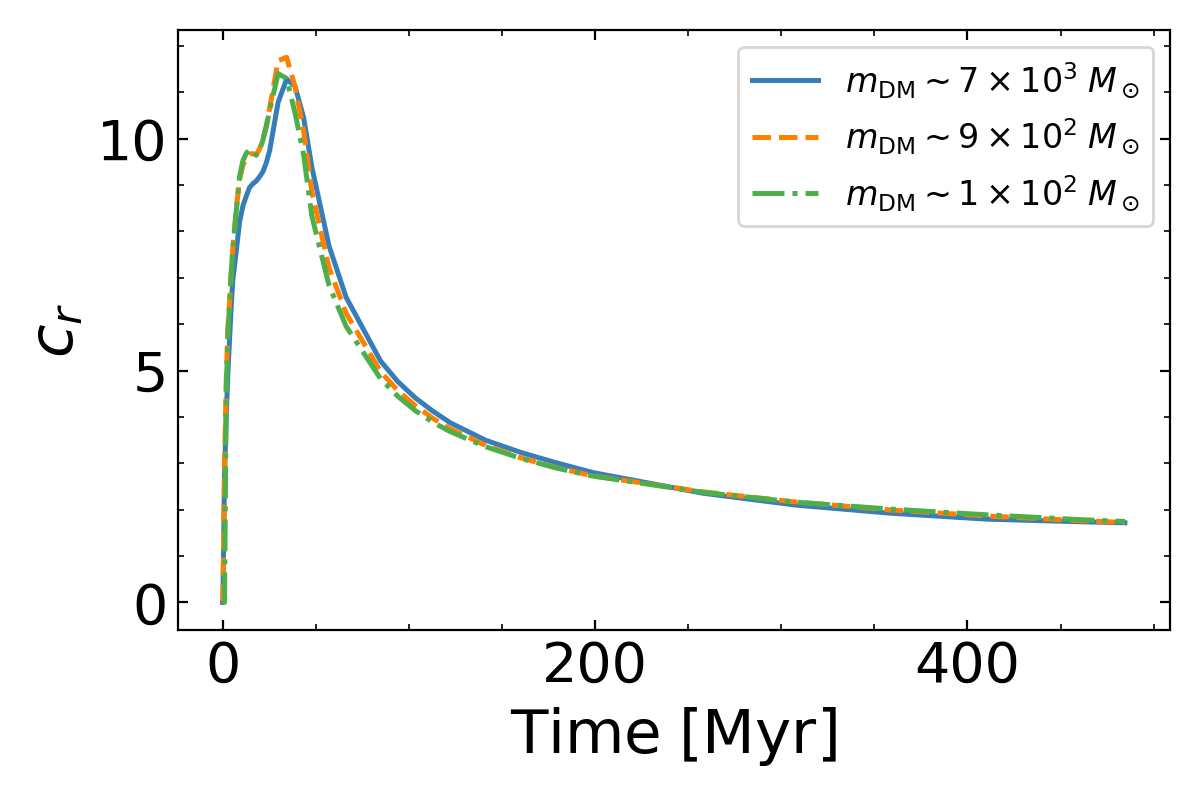}
\caption{As Fig.~\ref{fig:credtest_Hfrac}, but for simulations with different particles masses.
The simulation volume is 800~ckpc on a side, and we use $\Gamma_{-12}=0.3$, and $z_i=8$. The evolution of the clumping factor differs by less than 5~per cent between these runs.}
\label{fig:restest}
\end{figure}

\begin{figure}
\includegraphics[width=0.45\textwidth]{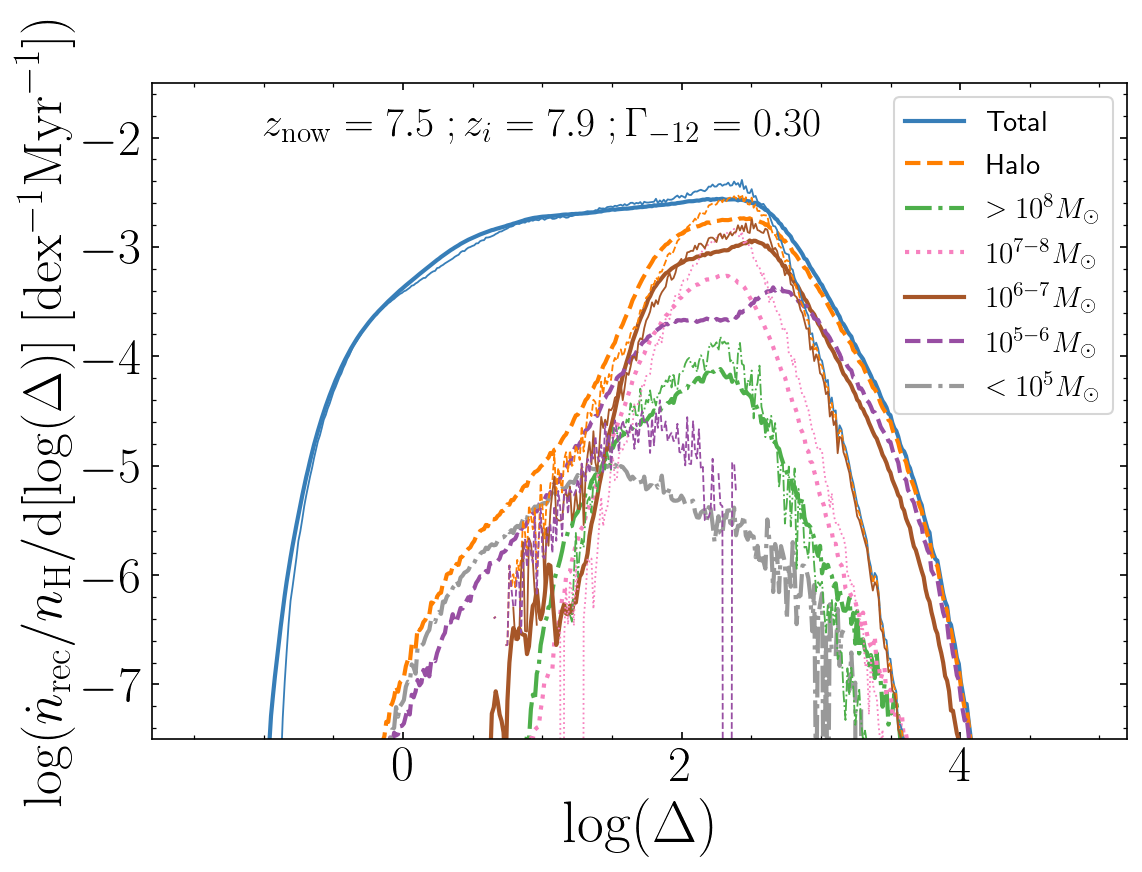}
\caption{As Fig.~\ref{fig:Recom_mbins}, showing the contribution to the recombination rate as a function of over-density with coloured lines depicting the net contribution of halos of different mass. Results are shown at the instance that $c_r=c_{r, {\rm peak}}$. The {\em thick lines} are for simulation M512z8G03 (particle mass $m_{\rm DM}=120\;\msun$), the {\em thin lines} are simulation M128z8G03 (with $m_{\rm DM}=9600\;\msun$). The simulations differ significantly for halos with with mass $M_h<10^6\msun$, but these do not contribute much to the over all recombination rate. For halos more massive than $M_h=10^6\msun$, the lower resolution simulation
yields {\em more} recombinations at higher density. This causes the over all recombination rate to be higher at lower resolution. }
\label{fig:restest_recom}
\end{figure}

We have performed simulations with different mass resolutions in the same cosmological volume to verify the level of numerical convergence, see Table~\ref{table:sim}. In Fig.~\ref{fig:restest_barfrac}, we plot the baryon fraction (in units of the cosmological mean value, $f_b$) as a function of halo mass, for simulations with different particle masses\footnote{The legend gives the dark matter particle mass, the gas-particle mass is $f_b/(1-f_b)$ time the dark matter particle mass.}. In our simulations, the baryonic mass fraction of halos converges to within $\sim 30$~per cent for halo mass $M_{\rm h} \gtrsim 10^{4.3}\msun$ provided that the dark matter particle mass $m_{\rm DM}\sim100\;\msun$ (200 dark matter particles per halo). This agrees well
with the findings of \cite{Naoz09mhgf}, who found that $\sim 300$ particles are needed to reach convergence at the 30~per cent level. Given our fiducial resolution of $m_{\rm DM}\sim100\;\msun$, our simulations resolve the 
baryon fraction of halos of mass $M_{\rm min}$ to within 30 per cent. Here, $M_{\rm min}$ is the minimum mass of halos that retain 50~per or more of their baryons, as discussed in \S\ref{sec:halocat}.

We examine the impact of finite mass resolution on the photo-evaporation of \mini\ in Fig.~\ref{fig:restest_Mfrac}. At a given halo mass, \mini\ evaporate faster at lower resolution.
This is partially because the central gas density is lower when the profile is not well-resolved, which enables the \Ifront\ to propagate deeper into the halo.  This effect is less pronounced
at higher halo masses,  $M_h>5\times 10^6\;\msun$, because these are reasonably well resolved
at our coarsest numerical resolution of $m_{\rm DM}\sim 6400\;\msun$.

The impact of resolution on the evolution of $c_r$ is illustrated in Fig.~\ref{fig:restest}. Simulations with different resolutions agree well and the clumping factors converge at $m_{\rm DM}\lesssim 800\;M_\odot$. Somewhat surprisingly, this conclusion is different from that reached by \cite{Embe13clumping}, who found that the clumping factor converges only at $m_{\rm DM}\lesssim100\;M_\odot$. We suspect that the reason is that \cite{Embe13clumping} did not account for the photo-evaporation of \mini. Lower-mass \mini\ halos contribute significantly to $c_r$ but only in the very early stages of the \eor. Once this photo-evaporate, $c_r$ is determined by more massive \mini\ halos which can be sufficiently resolved even when $m_{\rm DM}=800\;M_\odot$.

We verify this scenario in Fig.~\ref{fig:restest_recom} in which we plot the recombination rate as a function of over-density at the instance when $c_r$ is near its peak. Although the contribution of halos with mass $M_h<10^6\msun$ depends on resolution, they do not contribute significantly to the recombination rate at this stage of the evolution (This conclusion will depend on $z_i$, since lower-mass halos contribute more at higher $z$.). The value of $c_r$ is dominated by halos with mass in the range $10^{6-8}\msun$ halos, which can be resolved sufficiently well at a particle resolution of $m_{\rm DM}\lesssim10^3\msun$ (at least in the ionized outer regions)\footnote{The lower resolution run has slightly higher recombination rate because the ionization fronts propagate faster at lower resolution. It is because low-resolution runs have fewer structures to impede ionization fronts.}. At low $\Gamma_{-12}$ and/or late times, the clumping factor is dominated by 
gas at over-densities $\Delta<100$, and the recombination rate of this gas
is resolved well when $m_{\rm DM}< 10^3\;\msun$.

However, \mini\ need to be resolved spatially as well. A mini-halo
with $M_h=10^6\,\msun$ has a virial radius of $\sim 0.3~{\rm pkpc}$ at $z\sim 8$. For a quasi-Lagrangian code such as \swift,
the smoothing length of a gas particle is less than 0.05~kpc at the critical self-shielding density ($n_S\sim 0.1 \;{\rm cm^{-3}}$) and even smaller at higher density, when $m_{\rm DM}\lesssim 800\,\msun$ (and when using equal numbers of dark matter and gas particles). Therefore spatial resolution is not a limiting factor in our calculations.

In contrast, spatial resolution may be challenging in mesh-based simulations, e.g uniform mesh methods might have difficulties resolving the central parts of \mini\ (see e.g. the convergence test in the appendix of \citealt{DAlo20clumping}). 
Artificial diffusion can be problematic when the gas moves at high speed across a static mesh \citep{Robe10gridcodeerror}. This can suppress the formation of
structures that are not well resolved \citep{Pont20gridnoise}. Whether and how this affects the value of $c_r$ in mesh simulations may require further study.

In summary, a dark matter particle resolution of $m_{\rm DM}\lesssim 100 \;\msun$ is needed to model the photo-evaporation of \mini\ (provided that the adaptive spatial resolution is high enough)\footnote{Here we assume that the \ihm\ gas follows the adiabatic cooling limit (Eq.\ref{eq:Tihm}), which is a lower limit to the \ihm\ temperature. In reality the \ihm\ will be hotter than this lower limit due to, e.g., X-ray pre-heating as suggested by 21cm observations\citep{HERA23IGMheating}. Thus, $M_{\rm min}$ might be higher and the mass resolution requirement less stringent.}. This can be relaxed to
$m_{\rm DM}\lesssim 10^3~\msun$ for studies of the clumping factor, because the lower-mass halos that are not well resolved at this coarser resolution contribute little to $c_r$.

\subsection{Convergence with simulation volume}
\begin{figure}
\includegraphics[width=0.48\textwidth]{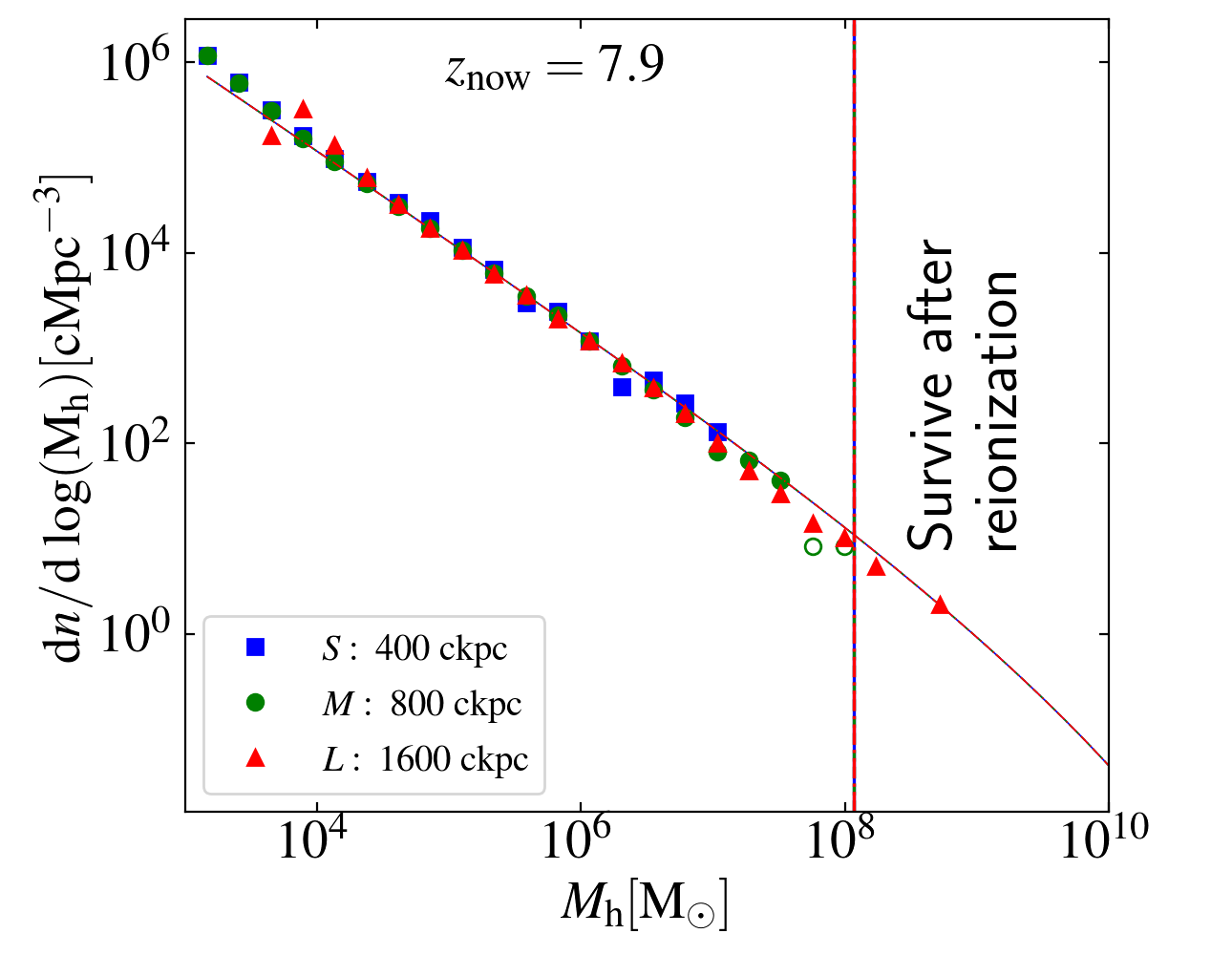}
\caption{Impact of the extent of the simulation volume on the halo mass function at redshift 7.9. {\em Colours and symbols} correspond to simulations performed in volumes of linear extent ranging from 400 to 1600~ckpc, as per the legend.
The empty symbols show bins with fewer than 5 halos per $\log_{10}(M)$. The {\em diagonal line} is the fit to the mass function from \protect\cite{Reed07}, the {\em vertical line}  shows $M_{\rm crit}$, the halo mass above which gas can be retained after reionization \protect\citep{Okam08}, as discussed in Section~\ref{sec:analytic}.  }
\label{fig:boxsizeNhalo}
\end{figure}

\begin{figure}
\includegraphics[width=0.48\textwidth]{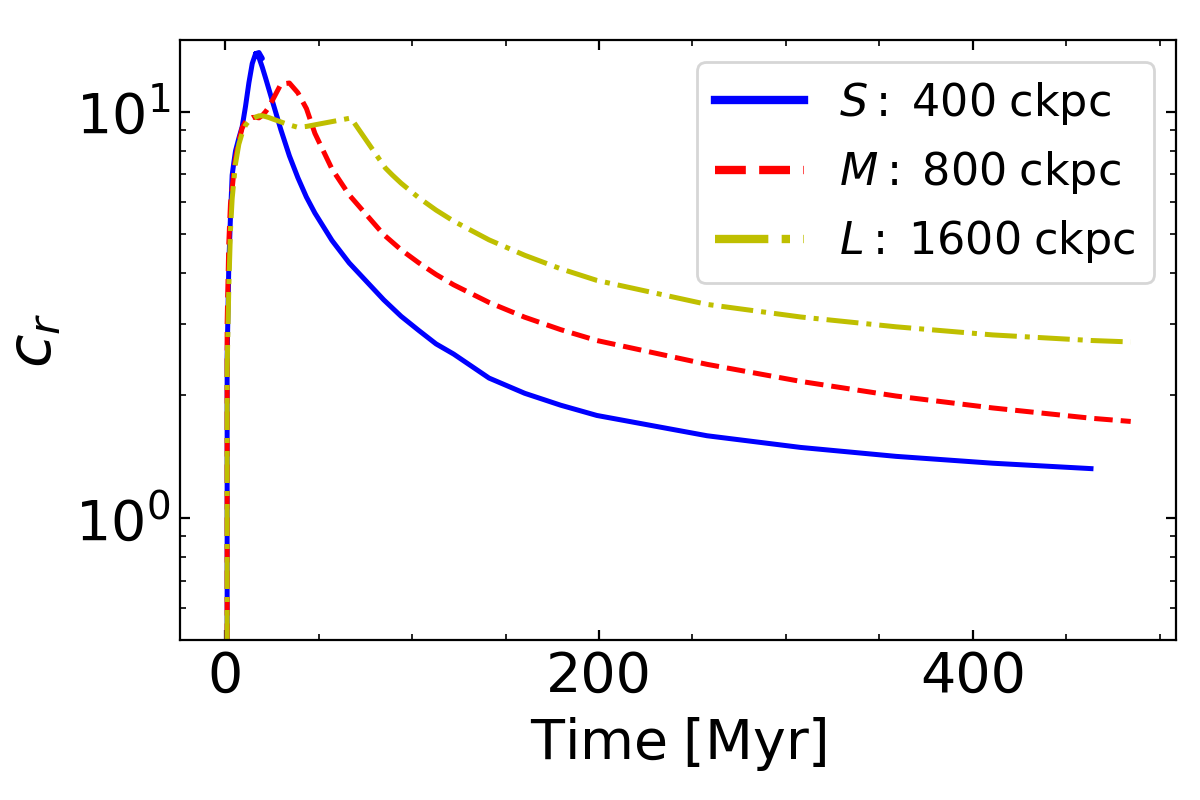}
\caption{Convergence of the clumping factor for different linear extents of the simulation volume. The mass resolution is $m_{\rm DM}=960\;\msun$, and we assumed $\Gamma_{-12}=0.3$, and $z_i=8$. The value of $c_r$ at its peak only depends weakly on volume, but the asymptotic value of $c_r$ increases from $\sim 2$ to $\sim 3$ when increasing the linear extent from 800 to 1600~pkpc. The figure also suggests that the asymptotic value of $c_r$ is underestimated, even in the largest volume.
}
\label{fig:boxsizetest}
\end{figure}

\begin{figure}
\includegraphics[width=0.48\textwidth]{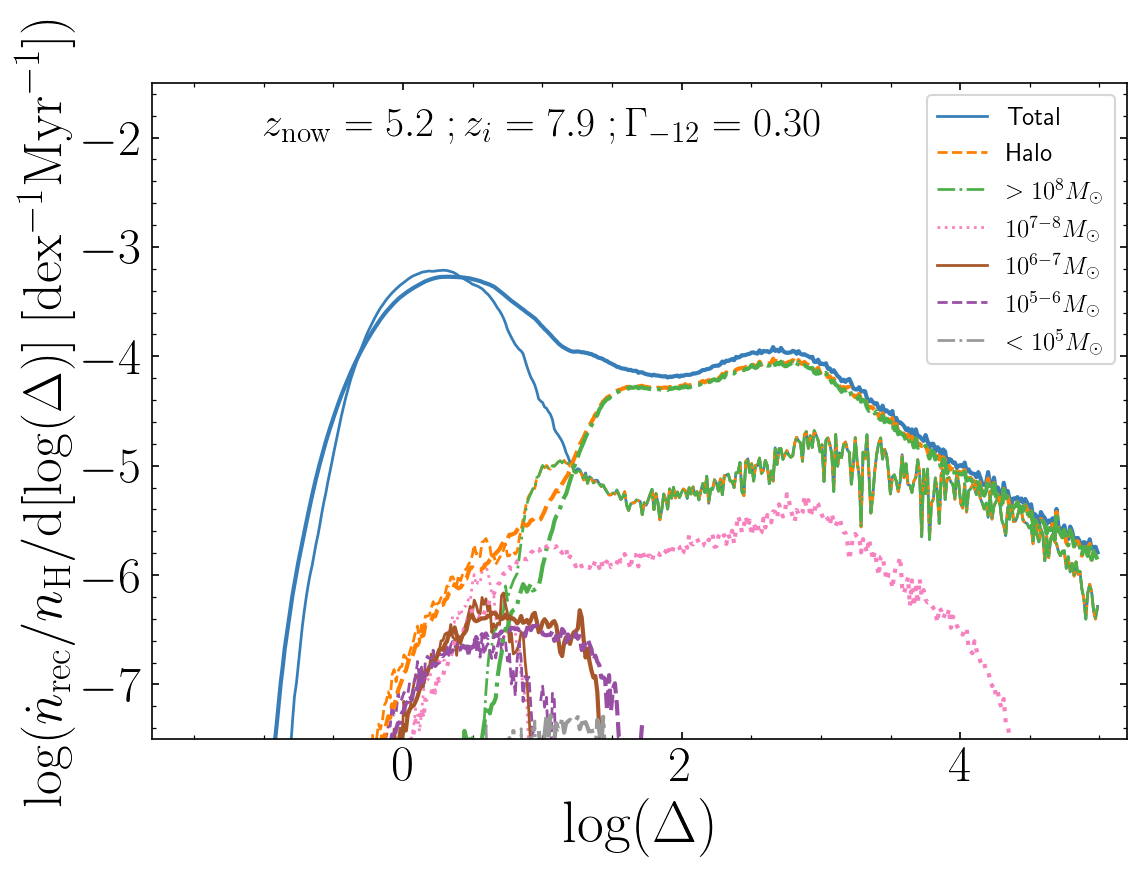}
\caption{As Fig.~\ref{fig:restest_recom}, but for simulations with different linear extent, $L_c$: 
$L_c=800$~ckpc (simulation $S$, {\em thin lines}) and
$L_c=1600$~ckpc (simulation $L$, {\em thick lines}).
The simulation particle mass is $m_{\rm DM}=800\;\msun$ for both. Simulation $L$ contains many more halos with mass $10^8~\msun$ compared to simulation $M$. Gas in those halos increases the recombination rate at $\Delta\sim 10^2$ significantly, as can be seen by comparing the {\em thick} and {\em thin green lines}.}
\label{fig:boxsizetest_recom}
\end{figure}

We quantify the impact of simulation volume on the halo mass function in Fig.~\ref{fig:boxsizeNhalo}. We consider three cubic volumes, $S$, $M$ and $L$, with linear extents $L_c\sim 400$, $\sim 800$ and $\sim 1600 \;{\rm ckpc}$ respectively. 
The numerical resolution is the same in these three runs.
Not surprisingly, the number of halos with mass $M_h>10^8~\msun $ is severely underestimated in simulation $S$. Such halos are significant photon sinks at the end of the \eor\ and later, and clearly the simulation volume needs to be large enough to sample them correctly (see also Appendix C of \citealt{Cain20streamingclumping}). Our production runs are performed in simulations with $L_c=800$~ckpc, (see Table~\ref{table:sim}) and therefore only begin to sample these more massive halos.

Fig. \ref{fig:boxsizetest} compares the evolution of $c_r$ for
these three simulations. The (first) value of $c_{r, {\rm peak}}$ is similar for all three runs (see also \citealt{Park16clumping}). This first maximum depends on \mini\ with mass $\gtrsim 10^6\;\msun$, which are well sampled even in simulation $S$. The second peak is higher when $L_c$ is smaller. The reason is that the ionization fronts that propagate from two opposing sides of the simulation cube overlap earlier for smaller values of $L_c$. At this earlier stage, $c_r$ is still higher, as \mini\ have had less time to photo-evaporate. The late time clumping factor is 50\% higher in the simulation $L$ compared to $M$, and a factor of two compared to $S$. This suggests that even our fiducial volume ($M$) does not contain enough (massive) collapsed structures after photo-evaporation to compute $c_r$ accurately.

We examine the reason for the relatively poor convergence with $L_c$ in more detail in Fig.~\ref{fig:boxsizetest_recom}. 
The ratio of more massive to lower mass halos is larger
in simulation $L$ compared to $M$ (Fig.\ref{fig:restest_recom}). The gas in and surrounding these more massive halos contributes significantly to the recombination rate and hence $c_r$ in simulation $L$. This effect causes the dependence of $c_r$ post-reionization on $L_c$.

After reionization, the minimum halo mass that can contribute significantly to $c_r$ is close to the critical mass $M_{\rm char}$ from \cite{Okam08} (as discussed in Section~\ref{sec:halocat}, the minimum mass above which halos retain 50~per cent or more of their cosmological baryon fraction). This mass is around $10^8\msun$ at $z=8$ and Fig.~\ref{fig:boxsizeNhalo} shows that only simulation $L$ samples a reasonable number of such halos. This is not surprising: the parameters for our production runs were a trade-off between mass resolution and simulation volume for a given number of simulation particles and hence computation time. Our value of
$c_r\lesssim 3$ post-reionization in simulation $L$ agrees well with the value of $c_r\sim 3$ in the  $L_c=2\;h^{-1}{\rm cMpc}$ simulation of \cite{DAlo20clumping}, and the range of $c_r=2.4-2.9$
found at $z=6$ in the simulations with $L_c=40\;h^{-1}{\rm cMpc}$ by \cite{McQu11LLS}.

Our convergence requirement on $L_c$ is more stringent than that of \cite{Embe13clumping}, who obtained differences of less than 5~per cent in $c_r$ for simulations with $L_c=1\;{\rm cMpc}$ and $L=0.5\;{\rm cMpc}$. We suggest that the difference stems from the fact that \cite{Embe13clumping} did not account for photo-evaporation and hence are unduly biased
to low-mass mini-halos, which dominate the recombination rate. Such low-mass halos are already sampled reasonably well in simulations with smaller $L_c$.

It would be interesting to extend the types of simulations performed here to larger volumes and investigate the impact of even more massive halos. Such simulations could then also be used to calculate the evolution of the attenuation length of ionizing photons. One aspect that such simulations should take into account is that these more massive halos may host a galaxy, and the feedback from the galaxy's stars may affect the distribution of neutral gas inside and outside the halo.

\section{I-front trapping in a minihalo}
\label{sec:minihalotest}
\begin{figure}
\includegraphics[width={0.45\textwidth}]{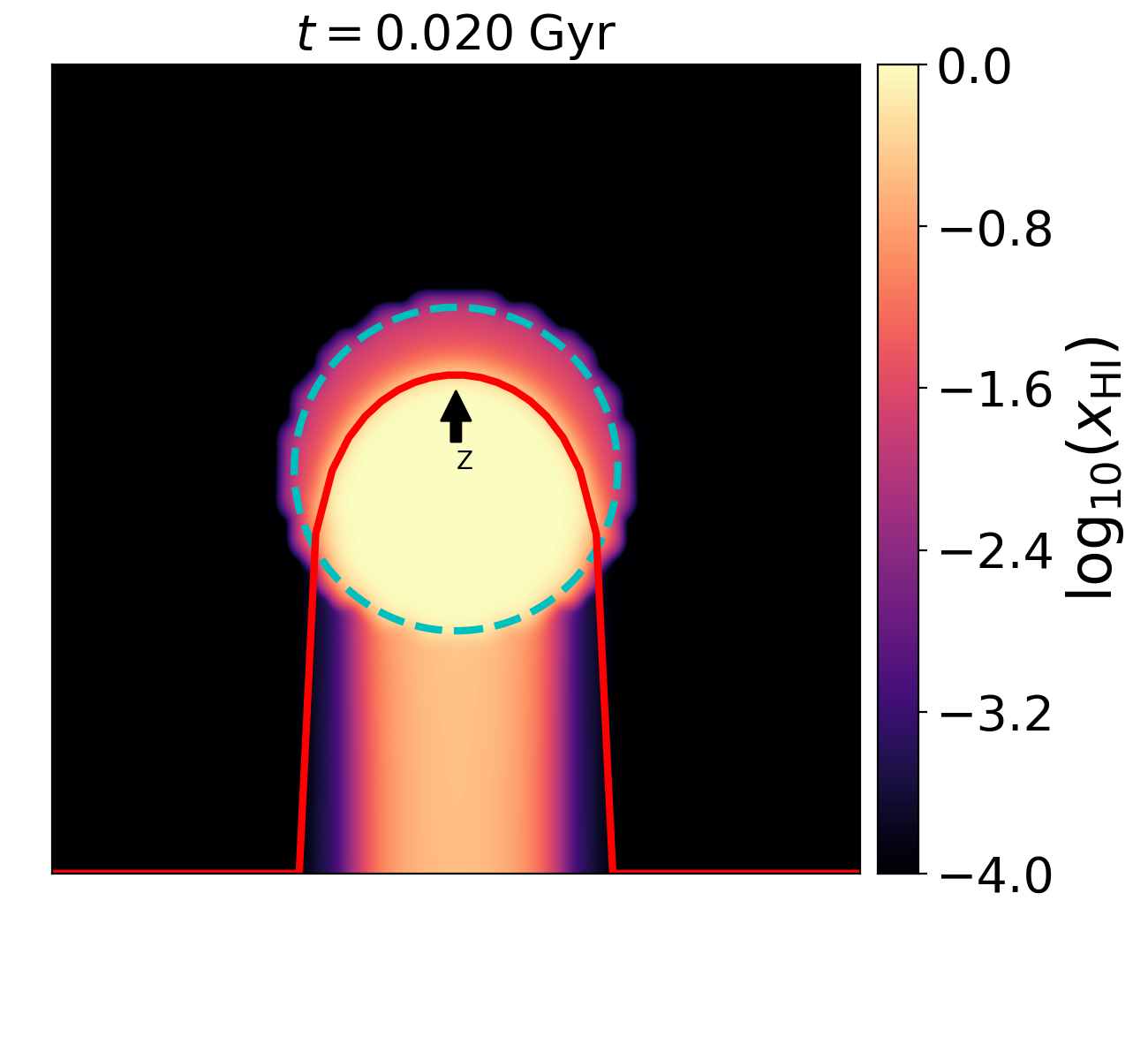}
\includegraphics[width={0.45\textwidth}]{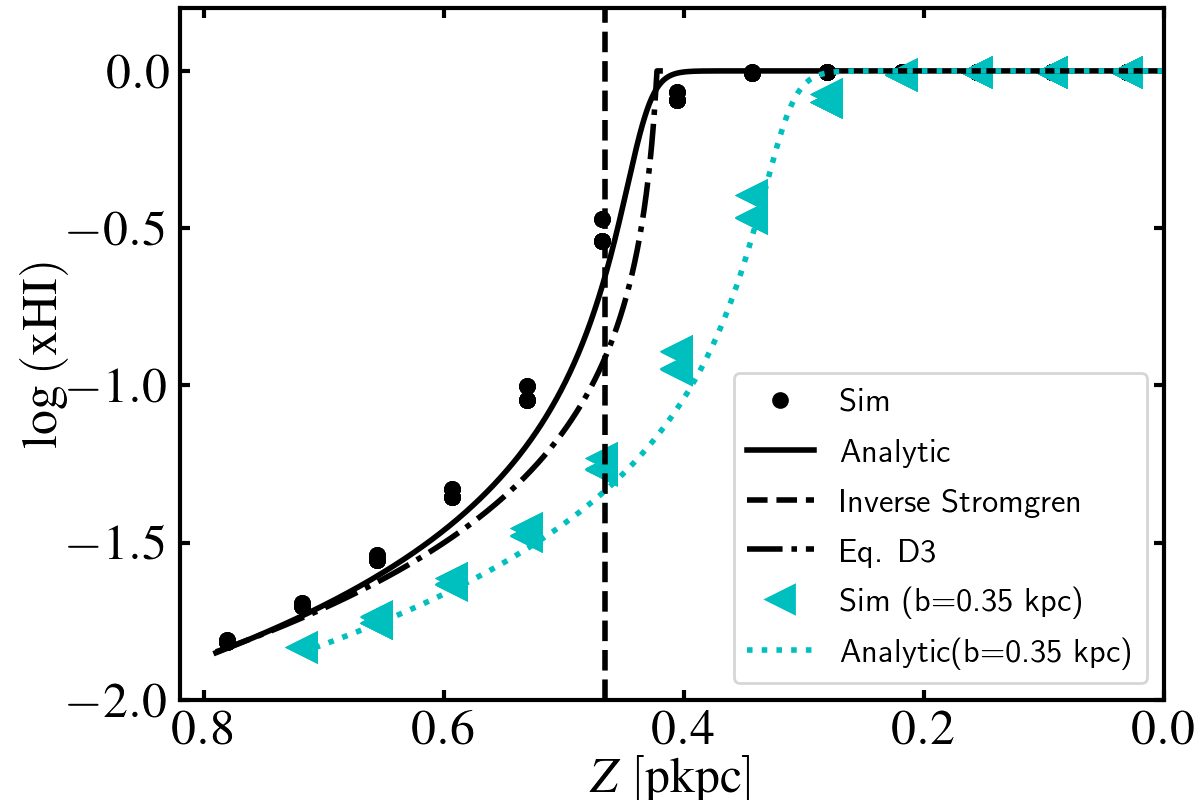}
\caption{Ionization front trapping by a static gas cloud with
a $1/R^2$ density distribution: the initially planar ionization front moves vertically down.{\em Upper panel}: slice through the centre of the sphere, with colours indicating neutral fraction. The {\em dashed blue line} is the radius of the sphere, the {\em solid red line} is the approximate location of the inverse Str\"omgren layer ({\sc isl}) from Eq.~(\ref{eq:mhpowerstromgren}). The arrow indicates the origin and the positive $Z$ direction.  {\em Lower panel}: neutral fraction $x_{\rm HI}$ as a function of $Z$, where the $Z$ axis runs vertically up from the centre of the sphere (as indicated in the upper panel). The {\em black circles} and {\rm cyan triangles} are obtained from the simulation with impact parameters $b=0~{\rm pkpc}$ and $b=0.35~{\rm pkpc}$ respectively. The {\em black solid line} and {\em cyan dotted line} are obtained by numerically integrating Eq.~(\ref{eq:mhpowerode}) with impact parameters $b=0~{\rm pkpc}$ and $b=0.35~{\rm pkpc}$ respectively. The vertical {\em dashed black line} is the location of the {\sc isl} from Eq.~(\ref{eq:mhpowerstromgren}) for an impact parameters $b=0~{\rm pkpc}$. Finally, the black dashdot line represents the approximate neutral fraction (Eq.~ \ref{eq:xb_approx}). The simulation follows the analytical result very closely. The text describes the simulation set-up in detail.}
\label{fig:minihalopowerlawstatic}
\end{figure}

Here we test the ability of our \rt\ implementation to trap an \Ifront\ in a halo by comparing to analytical solutions. These are based on the model presented in Section~\S\ref{sec:shielding}, except that here we assume that an initially planar \Ifront\ over runs a halo, rather than that the halo is illuminated by ionizing radiation from all directions.

Consider therefore an initially planar \Ifront, propagating along the $Z$ axis 
from $Z>0$ to $Z<0$, and encountering a mini-halo located at the origin of the coordinate system ($Z=0$). The mini-halo gas is modelled as a singular isothermal profile of Eq.~(\ref{eq:prof}). We assume that the gas density outside the halo's virial radius $R_h$ is low enough such that the attenuation is negligible (but see \S\ref{sec:shielding}). All of the gas is assumed to be static and has a uniform and constant temperature (hence, the recombination coefficient $\alpha_{\rm B}$ is also uniform and constant). 

There are two ways to calculate the equilibrium neutral hydrogen profile. First, we consider the {\it inverse Str\"omgren Layer (ISL) approximation} under which gas downstream from the \Ifront\ is fully ionized and upstream is fully neutral. With this approximation, we can compute the total recombination rate, ${\cal R}$, along a line parallel to the $Z$ axis at impact parameter $b$. Equating ${\cal R}$ to the flux incoming flux yields the location $Z_{\rm ISL}(b)$
of the inverse Str\"omgren layer, where gas transits from ionized to neutral. $Z_{\rm ISL}(b)$ can be solved through the following equation,
\begin{align}
F &=\alpha_{\rm B}n_{\rm H,h}^2\,R^4_h\int^{Z_h}_{Z_{\rm ISL}}\frac{{\rm d}Z'}{(b^2+Z'^2)^2}\nonumber\\
&=\alpha_{\rm B}n_{\rm H,h}^2\,\frac{R^4_h}{2b^3}
\left.\left(\arctan(z)+\frac{z}{1+z^2}\right)\right|^{z=Z_h/b}_{z=Z_{\rm ISL}/b}\,,
\label{eq:mhpowerstromgren}
\end{align}
where $Z_h=(R^2_h-b^2)^{1/2}$. The sign of $Z_{\rm ISL}$ can be positive or negative. 

Ignoring the optical depth of the neutral gas downstream, we can find an approximate expression for the neutral fraction $x\equiv n_{\rm HI}/n_{\rm H}$ in photo-ionization equilibrium, \begin{align}
\frac{x}{(1-x)^2} &= \frac{\alpha_{\rm B}\,n_{\rm H}(Z,b)}{F_c}\,\nonumber\\
F_c &=\alpha_{\rm B}n_{\rm H,h}^2\,\frac{R^4_h}{2b^3}
\left.\left(\arctan(z)+\frac{z}{1+z^2}\right)\right|^{z=Z/b}_{z=Z_{\rm ISL}/b}
\end{align}

\begin{align}
    x(b, Z)&\approx \frac{n_{\rm H, h}\alpha_{\rm B}}{\sigma_{\rm HI}F_c}\,\frac{R^2_h}{Z^2+b^2}\nonumber\\
    &=\frac{2b^3}{\sigma_{\rm HI}n_{\rm H,h}R_h^2}\frac{1}{Z^2+b^2}\nonumber\\
    &\times\left[\left.\left(\arctan(z)+\frac{z}{1+z^2}\right)\right|^{z=Z/b}_{z=Z_{\rm ISL}/b}\right]^{-1} \,   
    \label{eq:xb_approx}
\end{align}
where $x_h$ is the neutral fraction at $R_h$\,.

Second, we can {\it drop the ISL approximation} and calculate the neutral gas distribution considering the optical depth.
Assuming photo-ionization equilibrium, we calculate the photon number density $n_\gamma$:
\begin{align}
\frac{\partial n_\gamma}{\partial t}=-\frac{\partial f_\gamma}{\partial Z}-n_{\rm HI}c\sigma_{\rm HI} n_\gamma=0\,,
\end{align}
where $f_\gamma$ is the photon flux which is approximately $f_\gamma=n_\gamma c$ if we ignore scattering. Then we solve $n_\gamma$ by integration
\begin{align}
n_\gamma(Z)=\frac{F}{c}\exp\left (-\int_{Z}^{Z_h}
n_{\rm HI}\sigma_{\rm HI}{\rm d}Z' \right )\,.
\end{align}

Substituting this relation into the equation expressing photo-ionization equilibrium:
\begin{align}
n_{\rm HI}c\sigma_{\rm HI} n_\gamma=n_en_{\rm HII}\alpha_{\rm B}\,,
\end{align}
yields an equation for the neutral fraction, $x$,
\begin{align}
xc\sigma_{\rm HI}\frac{F}{c}\exp(-\tau)=(1-x)^2n_{\rm H}\alpha_{\rm B},
\end{align}
where the optical depth is given by 
\begin{align}
\tau=\int^{Z_h}_Zx n_{\rm H}\sigma_{\rm HI}{\rm d}Z'\,.
\end{align}
We follow \cite{Alta13revrt} and take the logarithm on both side and differentiate with respect to $Z$,
\begin{align}
\left(\frac{1}{x}+\frac{2}{1-x} \right )\frac{\mathrm{d} x}{\mathrm{d} Z}= - xn_{\rm H}\sigma_{\rm HI}+\frac{\mathrm{d} \ln n_{\rm H}}{\mathrm{d} Z}.
\end{align}
We then simplify and obtain the final differential equation
\begin{align}
\frac{\mathrm{d} x}{\mathrm{d} Z}=-\frac{x(1-x)}{1+x}\left (xn_{\rm H}\sigma_{\rm HI}+\frac{2Z}{b^2+Z^2} \right )\,.
\label{eq:mhpowerode}
\end{align}
with boundary condition $x\rightarrow x_h$ when $Z\rightarrow Z_h$. We can integrate this equation numerically to find $x(Z, b)$. This more accurate expression is actually quite similar to the approximate, analytical solution of
Eq.~(\ref{eq:xb_approx}) (see the lower panel of Fig.\ref{fig:minihalopowerlawstatic}).

We now perform a simulation with the following numerical set-up.
The computational volume has linear extent of $L_c=6$~pkpc, and is filled with particles such that the mean density is 
$\langle n_{\rm H}\rangle_{\rm out}=2\times 10^{-4}{\rm cm^{-3}}$. The sphere is located at the centre of the simulation volume, has radius $R_h=0.8$~pkpc and its density is normalized by $n_{\rm H, h}=(200/3)\langle n_{\rm H}\rangle_{\rm out}$ so that its mean density is 200 times the surrounding density (and hence mimicking a cosmological halo). This density profile is realised with $\approx 2000$ particles, this is equivalent to the resolution of a halo of mass $2.4\times 10^5~\msun$ in our main production runs (Table~\ref{table:sim}). We neglect
hydrodynamics and keep the temperature of the gas constant and uniform, using a recombination coefficient of $\alpha_{\rm B}=2.59\times10^{-13}{\rm cm^3~s}^{-1}$ everywhere.
We inject radiation from one side of the cubic simulation volume, with constant photon flux of $F=1.5\times10^5{\rm cm^{-2}s^{-1}}$, and use $\sigma_{\rm HI}=8.13\times10^{-18}{\rm cm}^2$. Finally, we impose that the optically thin direction of the Eddington tensor is parallel to the $Z$-axis \citep{Chan21SPHM1RT} (which is perpendicular to the side through which we inject radiation), and use the reduced speed of light value of $\tilde{c}=0.01c$.

The results of the simulation are compared to the analytical solution in Fig.~\ref{fig:minihalopowerlawstatic}, at $20~$~Myr after radiation was first injected. The upper panel is a slice through the centre showing the neutral fraction in the simulation, with the solid red line the approximate location of the inverse Str\"omgren layer from Eq.~(\ref{eq:mhpowerstromgren}). The lower panel compares the simulation results (black circle symbols) to the more accurate expression found by integrating Eq.~(\ref{eq:mhpowerode}) numerically (black solid line). The agreement between the simulation and the analytical solution is within a few percentages, demonstrating the accuracy of the numerical scheme. This close agreement also suggests that the simulation can accurately account for \Ifront\ trapping in \mini\ with mass $M_h=2.5\times 10^5~\msun$ and higher.

\section{Detailed response to the passing of ionisation fronts}
\label{partracking}

\begin{figure*}
\includegraphics[width={1.0\textwidth}]{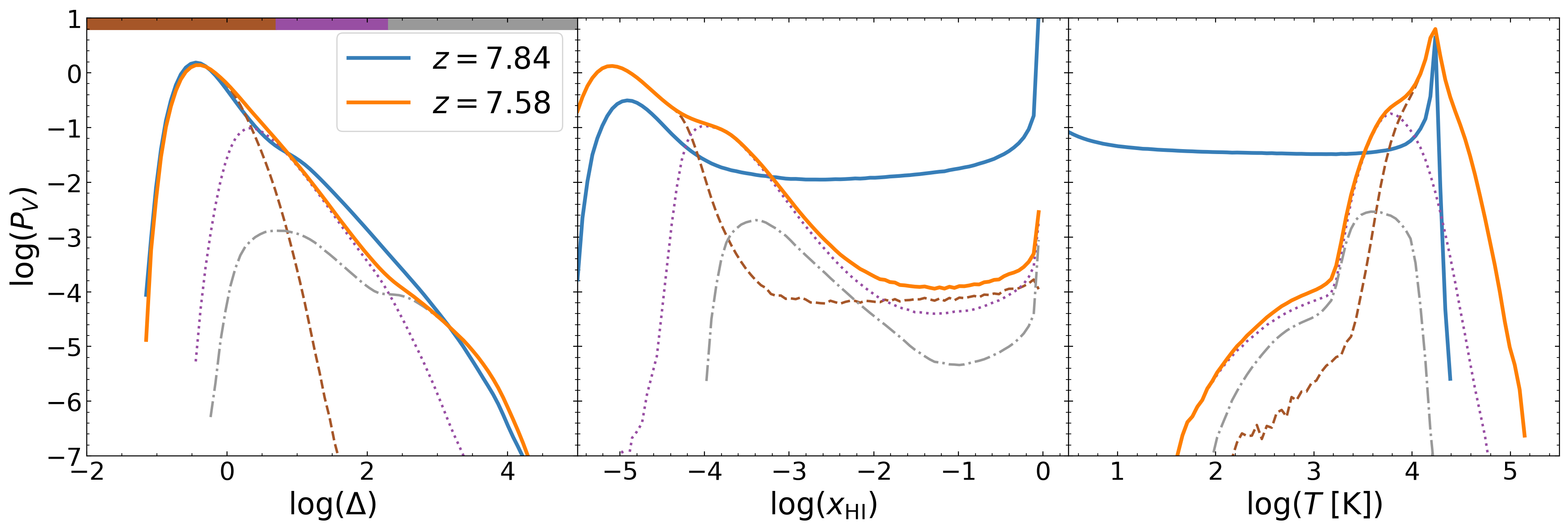}
\caption{Probability distribution by volume, ${\cal P}_V$, of the gas over-density, $\Delta(=\rho/\langle\rho\rangle)$, neutral fraction, $x_{\rm HI}$, and gas temperature, $T$ ({\em left to right}), for two different redshifts as labelled, for M512z8G03 (the simulation shown in Fig.~\ref{fig:simoverview}). We select gas at the higher redshift, $z=7.84$, in three density ranges $\Delta<5$, $5<\Delta<200$ and $\Delta>200$,
as illustrated by coloured top patches in the left panel, and trace it to redshift $z=7.58$. The probability distribution by volume of this gas is plotted as {\em brown dashed}, {\em purple dotted} and {\em grey dash-dotted} lines in all  panels.}
\label{fig:simoverview_his}
\end{figure*}

Fig. \ref{fig:simoverview_his} shows the response of the \ihm\ to the passing of \Ifront\ for the simulation M512z8G03 quantitatively. The probability density distribution by volume, ${\cal P}_V$, is plotted at $z=7.84$ (blue line) and $z=7.58$ (orange line), as a function of\footnote{Since the horizontal axis is $\log\Delta$ rather than $\Delta$, the area under the curve is not proportional the fraction of volume at a given density. Rather, that would be $d{\cal P}/d\log\Delta\propto \Delta\,d{\cal P}/d\Delta$.} $\log\Delta$, $\log x_{\rm HI}$, and $\log T$. Photo-ionization reduces ${\cal P}_V$ at intermediate densities, $\log\Delta\sim 1-3$, due to photo-evaporation of the outskirts of \mini.

To clarify better how regions of different densities react, we select particles in three ranges of over-density  $\Delta\equiv\rho/\langle\rho\rangle$ at $z=7.84$, and track them to $z=7.58$.
We plot their \pdf's as a function of $\Delta$, $x_{\rm HI}$ and $T$ in Fig.~\ref{fig:simoverview_his}. 

The initially lowest-density gas ($\Delta<5$, brown dashed line) changes little in volume density due to reionization but becomes mostly highly ionized, $x_{\rm HI}\le 10^{-4}$, and is photo-heated to a temperature, $T\sim 2.2\times 10^4~{\rm K}$. A small fraction of this gas is only partially ionized and partially heated ($T<10^4~{\rm K}$). The initially highest density gas ($\Delta>200$, grey dot-dashed line) also changes relatively little, staying mostly dense, neutral and cold, as it self-shields from the \Ifront.
A small fraction by mass becomes ionized and heated, expanding to lower densities. The passage of the \Ifront\ impacts predominantly gas initially at intermediate densities ($5<\Delta<200$, purple dotted line): this gas expands significantly to lower densities as it is ionized and photo-heated.

Some further features of Fig.~\ref{fig:simoverview_his} are worth noting. Although most of the gas is photo-heated to $T\sim 2.2\times 10^4{\rm K}$, a fraction is hotter, reaching $T\sim 10^5{\rm K}$. This corresponds to initially low-density gas ($\Delta<5$) that is first photo-heated
by \Ifront\ and subsequently further adiabatically or shock-heated by the expansion of a nearby filament. This can be seen in Fig.~\ref{fig:simoverview}, where the hot gas appears as red lines delineating filamentary structures. This comparison also shows the origin of the slightly cooler gas of $\log T[{\rm K}]\sim 3.8$: this is initially intermediate density gas in filaments that cools adiabatically as the filaments expand.
Finally, there is a \lq knee\rq\ of gas at temperature $T\le 10^3{\rm K}$ below which gas has not been ionized and photo-heated.
\label{lastpage}
\bsp
\end{document}